\documentclass[fleqn,usenatbib]{mnras}

\usepackage[T1]{fontenc}
\usepackage{ae,aecompl}

\DeclareRobustCommand{\VAN}[3]{#2}
\let\VANthebibliography\thebibliography
\def\thebibliography{\DeclareRobustCommand{\VAN}[3]{##3}\VANthebibliography}

\usepackage{graphicx}	
\usepackage{amsmath}	
\usepackage{amssymb}	
\usepackage{color}
\usepackage{algorithm}
\usepackage{booktabs}
\usepackage[noend]{algpseudocode}
\usepackage{pifont}
\usepackage{soul}
\usepackage{float}
\usepackage{multibib}
\newcites{Appendix}{Appendix References}
\usepackage{natbib}
\usepackage[flushleft]{threeparttablex}
\usepackage{mathrsfs}
\usepackage{widetext}
\usepackage{mathtools}
\usepackage{tikz}
\usepackage{esint}
\usepackage[draft]{todonotes}
\usepackage{lipsum}
\usepackage{longtable}
\hypersetup{breaklinks=true}

\usetikzlibrary{calc,shapes}

\graphicspath{{Figures/}}

\newcommand{\appropto}{\mathrel{\vcenter{
  \offinterlineskip\halign{\hfil$##$\cr
    \propto\cr\noalign{\kern2pt}\sim\cr\noalign{\kern-2pt}}}}}

\newcommand{\vecB}[1]{\mathrm{{\bmath{\mathit{#1}}}}}
\newcommand{\Exp}[1]{\left\langle #1 \right\rangle}
\renewcommand{\cor}[1]{\ell_{\text{cor},#1}}

\renewcommand{\d}[1]{\ensuremath{\operatorname{d}\!{#1}}}

\DeclareMathOperator\V{\mathcal{V}}
\newcommand{\M}{\mathcal{M}}
\DeclareMathOperator\Ma{\mathcal{M}_{\text{A}}}
\DeclareMathOperator\Mao{\mathcal{M}_{\text{A0}}}
\DeclareMathOperator\Maturb{\mathcal{M}_{\text{A,turb}}}
\DeclareMathOperator\Matot{\mathcal{M}_{\text{A,total}}}
\newcommand{\MaO}[1]{\mathcal{M}^{#1}_{\text{A0}}}

\DeclareMathOperator\Bo{\vecB{B}_0}
\newcommand{\dB}{\delta\vecB{B}}
\DeclareMathOperator\dbBo{\dB\cdot\Bo}
\DeclareMathOperator\dbBovol{\Exp{(\dB\cdot\Bo)^{2}}^{1/2}_{\mathcal{V}}}
\newcommand{\dBpar}{\delta B_{\parallel}}
\newcommand{\dBperp}{\delta B_{\perp}}
\DeclareMathOperator\ekin{e_{\rm kin}}
\DeclareMathOperator\emag{e_{\rm mag}}

\DeclareMathOperator\sat{\alpha_{\rm sat}}
\DeclareMathAlphabet\mathbfcal{OMS}{cmsy}{b}{n}

\newcommand{\tens}{(\vecB{B}\cdot\nabla)\vecB{B}}

\defcitealias{Brunt2010a}{BFP2010}
\defcitealias{Beattie2020}{BF2020}
\defcitealias{Kolmogorov1941}{K41}
\defcitealias{Menon2020}{MFK2020}
\defcitealias{Hopkins2013}{H2013}
\defcitealias{Skalidis2021}{S+2021}

\usepackage{scalerel,tikz}
\usetikzlibrary{svg.path}
\definecolor{orcidlogocol}{HTML}{A6CE39}
\tikzset{orcidlogo/.pic={\fill[orcidlogocol] svg{M256,128c0,70.7-57.3,128-128,128C57.3,256,0,198.7,0,128C0,57.3,57.3,0,128,0C198.7,0,256,57.3,256,128z}; \fill[white] svg{M86.3,186.2H70.9V79.1h15.4v48.4V186.2z} svg{M108.9,79.1h41.6c39.6,0,57,28.3,57,53.6c0,27.5-21.5,53.6-56.8,53.6h-41.8V79.1z M124.3,172.4h24.5c34.9,0,42.9-26.5,42.9-39.7c0-21.5-13.7-39.7-43.7-39.7h-23.7V172.4z} svg{M88.7,56.8c0,5.5-4.5,10.1-10.1,10.1c-5.6,0-10.1-4.6-10.1-10.1c0-5.6,4.5-10.1,10.1-10.1C84.2,46.7,88.7,51.3,88.7,56.8z};}}
\newcommand\orcidicon[1]{\href{https://orcid.org/#1}{\mbox{\scalerel*{
\begin{tikzpicture}[yscale=-1,transform shape]\pic{orcidlogo};
\end{tikzpicture}}{|}}}}

\newcommand{\aref}[1]{\hyperref[#1]{Appendix~\ref{#1}}}

\title[MHD Energy balance]{Energy balance and Alfv\'en Mach numbers in compressible magnetohydrodynamic turbulence with a large-scale magnetic field}

\author[Beattie, et al., 2022]{
James R. Beattie$^{\orcidicon{0000-0001-9199-7771}\,1}$\thanks{E-mail: james.beattie@anu.edu.au}, 
Mark R. Krumholz$^{\orcidicon{0000-0003-3893-854X}\,1,2}$,
Raphael Skalidis$^{\orcidicon{0000-0003-2337-0277}\,3,4}$,
Christoph Federrath$^{\orcidicon{0000-0002-0706-2306}\,1,2}$,
\newauthor
Amit Seta$^{\orcidicon{0000-0001-9708-0286}\,1}$
Roland M. Crocker$^{\orcidicon{0000-0002-2036-2426}\,1}$,
Philip Mocz$^{\orcidicon{0000-0001-6631-2566}\,5}$,
and Neco Kriel$^{\orcidicon{0000-0002-3558-3926}\,1}$
\\
$^{1}$Research School of Astronomy and Astrophysics, Australian National University, Canberra, ACT 2611, Australia \\
$^{2}$Australian Research Council Centre of Excellence in All Sky Astrophysics (ASTRO3D), Canberra, ACT 2611, Australia \\
$^{3}$Institute of Astrophysics, Foundation for Research and Technology-Hellas, Vasilika Vouton, GR-70013 Heraklion, Greece\\
$^{4}$Department of Physics \& ITCP, University of Crete, GR-70013, Heraklion, Greece \\
$^{5}$Lawrence Livermore National Laboratory, 7000 East Ave, Livermore, CA 94550, USA \\
}

\date{Accepted XXX. Received YYY; in original form ZZZ}

\pubyear{2022}

\begin{document}
\label{firstpage}
\pagerange{\pageref{firstpage}--\pageref{lastpage}}
\maketitle

\begin{abstract}
    Energy equipartition is a powerful theoretical tool for understanding astrophysical plasmas. It is invoked, for example, to measure magnetic fields in the interstellar medium (ISM), as evidence for small-scale turbulent dynamo action, and, in general, to estimate the energy budget of star-forming molecular clouds. In this study we motivate and explore the role of the volume-averaged root-mean-squared (rms) magnetic coupling term between the turbulent, $\dB$ and large-scale, $\Bo$ fields, $\dbBovol$. By considering the second moments of the energy balance equations we show that the rms coupling term is in energy equipartition with the volume-averaged turbulent kinetic energy for turbulence with a sub-Alfv\'enic large-scale field. Under the assumption of exact energy equipartition between these terms, we derive relations for the magnetic and coupling term fluctuations, which provide excellent, parameter-free agreement with time-averaged data from 280 numerical simulations of compressible MHD turbulence. Furthermore, we explore the relation between the turbulent, mean-field and total Alfv\'en Mach numbers, and demonstrate that sub-Alfv\'enic turbulence can only be developed through a strong, large-scale magnetic field, which supports an extremely super-Alfv\'enic turbulent magnetic field. This means that the magnetic field fluctuations are significantly subdominant to the velocity fluctuations in the sub-Alfv\'enic large-scale field regime. Throughout our study, we broadly discuss the implications for observations of magnetic fields and understanding the dynamics in the magnetised ISM. 
\end{abstract}


\section{Introduction}\label{sec:intro}

    Magnetohydrodynamic (MHD) turbulence is pervasive across the Universe, and for this reason the study of MHD turbulence is a necessary prerequisite for understanding a broad range of astrophysical processes. For example, each of the planets in our Solar System probably assembled as the protoplanetary disc underwent hydrodynamical and magnetohydrodynamical (MHD) instabilities, driving turbulence and establishing the initial conditions for planet formation \citep[][and references therein]{Wladimir2019_planet_turb_review}. The Sun maintains a magnetised and turbulent heliosphere, with decades of scale-free velocity and magnetic fluctuations that play an important role in the generation of solar winds, plasma heating, and particle acceleration \citep[][and references therein]{Roberto2013_solar_turb_review}. Just like the planets, the Sun was born in a turbulent plasma environment.  
    
    In the context of star formation, turbulent density fluctuations in the cool molecular gas clouds of galaxies seed the over-densities that fragment, become gravitationally unstable, and collapse to form stars  \citep{Krumholz2005,Padoan2011,Hennebelle2011,Federrath2012,Federrath2015_inefficient_SFR,Hopkins2013a,Burkhart2018,Mocz2018}. The turbulent motions themselves steepen or flatten the initial mass function (IMF) of stars \citep{Padoan1997_imf,Hennebelle2009,Hopkins2012_imf,federrath2017_imf,Nam2021} and potentially underlie the universality that we observe for the IMF by setting the density correlation scale for star-forming regions \citep{Jaupart2021}, or more generally, from the universality of the supersonic turbulence energy cascade in the interstellar medium (ISM) \citep{Padoan1997_imf,Federrath2013_universality}. On scales above the neutral-ion decoupling scale, magnetic fields are approximately flux-frozen into the gas, and fluctuate, tangle and become turbulent with the gas velocities, hence magnetic fields also play an important role in all these processes \citep{Hennebelle2019,Krumholz2019}.
    
    In the ISM magnetic fields and turbulence coexist in a partnership. Extremely weak, primordial magnetic fields were potentially formed through a battery process \citep[e.g.,][]{Biermann1950_battery}, or a phase transition in the early Universe \citep{Subramanian2016_origins_review,Subramanian2019_origins}, and, once generated, they are hard to destroy due to the lack of magnetic monopoles \citep{Parker1970_monopoles_origin,Beck_2013_Bfield_in_gal,Acharya2022_no_monopoles}. Instead, turbulent motions of gas exponentially amplify the weak seed fields, growing them through the turbulent dynamo and magnetising the plasma (see \citealt{McKee2020} for a recent review). Turbulence, through the dynamo process, likely continues to maintain the magnetic fields found in the present-day Universe, ensuring they are \textit{roughly} in energy equipartition with the turbulent motions, i.e., the saturated state of the turbulent dynamo \citep{Federrath2014,Schober2015,Federrath2016_dynamo,Xu2016_dynamo,McKee2020,Seta2020, Seta2020b,Chirakkara2021,Seta2021}. 
    
    Given the importance of MHD turbulence in the ISM, it is not surprising that the ISM community has built numerous tools for measuring turbulent properties (for a recent review, see \citealt{Burkhart2021_stats_turb_review}). These include, but are certainly not limited to: techniques that relate starlight polarisation dispersion to plane-of-sky magnetic fields strengths, such as the methods described in \citet{Skalidis2020}, \citet{Davis1951} and \citet{Chandrasekhar_fermi_1953} (for recent reviews, extensions and modifications, see \citealt{Lazarian2020_diff_measure_VCM,Lazarian2022_diff_measure}); inference of magnetic field strengths and plasma energetics from local velocity centroids or intensity fluctuations \citep{Lazarian2018_VGT} and density gradients of dust continuum maps \citep{Soler2013}; ascertaining the ratio of compressive to solenoidal modes of turbulent driving sources from the deprojected column density \citep{Federrath2009,Brunt2010a,Brunt2010b,Brunt2014,Kortgen2020,Menon2020,Sharda2021_driving_mode}; and data-driven statistical techniques that can capture abstract features of sub- or super-Alfv\'enic turbulence using wavelet scattering transforms \citep{Allys2019_scatteringtransforms,Saydjari2021_scattering_transform} or deep convolutional neural networks \citep{Peek2019_CNN}. All of these diagnostics rely upon a thorough, physical understanding of the underlying phenomenology of MHD turbulence. 
    
    However, the parameter space of MHD turbulence is large, and there need not be a universal phenomenology that captures the richness of the topic (see the eloquent review from \citealt{Schekochihin2020_bias_review} about phenomenologies and two-point statistical models for incompressible MHD turbulence, which continue to be subject to debate; \citealt{Iroshnikov_1965_IK_turb,Kraichnan1965_IKturb,Sridhar1994_weak_turbulence,Goldreich1995,Boldyrev2006}). In this study we aim to explore energy balance in a particular part of the parameter space relevant to the ISM: isothermal, highly-compressible MHD turbulence, driven with a mixture of compressible and solenoidal modes, and that is threaded by a large-scale magnetic field, $\Bo$, flux-frozen on the system scale. Such a description is potentially applicable to any of the approximately isothermal phases of the ISM \citep{Wolfire1995_isothermal_ISM,Omukai2005_isothermal_ism}. In this context, we can identify several distinct energy reservoirs, but in this study our main aim is to understand the correlation between the large-scale and turbulent magnetic field; mathematically, this term takes the form $\dbBo$, where $\dB$ and $\Bo$ are the fluctuating and large-scale fields, respectively. We henceforth refer to this as the ``magnetic coupling term'', or simply the ``coupling term''. This term has been neglected previously in the literature \citep[e.g.,][]{Zweibel1995_energy_equipartition} because when averaging over a volume $\V$ that contains a few turbulent correlation scales, $\Exp{\dbBo}_{\V} = 0$. However, we show that when one instead considers the $2^{\rm nd}$ moments of the energy equation (the fluctuations of energy), which maintains the positivity for all of the contributions to the energy, including $\dbBo$, the coupling term plays a leading order role in the energy balance when $\Bo$ is strong, corresponding to sub-to-trans-Alfv\'enic turbulence.   
    
    \citet{Skalidis2020} and \citet[][hereafter \citetalias{Skalidis2021}]{Skalidis2021} recently showed that the coupling term is important for measuring the plane-of-sky magnetic field using polarisation dispersion techniques for interstellar gas, especially in highly-magnetised regions of the ISM \citep{HuaBai2013,Federrath2016_brick,Hu2019,Heyer2020,Hwang2021,Hoang2021,Skalidis2021_obs_sub_alf}. In this paper, we show that by constructing a set of analytical models for the coupling term and turbulent magnetic fluctuations, based on kinetic and magnetic energy balance, one can derive strong constraints on the magnetic fluctuations and Alfv\'en Mach numbers $\Ma$ in the plasma. We also study the impact of a large-scale magnetic field on the turbulence by analysing the turbulent, total and mean-field $\Ma$, and the relationships between them. Beyond significantly suppressing the turbulent component of the magnetic field as the large-scale field grows in energy in a power-law fashion, $\delta B \propto B_0^{-1}$, we show that having a strong large-scale field is a necessary prerequisite for sub-Alfv\'enic turbulence, i.e., a plasma can only be in the sub-Alfv\'enic regime when the large-scale, ordered field contains almost all of the magnetic energy, making the magnetic fluctuations highly super-Alfv\'enic and hence dynamically sub-dominant.     
    
    This study is organised as follows: in \autoref{sec:sims} we outline the compressible MHD turbulence simulations that we will use. In \autoref{sec:energy_balance} we review the basics of energy balance between magnetic and kinetic energy in MHD turbulence. We focus upon the coupling term, justify why it ought to be considered in the energy balance equation, and in \autoref{sec:model_for_dbBo} we provide analytical models for this term in both the super- and sub-Alfv\'enic regimes. In \autoref{sec:alfven_mach} we turn our attention to the fluid energetics in the context of the Alfv\'en Mach number, highlighting the difference between the turbulent, mean-field, and total Alfv\'en Mach numbers and deriving relationships between them. Next, in \autoref{sec:consideration} we discuss the role of the turbulent correlation scale for measuring magnetic field statistics in simulations and observations. Finally, in \autoref{sec:conclusion} we summarise the key results of this study.
    
\section{Numerical simulations}\label{sec:sims}
    To test our energy balance models, we use a modified version of the \textsc{flash} code \citep{Fryxell2000,Dubey2008}, utilising a second-order conservative MUSCL-Hancock 5-wave approximate Riemann scheme \citep{Bouchut2010,Waagan2011,Federrath2021} to solve the 3D, ideal, isothermal, compressible MHD equations with a stochastic acceleration field acting to drive non-helical turbulence,
    \begin{align}
        \frac{\partial \rho}{\partial t} + \nabla\cdot(\rho \vecB{v}) &= 0 \label{eq:continuity}, \\
        \rho\frac{\partial\vecB{v}}{\partial t}  - \nabla\cdot\left[ \frac{1}{4\pi}\vecB{B}\otimes\vecB{B} - \rho \vecB{v}\otimes\vecB{v} - \left(c_s^2 \rho + \frac{B^2}{8\pi}\right)\mathbb{I}\right] &= \rho \vecB{f},\label{eq:momentum} \\
        \frac{\partial \vecB{B}}{\partial t} - \nabla \times (\vecB{v} \times \vecB{B}) &= 0,\label{eq:induction}\\
        \nabla \cdot \vecB{B} &= 0, \label{eq:div0}
    \end{align}
    where $\otimes$ is the tensor product and $\mathbb{I}$, the identity matrix. We solve the equations on a periodic domain of dimension $L^3\equiv \V_L$, discretised with between $16^3-1152^3$ cells, where $\vecB{v}$ is the fluid velocity, $\rho$ is the gas density, $\vecB{B} = B_0 \vecB{\hat{e}}_{\parallel} + \delta\vecB{B}(t)$ is the magnetic field, with mean-field $B_0 \vecB{\hat{e}}_{\parallel}$\footnote{Note that we refer to a mean-field coordinate system, as adopted in \citet{Hartlep2000_b_field_pdf}, where $\Bo$ always points along $\vecB{\hat{z}} = \vecB{\hat{e}}_{\parallel}$ and hence $\vecB{\hat{x}} = \vecB{\hat{e}}_{\perp,1}$ and $\vecB{\hat{y}} = \vecB{\hat{e}}_{\perp,2}$. The plasma is statistically symmetric in the $(\vecB{\hat{e}}_{\perp,1},\vecB{\hat{e}}_{\perp,2})$ plane, so we will regularly state quantities for $\vecB{\hat{e}}_{\perp}$.} and turbulent field, $\delta\vecB{B}(t)$, where $\Exp{\delta\vecB{B}(t)}_{\V_L}=0$, $c_s$ is the sound speed and $\vecB{f}$, the stochastic turbulent acceleration source term that drives the turbulence, which, in the ISM could be from, for example, supernova shocks, internal instabilities in the gas, gravity, galactic-scale shocks and shear, or ambient pressure from the galactic environment \citep{Brunt2009,Elmegreen2009IAUS,Federrath2015_inefficient_SFR,Krumholz2016,Padoan2016_supernova_driving,Grisdale2017,Jin2017,Kortgen2017,Federrath2017IAUS,Colling2018,Schruba2019,Lu2020}. Here, and throughout this paper, we use the notation $\Exp{\hdots}_{\V}$ to indicate the mean value of some quantity within a specified volume $\V$ (which can be the entire simulation volume $\V_L$, but need not be). We discuss the resolution of our simulations, and demonstrate that the quantities of interest for us are converged in them, in \aref{app:convergence}.
    
    The forcing term $\vecB{f}$ follows an Ornstein-Uhlenbeck process with finite correlation time, $\tau = \ell_0/\Exp{v^2}_{\V_{L}}^{1/2} = L/(2 c_s \M)$, where $\M$ is the sonic Mach number, such that $\ell_0 = L/2$ is the driving, or energy injection scale, and $\vecB{f}$ is constructed so that we are able to set $0.5 \lesssim \M \lesssim 10$, encapsulating the $\M$ values of supersonic molecular gas clouds in the interstellar medium \citep[e.g.,][]{Schneider2013,Federrath2016_brick,Orkisz2017,Beattie2019b} as well as the subsonic, diffuse, warm medium \citep[e.g.,][]{Kritsuk2017}. We force with equal energy in both compressive $(\nabla \times \vecB{f}=0)$ and solenoidal $(\nabla \cdot \vecB{f}=0)$ modes. The energy injection is isotropic, centred on $|\vecB{k}L/2\pi|=2$ and falling off to zero with a parabolic spectrum within $1 \leq |\vecB{k}L/2\pi| \leq 3$ (see \citealt{Federrath2008,Federrath2009,Federrath2010,Federrath2022_turbulence_driving_module} for turbulence driving details). $\Mao$ is set by fixing $B_0$ and using the definition of the mean-field Alfv\'en velocity and $\M$, $\Mao = 2c_s\sqrt{\pi\rho_0}\M/B_0$. We vary this value for each of the simulations between $10^{-2} \lesssim \Mao \lesssim 10^{3}$, resulting in a total of 280 simulations across different grid resolutions, with 56 unique simulations, which we list in \autoref{tb:simtab}. The initial velocity field is set to $\vecB{v}(x,y,z,t=0)=\vecB{0}$, with units $c_s=1$, the density field $\rho(x,y,z,t=0)=\rho_0$, with units $\rho_0=1$, and $\delta\vecB{B}(t=0) = \vecB{0}$, with units $c_s\rho_0^{1/2} = 1$. 
    
\begin{figure}
        \centering
        \includegraphics[width=\linewidth]{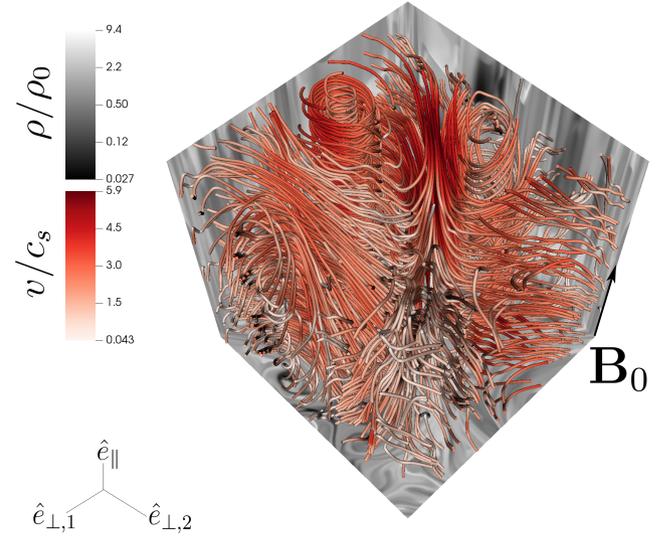}
        \caption{Typical velocity (red) streamline structure in sub-Alfv\'enic mean-field turbulence, with $\M = 2$ and $\Mao = 0.1$. The direction of the mean-field, $\Bo$, is shown in the bottom right corner of the box. Slices of the density are shown in grey scale at the box edges. Ordered vortex structures occupy the full extent of the box along $\hat{e}_{\parallel}$, and out to the driving scale in $\hat{e}_{\perp}$.}
        \label{fig:k2_vortex}
    \end{figure}

    We run the simulations for 10 correlation times of $\vecB{f}$, and report statistics from time-averages over the last 5 correlation times to ensure that the sub-Alfv\'enic mean-field simulations are statistically stationary \citep{Beattie2021b}. After 5 correlation times large vortical structures develop in the sub-Alfv\'enic mean-field simulations, extending along the strong large-scale field and out to the driving scale perpendicular to the field, which we show, as an example, in \autoref{fig:k2_vortex} for the \texttt{M2MA01} simulation.\footnote{We use a naming convention for our simulations whereby the value following the \texttt{M} gives the target sonic Mach number $\mathcal{M}$ (with decimal points omitted) and the value following \texttt{MA} gives the target Alfv\'en Mach number $\mathcal{M}_{\rm A0}$ -- thus run \texttt{M2MA01} is one where we set the mean magnetic field and tune the forcing to produce $\mathcal{M}=2$ and $\mathcal{M}_{\rm A0}=0.1$.} For more details about the current simulations, we refer the readers to \citet{Beattie2020} for the anisotropy in $\rho/\rho_0$, \citet{Beattie2020c} for a detailed analysis of $\dB$, \citet{Beattie2021} for an anisotropic model of the $\rho/\rho_0$ variance, and \citet{Beattie2021b} for the density intermittency and the $\ln(\rho/\rho_0)$-PDF.
    
    \begin{figure}
        \centering
        \includegraphics[width=\linewidth]{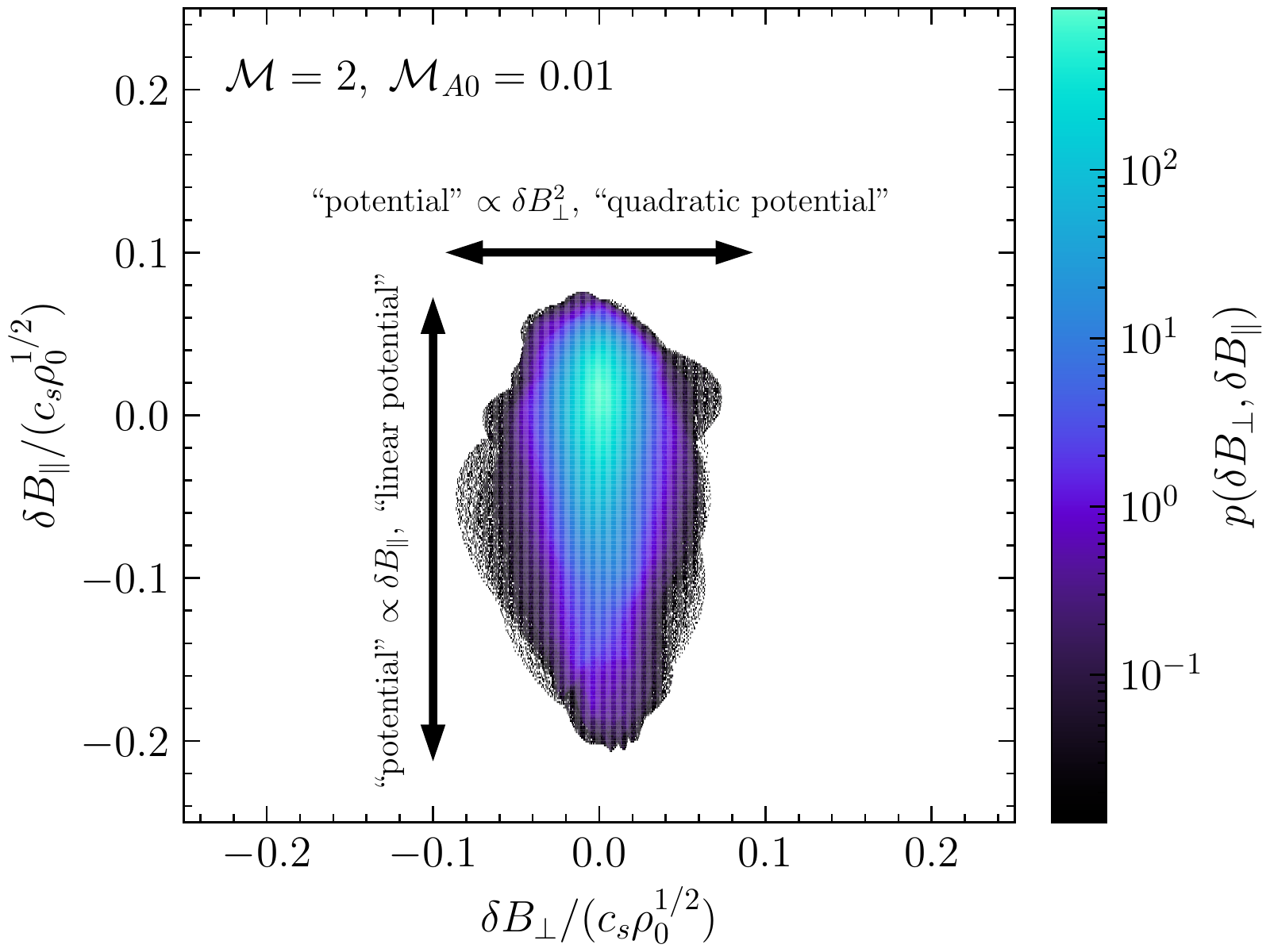}
        \caption{The joint PDF for $\dBpar$ and $\dBperp$ for the \texttt{M2MA001} simulation, showing a long, asymmetric tail into the negative values for $\dBpar$. As demonstrated in \citetalias{Skalidis2021}, the $\dBperp$ fluctuations are symmetric about $\dBperp=0$, and are analogous to a harmonic oscillator (in magnetic amplitude space) that is restored by the magnetic tension $\propto \dBperp^2$ with a quadratic potential. On the other hand, the $\dBpar$ amplitude fluctuates anharmonically, $\propto \dBpar$, with a linear potential. In sub-to-trans-Alfv\'enic compressible MHD the anharmonic, parallel magnetic field fluctuations contain most of the magnetic energy.}
        \label{fig:joint_mag_pdf}
    \end{figure} 
    
\section{Energy balance}\label{sec:energy_balance}
    \subsection{Energy balance basics \& averaging}
    Recent studies have shown that one can use energy balance arguments that include the large-scale magnetic field, $\Bo$, to derive scaling laws between the Alfv\'enic and kinetic fluid quantities (\citealt{Federrath2016_dynamo,Beattie2020c,Skalidis2020}, \citetalias{Skalidis2021}). The dimensionless magnetic energy density, by which we mean the magnetic energy density normalised to the mean thermal pressure\footnote{A natural normalisation for an isothermal plasma because $\rho_0$ and $c_s$ are both constant, and problem dependant.} $\rho_0 c_s^2$, is
    \begin{align}
        \emag = \frac{B^2}{8\pi c_s^2 \rho_0} = \frac{1}{8\pi c_s^2 \rho_0}\Big( B_0^2+ \underbrace{2\delta\vecB{B}\cdot\vecB{B}_0}_{\mathclap{\substack{\text{coupling} \\ \text{term}}}} {} + \delta B^2 \Big),
        \label{eq:emag}
    \end{align}
    where $B_0^2$ is the large-scale field contribution to the total energy, $\delta B^2$ is the turbulent field contribution and $2\delta\vecB{B}\cdot\vecB{B}_0$ is the coupling term between the two field components. In the linear perturbation theory limit of the MHD equations, $\delta B^2$ includes contributions from shear Alfv\'en, fast and slow magnetosonic compressive eigenmodes \citep[e.g.,][]{Landau1959}. Because $\delta\vecB{B}\cdot\vecB{B}_0 = \dBpar B_0$, the coupling term only contains the component of magnetic field fluctuations that are parallel to the large-scale field. In linear theory, both fast and slow magnetosonic compressible modes are able to perturb the field variables parallel to $\Bo$, so under the lens of linear theory, the coupling term is the fluctuation contribution from the compressible modes in the turbulence scaled by $B_0$ \citep{Bhattacharjee1998}. Furthermore, for sub-Alfv\'enic turbulence \citet{Beattie2021b} showed that converging, shocked flows along magnetic field lines excite strong $\dBpar$ fluctuations, which travel roughly at the theoretical fast Alfv\'en mode speed. Therefore, it is likely, assuming that $\dBpar/B_0 \ll 1$ (this is indeed the case for $\Mao<1$ plasmas; see left panel of Figure~5 in \citealt{Beattie2022_va_fluctuations}) where a linear theory may be valid for the magnetic field, the coupling term contains significant energy contributions from fast magnetosonic modes excited by shocked gas that converges and forms dense filaments perpendicular to magnetic field lines. 
    
    The excitation of a dominating $\dBpar$ is something characteristic of sub-Alfv\'enic compressible turbulence. We demonstrate this by plotting the time-averaged joint $\dBpar-\dBperp$ PDF in \autoref{fig:joint_mag_pdf}, for highly-sub-Alfv\'enic turbulence $(\Mao = 0.01)$. It is evident that the distributions of $\dBpar$ and $\dBperp$ are not the same\footnote{We show the super-Alfv\'enic version of this plot, which admits to isotropic fluctuations, in \autoref{fig:m2ma10_b_pdf}.}. The reason is straightforward: $\dBperp$ (fluctuations from shear Alfv\'en waves) is subject to a quadratic restoring force via the magnetic tension\footnote{When $B_0 \gg \delta B$, $\tens \approx -\kappa B_0^2\vecB{\hat{e}}_{\perp}$, where $\kappa$ is the field line curvature. Hence $\tens$ acts to strongly dampen shear Alfv\'en waves. This approximation for $\tens$ is most appropriate for regions of the plasma where $\nabla_{\parallel} \cdot v_{\parallel} \approx 0$, because compressions can excite $\dBpar$, creating parallel gradients in the magnetic field that also act to increase the tension \citep{Beattie2021b}} \citep{Yuen2020,Beattie2021b}, which results in a symmetry about $\dBperp=0$. However, $\dBpar$ has a linear restoring force and is forced out of the minimum energy state $\dBpar = -B_0$ to conform to $\Exp{\dBpar}_{\V}=0$\footnote{Note that in the language of solid state physics, we may consider $\dBpar$ to be a topologically frustrated field, because the minimum energy state is $\dBpar=-B_0$, but conservation of total magnetic flux requires $\Exp{\dBpar}_{\V}=0$. Hence, populations of parallel magnetic fluctuations can be imagined to compete to get to $\dBpar=-B_0$, but for every $\dBpar$ that comes close to $-B_0$ there must be either another fluctuation that comes close to $+B_0$ or a population of fluctuations that in total add to $+B_0$, ensuring globally that $\Exp{\dBpar}_{\V}=0$. We do not take this analogy any further in this study but it may stimulate future works on magnetic field fluctuation PDFs.} \citepalias{Skalidis2021}. This gives rise to a skewed distribution in $\delta B_{\parallel}$, with a long extended tail of negative $\delta B_{\parallel}$ values. We 
    will show below that $\dBpar$ contains almost all of the turbulent magnetic energy in the compressible plasma. Now we turn our attention to what feeds the magnetic field fluctuations. 

    The dimensionless turbulent kinetic energy, normalised by the mean thermal pressure (similarly to $\emag$; see \autoref{eq:emag}), is
    \begin{align} \label{eq:ekin}
        \ekin = \frac{1}{2}\left(\frac{\delta v}{c_s}\right)^2,
    \end{align}   
    which acts as an energy reservoir for the magnetic field fluctuations via the velocity term in the induction equation, \autoref{eq:induction}. Considering our ideal, isothermal (in our units, the thermal energy is $e_{\rm thermal} = 3/2$), MHD system, the total energy is then
    \begin{align}
        e_{\rm tot} &= \frac{1}{2}\left(\frac{\delta v}{c_s}\right)^2 + \frac{1}{8\pi c_s^2 \rho_0}\left( B_0^2+ 2\delta\vecB{B}\cdot\vecB{B}_0 + \delta B^2 \right) + \frac{3}{2},
    \end{align}
    and for just the `total' turbulent energy, 
    \begin{align}\label{eq:total_energy}
        e_{\rm turb} &= \frac{1}{2}\left(\frac{\delta v}{c_s}\right)^2 + \frac{1}{8\pi c_s^2 \rho_0}\left( 2\delta\vecB{B}\cdot\vecB{B}_0 + \delta B^2 \right),
    \end{align} 
    where only the $\delta B^2$ and $\dbBo$ terms are retained in the magnetic energy, because they contain the turbulent contribution. 
    
    In a fluid with initially weak magnetic fluctuations and $B_0 = 0$, $\ekin$ (\autoref{eq:ekin}) will transfer energy and enhance $\emag$ (\autoref{eq:emag} with $B_0$ set to 0) via the small-scale turbulent dynamo \citep[for a recent review see ][]{McKee2020}. A standard ansatz of dynamo theory is that saturation will be reached between the turbulent fields, such that,
    \begin{align} \label{eq:energy_balance_1}
        \frac{\Exp{\emag}_{\V}}{\Exp{\ekin}_{\V}} = \sat,
    \end{align}
    where $0 \leq \sat \leq 1$. The value of $\sat$ is a function of $\M$, the Alfv\'en Mach number $\Ma$ (a precise definition for which we defer to \autoref{sec:alfven_mach}), the nature of the driving mechanism, $\vecB{f}$, in particular if it is compressive $\nabla\times\vecB{f}=0$ or solenoidal $\nabla\cdot\vecB{f}=0$, and the Prandtl and Reynolds numbers of the fluid \citep{Federrath2011_mach_dynamo,Schober2012,Federrath2014,Schober2015,Federrath2016_dynamo,Chirakkara2021,Kriel2022_turb_dynamo}. The exact physics of the saturation is still an open problem in dynamo theory, but most likely the saturation develops due to the effect of strong magnetic fields on both the amplification (via field line stretching), diffusion of magnetic fields and instabilities caused by tearing and magnetic reconnection \citep{Schekochihin2002_saturation_evolution,Xu2016_dynamo,Seta2021,Galishnikova2022_tearing_instability_sat}; however, the exact value of $\sat$ and its dependence on other parameters is not important for our purposes. What is significant is that, assuming that the energy transfer from $\ekin$ to $\emag$ is solely through the turbulent components of the respective fields, including the turbulent and large-scale field coupling term for the more general case where $B_0 \neq 0$, from \autoref{eq:total_energy}, \autoref{eq:energy_balance_1} becomes
    \begin{align} \label{eq:energy_balance}
        \frac{1}{8\pi c_s^2 \rho_0}\Exp{2\dB\cdot\Bo + \delta B^2}_{\V} = \frac{\sat}{2}\Exp{\left(\frac{\delta v}{c_s}\right)^2}_{\V},
    \end{align}
    which naively reduces to 
    \begin{align}
        \frac{1}{8\pi c_s^2 \rho_0}\Exp{\delta B^2}_{\V} = \frac{\sat}{2}\Exp{\left(\frac{\delta v}{c_s}\right)^2}_{\V},
    \end{align}    
    if $\Exp{\dB\cdot\Bo}_{\V} = \Exp{\dBpar}_{\V}B_0 = 0$ because $\Exp{\dBpar}_{\V}=0$ when $\V$ captures a few correlation lengths of the turbulence, for the regular \citet{Reynolds1895_averaging} decomposition of a stochastic field\footnote{\citetalias{Skalidis2021} showed that even this leads to complications because $\dbBo$ is analogous to a potential energy, which does not make sense to average because it is invariant to gauge transforms, not positive definite, nor symmetric around the minimum energy state for $\dBpar$. See discussion in \S4 of \citetalias{Skalidis2021} for more details.}. But this is not necessarily a sensible result because when the large-scale field is strong the coupling term is leading order in the turbulent magnetic energy, and all energy reservoirs should be strictly positive. Because the coupling term is the only term that is not positive semi-definite in \autoref{eq:energy_balance} we may want to treat averaging the equation with more care. 
    
    These considerations lead us to consider an alternative ansatz, one that enforces the positivity of all terms. Our approach is to take the $2^{\rm nd}$ moments of \autoref{eq:energy_balance}, but also taking the square root to ensure that the units are appropriate for an energy balance,
    \begin{align} \label{eq:energy_balance_rms}
        \frac{1}{8\pi c_s^2 \rho_0}\Exp{\left(2\dB\cdot\Bo + \delta B^2\right)^2}_{\V}^{1/2} = \frac{\sat}{2}\Exp{\left(\frac{\delta v}{c_s}\right)^4}^{1/2}_{\V}.
    \end{align}
    The physical interpretation of this balance is that instead of balancing the means of the energy distributions, we balance the root-mean-squared values, which are a measure of the typical local fluctuations in energy of the plasma. This method of volume-averaging \autoref{eq:energy_balance} gives rise to $4^{\rm th}$ order terms in velocity and magnetic field fluctuations ($2^{\rm nd}$ order in energies). Note that for finite $\Mao$, $\sat$ is now different from $\sat$ in small-scale dynamo experiments (where $B_0 = 0$) because it is now  sensitive to the large-scale field through the energy contribution of the coupling term.  
    
\begin{figure*}
        \centering
        \includegraphics[width=\linewidth]{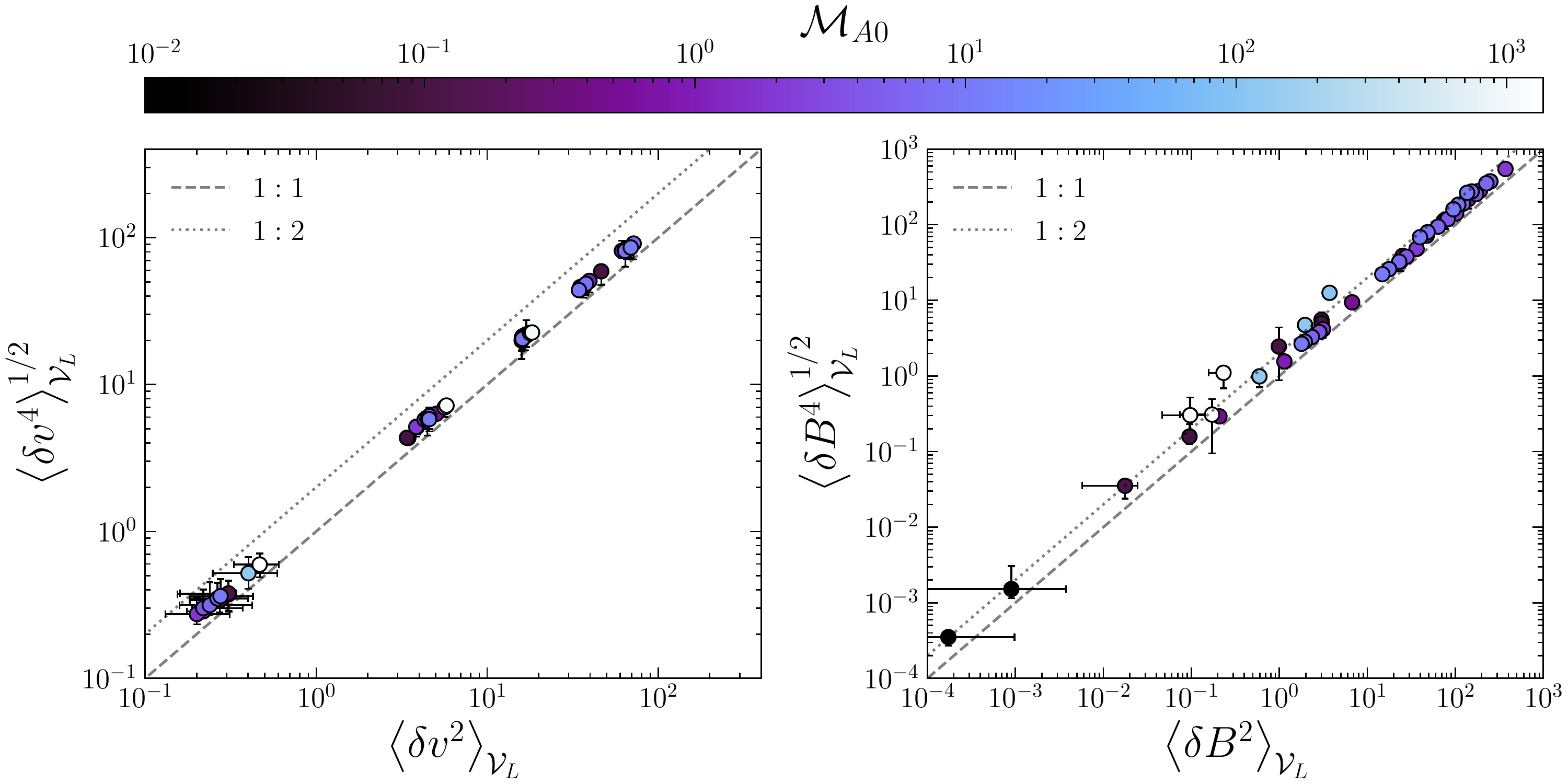}
        \caption{The square root of the $4^{\rm th}$ order turbulent velocity (left) and magnetic field (right) moments as a function of $2^{\rm nd}$ order moments, coloured by $\Mao$, for all of the simulations listed in \autoref{tb:simtab}. The $2^{\rm nd}$ order moments of the turbulent fields are approximately equivalent to the scaled $4^{\rm th}$ order moments, $\Exp{\delta v^4}^{1/2}_{\V} = (1.3\pm 0.3) \Exp{\delta v^2}_{\V}$, and $\Exp{\delta B^4}^{1/2}_{\V} = (1.5\pm0.4)\Exp{\delta B^2}_{\V}$, rarely deviating by
        more than a factor of 2 (dotted line) from the 1:1 dashed line. This means that the standard deviation of the magnetic and kinetic energy distributions scale with the mean.}
        \label{fig:second_fourth_order}
    \end{figure*}

    To better understand the $4^{\rm th}$ order terms we plot them as a function of $2^{\rm nd}$ order terms in \autoref{fig:second_fourth_order} and show the 1:1 and 1:2 lines with dashes and dots, in each of the plots, respectively. Using least-squares fitting we find that $\Exp{\delta v^4}^{1/2}_{\V} = (1.3\pm0.3) \Exp{\delta v^2}_{\V}$, and $\Exp{\delta B^4}^{1/2}_{\V} = (1.5\pm0.4) \Exp{\delta B^2}_{\V}$, hence, within $\approx1\sigma$, the proportionality constants are approximately unity. Physically, this means that as the mean of the energy distributions increase, so does the root-mean-squared, or spread of the distributions. This has been shown before, for example, in \citet{schekochihin2004simulations}, where they found $\Exp{\delta B^4}^{1/2}_{\V} \approx \sqrt{2} \Exp{\delta B^2}_{\V}$ (see Fig~11, saturated regime). This is an important point, because it means that the contributions from the turbulent fields remain approximately the same in both averaging schemes, \autoref{eq:energy_balance} and \autoref{eq:energy_balance_rms}, but now we are able to properly include the energy contribution from the coupling term. 
    
    \subsection{Weak and strong B-field limits for rms energy balance}
    
    Consider now \autoref{eq:energy_balance_rms} in the weak $\Bo$ regime, such that $B_0 \ll \delta B$, averaged over $\V$. This means
    \begin{align}
        \lefteqn{\Exp{\left(2\dB\cdot\Bo + \delta B^2\right)^2}_{\V}} \qquad \nonumber \\ 
        = & \Exp{(2\dB\cdot\Bo)^2}_{\V} + \Exp{4(\dB\cdot\Bo)\delta B^2}_{\V} + \Exp{\delta B^4}_{\V} \label{eq:average_expansion} \\
        \sim & \Exp{\delta B^4}_{\V},
    \end{align}
    to leading $\delta B^4$ order, and therefore
    \begin{align}
        \frac{\Exp{\delta B^4}_{\V}^{1/2}}{8\pi c_s^2 \rho_0} \approx \frac{\sat}{2}\Exp{\left(\frac{\delta v}{c_s}\right)^4}^{1/2}_{\V}.
    \end{align}    
    Based upon our numerical results this equation can be re-written in terms of $2^{\rm nd}$ order terms (see \autoref{fig:second_fourth_order}),
    \begin{align}\label{eq:e_balance_super_limit}
        \frac{\Exp{\delta B^2}_{\V}}{8\pi c_s^2 \rho_0} \approx \frac{\sat}{2}\Exp{\left(\frac{\delta v}{c_s}\right)^2}_{\V},
    \end{align}
    with the $\Exp{\delta B^2}_{\V}$ dominating the balance between the kinetic turbulent energy. Likewise, as \citet{Federrath2016_dynamo} framed the relation, the kinetic energy is feeding the magnetic field through the $\delta B^2$ term in this regime.
    
    In the strong $\Bo$ regime we have $B_0 \gg \delta B$, and hence, to leading $B_0^2$ order \autoref{eq:average_expansion} becomes
    \begin{align}
        \Exp{\left(2\dB\cdot\Bo + \delta B^2\right)^2}_{\V} \sim 2\Exp{\left(\dB\cdot\Bo\right)^2}_{\V},
    \end{align}
    with the $2\dbBovol$ term dominating the balance. Hence the energy balance must be between
    \begin{align}
        \frac{\Exp{\left(\dB\cdot\Bo \right)^2}_{\V}^{1/2}}{4\pi c_s^2 \rho_0} \approx \frac{\sat}{2}\Exp{\left(\frac{\delta v}{c_s}\right)^4}^{1/2}_{\V},
    \end{align}
    which we can similarly reduce to $2^{\rm nd}$ order terms,
    \begin{align}\label{eq:e_balance_sub_limit}
        \frac{\Exp{(\dB\cdot\Bo)^2}^{1/2}_{\V}}{4\pi c_s^2 \rho_0} \approx \frac{\sat}{2}\Exp{\left(\frac{\delta v}{c_s}\right)^2}_{\V}.
    \end{align}
    Note now this is the same relation derived in \citetalias{Skalidis2021}, but it comes from directly considering the rms balanced energy equations, and then invoking the numerical result that the square root of the $4^{\rm th}$ order velocity and magnetic terms scale almost perfectly with the $2^{\rm nd}$ order terms. The $2^{\rm nd}-4^{\rm th}$ moment relation should be accurate to a factor less than 2, as indicated in \autoref{fig:second_fourth_order}. Establishing the strong mathematical footing for this relation is a key result from our study.

    We will return to \autoref{eq:energy_balance} and the two limiting cases, \autoref{eq:e_balance_super_limit} and \autoref{eq:e_balance_sub_limit}, throughout this study. Specifically, we will show that by using this simple energy balance model that includes $\dbBovol$, we can learn a great deal about the magnetic and velocity field fluctuations. First, we start by understanding the nature of the coupling term.    
    
    \begin{figure}
        \centering
        \includegraphics[width=\linewidth]{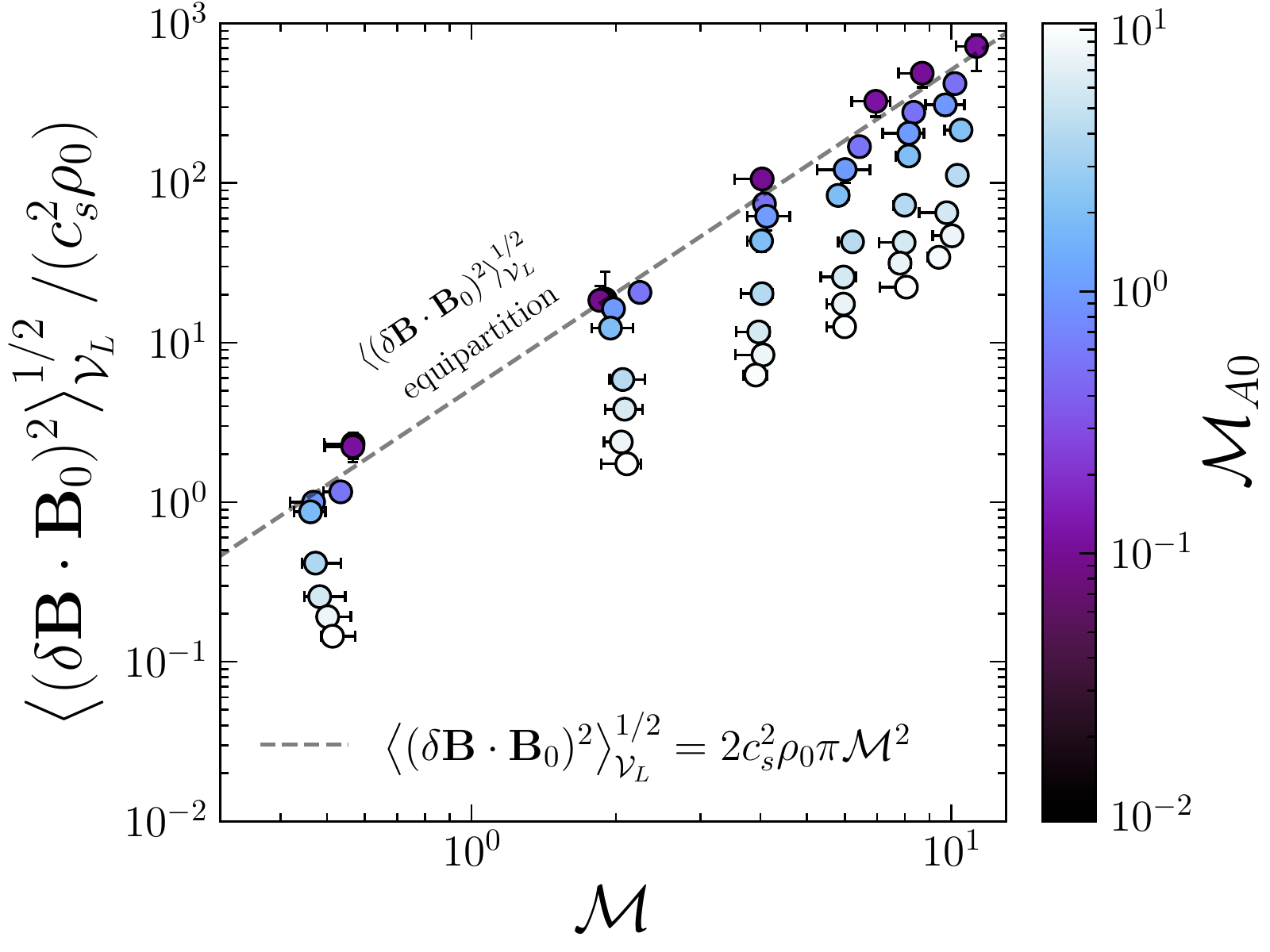}
        \caption{The magnetic coupling term, $\dbBovol$, as a function of the $\M$, coloured by $\Mao$, for all of the simulations up to $\Mao = 10$. We show the strong-field model, \autoref{eq:strong_B_coupling_term}, for the coupling term, indicated with the grey dashed line, which is valid for the simulations with dark shading, assuming exact energy equipartition between the turbulent kinetic and the coupling term energy.}
        \label{fig:B_coupling}
    \end{figure}
    
    \begin{figure}
        \centering
        \includegraphics[width=\linewidth]{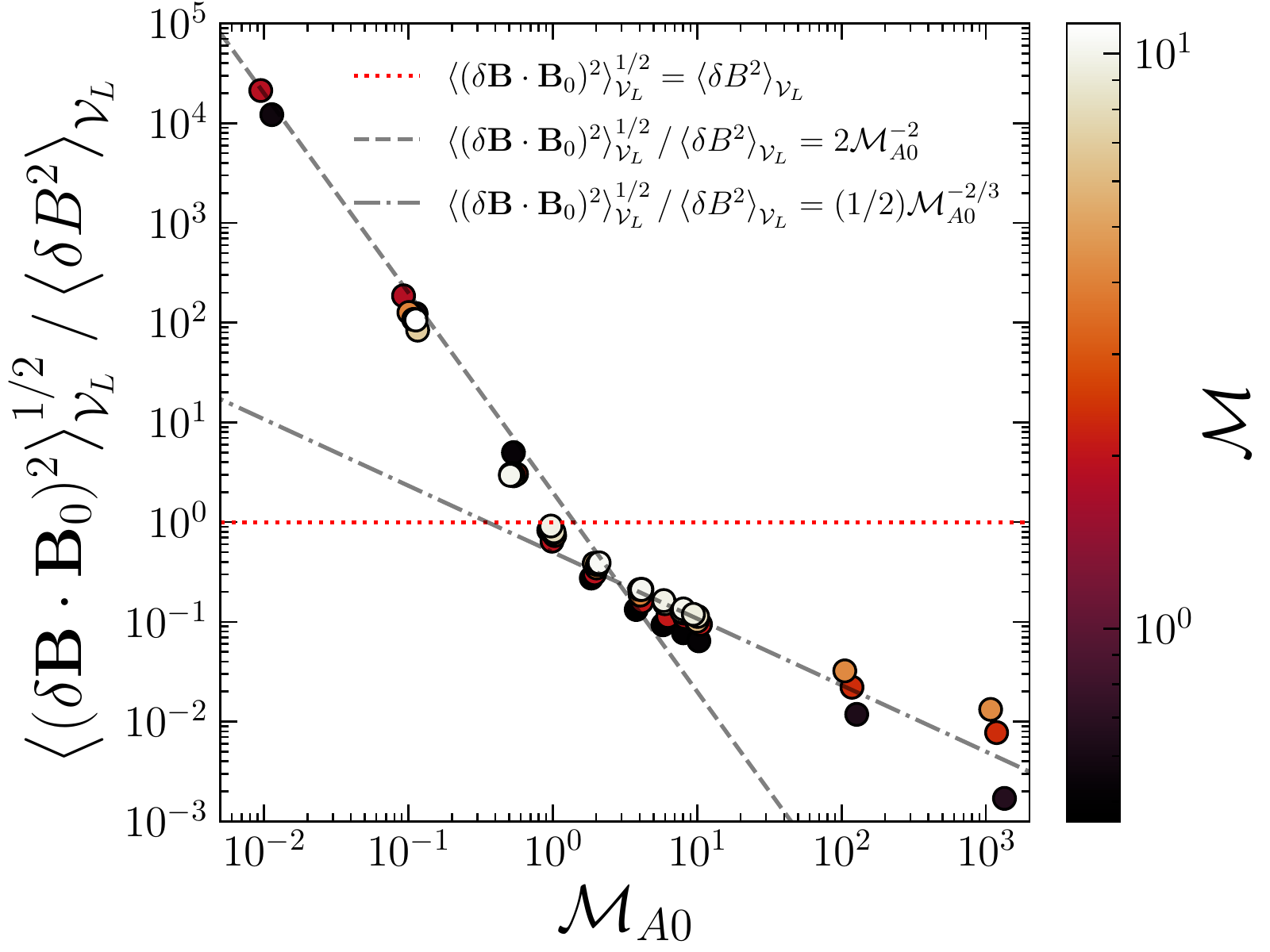}
        \caption{The ratio of the magnetic coupling term, $\dbBovol$, to the turbulent magnetic energy, $\Exp{\delta B^2}_{\V}$ as a function of $\Mao$, coloured by the $\M$, for all simulations. We show in red dots the equipartition between the two terms. The grey dashed line shows the strong-field model, \autoref{eq:strong_B_coupling_ratio}, which is valid for $B_0 \gg \delta B$, or $\Mao \lesssim 2$, and the grey dot-dashed line for the weak-field model, \autoref{eq:weak_B_coupling_ratio}, valid for $\Mao > 2$.}
        \label{fig:coupling_ratio_ma0}
    \end{figure}  
    
    \begin{figure}
        \centering
        \includegraphics[width=\linewidth]{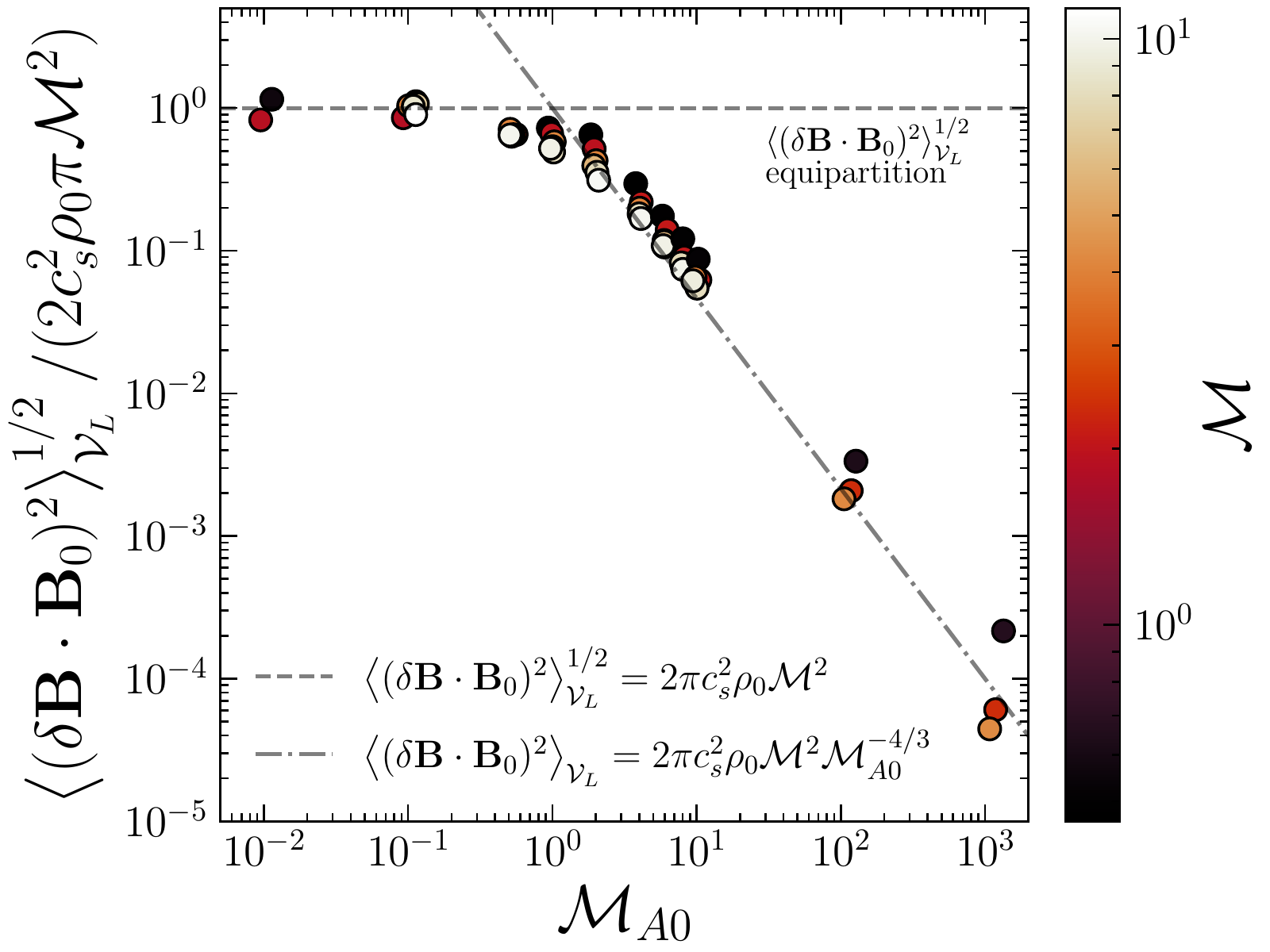}
        \caption{The magnetic coupling term, $\dbBovol$, compensated by \autoref{eq:strong_B_coupling_term}, as a function $\Mao$, coloured by $\M$, for all simulations. This choice of compensation reveals the $\Mao$ dependency in the super-Alfv\'enic turbulence regime, which we provide a model for in \autoref{sec:weak_field_limit}, shown with the grey dot-dashed line.}
        \label{fig:dbBo_ma}
    \end{figure}  

    \section{Models for \texorpdfstring{$\dbBovol$}{TEXT}}\label{sec:model_for_dbBo}
    
    \subsection{Strong mean-field, \texorpdfstring{$B_0 \gg \delta B$}{TEXT}}\label{sec:strong-field_dbBo}
    
    Assuming that the kinetic energy fluctuations are in energy equipartition with the coupling term ($\sat = 1$) it immediately follows from \autoref{eq:e_balance_sub_limit} that in the strong-mean-field regime the coupling term is,
    \begin{align}\label{eq:strong_B_coupling_term}
        \dbBovol = 2c_s^2\rho_0\pi\M^2.
    \end{align}
    We plot this predicted relationship, along with the values measured from our simulations, in \autoref{fig:B_coupling}. The plot is consistent with our expectations: simulations in the strong-mean-field regime, $\Mao < 1$, sit very close to the equipartition line, while those with $\Mao > 1$ sit below it, indicating that the $\Exp{\delta B^2}_{\V}$ term is playing an increasingly large role in the energy balance as we transition to the weak mean-field regime. 
    
    Even in the strong mean-field regime, we see weak variation with $\M$ in how closely the simulations follow the prediction of \autoref{eq:strong_B_coupling_term}. For low-$\M$, the strong-field model works best, but as $\M$ gets larger there is some scatter to lower values of $\dbBovol$, even in the sub-Alfv\'enic simulations. This suggests that for $\Mao \gtrsim 1$ there are some contributions to the magnetic energy through the $\Exp{\delta B^2}_{\V}$ term, which we neglect in our model, i.e., the shear Alfv\'en waves and fast modes that perturb the magnetic field perpendicular to $\Bo$. Of course, the turbulence naturally excites such modes but it is plausible that the magnetic tension significantly suppresses them when $B_0/\delta B$ is large.
    
    To further quantify when each of the magnetic terms in \autoref{eq:energy_balance_rms} contribute the most to the energy balance we examine the ratio of the two magnetic energy reservoir terms, $\dbBovol/\Exp{\delta B^2}_{\V}$. We estimate $\Exp{\delta B^2}_{\V}$ following the fluctuation models in \citet{Federrath2016_dynamo} and \citet{Beattie2020c},\footnote{Note that anisotropy in the magnetic and velocity fluctuations (decomposing to perpendicular and parallel field components) was ignored in these studies, as pointed out by \citetalias{Skalidis2021}, but the corrections are of order unity, which we show in \autoref{app:anisotropy}, and only become important for more sensitive calculations, which we discuss later in \autoref{sec:alfven_mach}.} which leads to a predicted relationship
    \begin{align}
        \Exp{\delta B^2}^{1/2}_{\V} = c_s \sqrt{\pi \rho_0}\M\Mao.
    \end{align}    
    The ratio between the coupling term to the energy from the above equation squared is then,
    \begin{align}\label{eq:strong_B_coupling_ratio}
        \frac{\dbBovol}{\Exp{\delta B^2}_{\V}} = \frac{2}{\MaO{2}} = 2\frac{e_{\rm mag,0}}{\ekin},
    \end{align}
    where $\MaO{-2} = e_{\rm mag,0}/\ekin$. This means at $\Mao = e_{\rm mag,0}/\ekin = 1$, i.e. when the turbulent and $\Bo$ energy are in equipartition, we expect $\dbBovol/\Exp{\delta B^2}_{\V} = 2$. We plot the relation measured in the simulations in \autoref{fig:coupling_ratio_ma0}, showing our predicted scaling in the strong mean-field regime with the dashed, grey line. Again, the plot shows excellent agreement between the model and the MHD data between $0.01 \leq \Mao \leq 2$, indicating a perfect balancing act between $\dbBovol$ and $\Exp{\delta v^2}_{\V}$. The $\Mao \approx 2$ transition is where $\Exp{\delta B^2}_{\V} / B_0^2 \approx 1$, and the turbulent magnetic field starts to dominate the magnetic energy reservoir. We will find that this is a reoccurring transition phase for compressible MHD turbulence. 
    
    \subsection{Weak large-scale field, \texorpdfstring{$B_0 \lesssim \delta B$}{TEXT}}\label{sec:weak_field_limit}
    
    Beyond $\Mao \gtrsim 2$ energy balance arguments only work if the saturation level changes as a function of plasma parameters, because $\dB$ is not constant with $\Bo$ \citep{Federrath2016_dynamo,Beattie2020c}. We have one free parameter, $\sat$, which need not be constant for all $\Mao$ and $\M$ \citep{Federrath2016_dynamo,Chirakkara2021,Seta2021}. To extract $\sat$, we model the coupling term in the super-Alfv\'enic regime by starting with an empirical model that \citet{Beattie2020c} found held universally for $\M$ in the $\Mao \gtrsim 2$ regime, $\Exp{\delta B^2}^{1/2}_{\V} / B_0 = \mathcal{M}_{A0}^{2/3}$. This provides an independent estimate of the super-Alfv\'enic turbulent magnetic fluctuations, 
    \begin{align} \label{eq:super_alfvenic_db}
        \Exp{\delta B^2}^{1/2}_{\V} & = \MaO{2/3}B_0 = 2 c_s\sqrt{\pi\rho_0}\M\MaO{-1/3} , 
    \end{align}
    and therefore, equating \autoref{eq:e_balance_super_limit} with the square of \autoref{eq:super_alfvenic_db}, $\sat = \MaO{-2/3}$, which implies that as the large-scale field becomes weaker, the turbulent magnetic field saturates to smaller and smaller values because there is less total magnetic energy, consistent with what was qualitatively found in \citet{Beattie2020c}. Following the same steps as in \autoref{sec:strong-field_dbBo}, additionally using $B_0 / (c_s \rho_0^2) = 2\sqrt{\pi}\M\MaO{-1}$, the definition of the large-scale field Alfv\'en Mach number, the coupling term in the super-Alfv\'enic regime becomes,
    \begin{align}\label{eq:weak_B_coupling_term}
        \dbBovol & = 2\pi c_s^2 \rho_0 \M^2\MaO{-4/3}, 
    \end{align}
    and ratio $\dbBovol / \Exp{\delta B^2}_{\V}$,
    \begin{align}\label{eq:weak_B_coupling_ratio}
        \frac{\dbBovol}{\Exp{\delta B^2}_{\V}} = \frac{1}{2}\MaO{-2/3}.
    \end{align}
    We plot \autoref{eq:weak_B_coupling_term} and \autoref{eq:weak_B_coupling_ratio}, alongside their strong-field counterparts, in \autoref{fig:coupling_ratio_ma0} and \autoref{fig:dbBo_ma}, respectively. Note that \autoref{fig:dbBo_ma} shows the same information as \autoref{fig:B_coupling}, but we have normalised by \autoref{eq:strong_B_coupling_term} to remove the $\M^2$ dependency. This allows us to better observe the dependence on $\Mao$ in the $B_0 \ll \delta B$ regime. Astonishingly, through this relatively simple analysis both of the theoretical models describe the data very well, providing excellent agreement over 3 orders of magnitude in $\Mao$, with no free parameters. 
    
    \subsection{Discussion of \autoref{sec:model_for_dbBo} and caveats}
    
    We have established that for sub-to-trans-Alfv\'enic turbulence, the volume-averaged turbulent kinetic energy is in \textit{exact} energy equipartition with the rms $\dbBo$ field. Each of the models in \autoref{sec:strong-field_dbBo} relies on this assumption $(\sat = 1)$, and without any further free parameters, with such a simple model, the agreement to the numerical data in \autoref{fig:B_coupling}, \autoref{fig:coupling_ratio_ma0} and \autoref{fig:dbBo_ma} is remarkable. The models in the super-Alfv\'enic regime critically rely on the empirical result from \citet{Beattie2020c}, $\Exp{\delta B^2}^{1/2}_{\V_L} / B_0 = \MaO{2/3}$, but likewise, the models for the coupling and fluctuation terms outlined in \autoref{sec:weak_field_limit}, are in excellent agreement with the data, again, with no free parameters. 
    
    Our results therefore strongly support the \citet{Skalidis2020} and \citetalias{Skalidis2021} model for relating the balanced rms magnetic field coupling term to the volume-averaged turbulent kinetic energy (the energy balance arguments). We hope that our treatment satisfies other authors' concerns about the \citet{Skalidis2020} coupling term method. These concerns have taken two forms; one is that the energy contribution from $\Exp{\dbBo}_{\V} \stackrel{\rm must}{=} 0$ because $\Exp{\dB}_{\V}=0$, as highlighted in the appendix of \citet{Li2021_dbB0_foundations}. But as has been extensively discussed in \citetalias{Skalidis2021}, and in \autoref{sec:energy_balance} of the current study, $\Exp{\dbBo}_{\V}=0$ is not a valid way of understanding the contribution from the coupling term. The other concern raised in the appendix of \citet{Li2021_critiqueSK2021}, is that energy balance only involves $2^{\rm nd}$ order quantities; this approach by definition omits the coupling term contribution, which is first order, in the energetics. Omission of the coupling term leads to significantly underestimating the magnetic energy in sub-Alfv\'enic turbulence\footnote{\citetalias{Skalidis2021} showed that omitting the coupling term in the estimation of the magnetic field strength in sub-Alfv\'enic turbulence, or equally applying the \citet{Davis1951} and \citet{Chandrasekhar_fermi_1953} method, can produce estimates which can be up to an order of magnitude larger than the actual values.}; this is strongly supported by our numerical results in \autoref{fig:coupling_ratio_ma0} and~\ref{fig:dbBo_ma}. In the same figures we show that $2^{\rm nd}$ order terms become significant only in super-Alfv\'enic turbulence. Therefore, one should clearly state the magnetisation level of turbulence (sub- or super-Alfv\'enic) before arguing about the relative contribution of the various terms in the energy balance. 
    
    \citet{Liu2021_cor_scales} further argues that self-gravity may modify this energy balance. We do not include gravity in this study, but it is possible that gravity may collapse locally bound (by self-gravity) regions in the ISM, enhancing and creating strong magnetic fields \citep{Sur2010}. This may make the coupling term even more relevant as the regions collapse, forming convergent flows parallel to the field lines, and strengthening the magnetic field and hence the local affect of $\dbBo$. This is speculative, and the exact effects of gravity are unclear; we will return to this topic in future work. Of course, all of this work is done in the isothermal context, so our relations are only relevant to individual phases of the ISM, which are well approximated by an isothermal equation of state \citep[e.g.,][]{Wolfire1995_isothermal_ISM}.
    
    \begin{figure*}
        \centering
        \includegraphics[width=\linewidth]{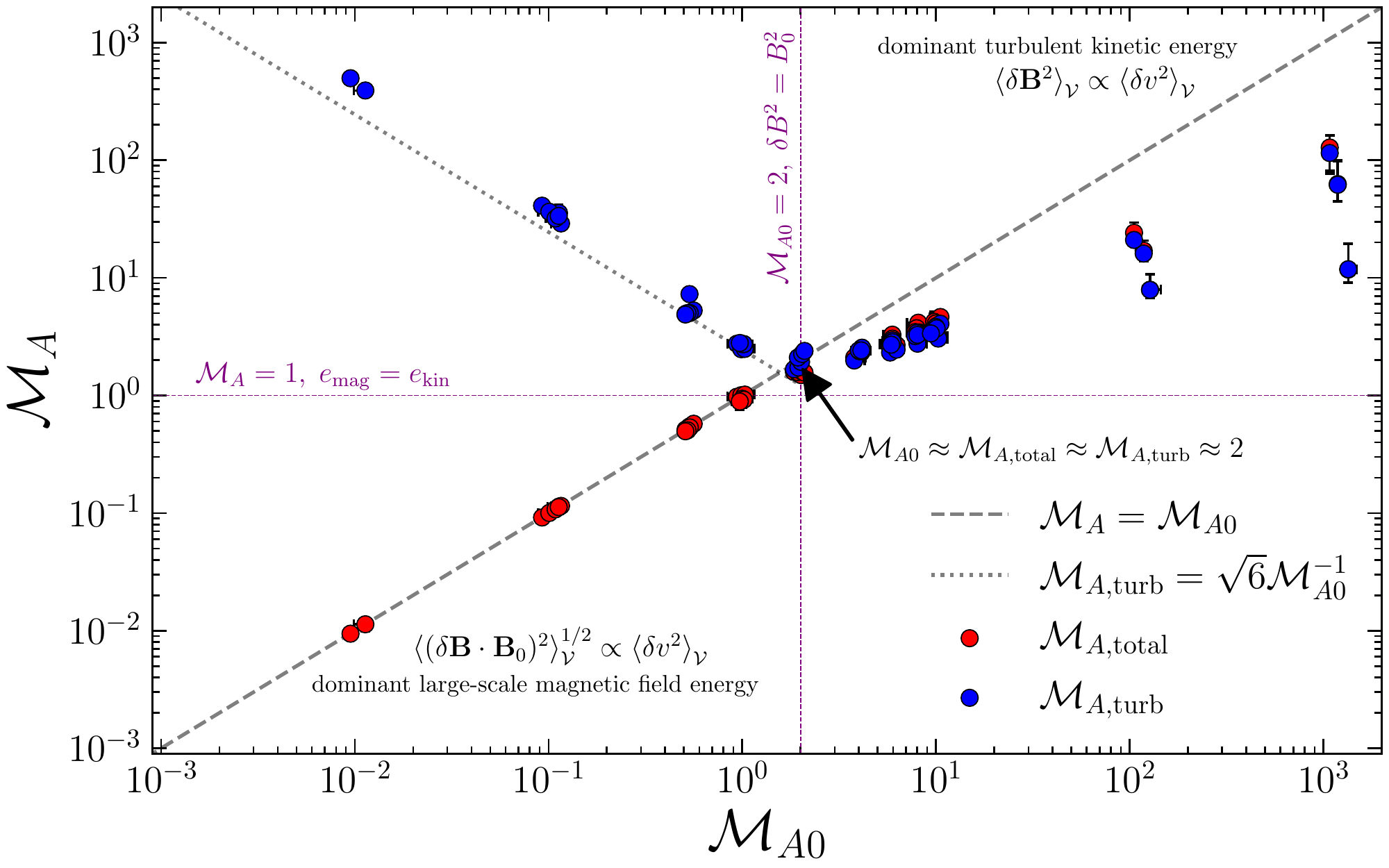}
        \caption{The total Alfv\'en Mach number, $\mathcal{M}_{A,\text{total}}$ (\autoref{eq:totalMa}), shown in red, and the turbulent Alfv\'en Mach number, $\mathcal{M}_{A,\text{turb}}$ (\autoref{eq:turbMa}), shown in blue, as a function of large-scale field Alfv\'en Mach number, $\Mao$ (\autoref{eq:meanMa}), for all simulations. The grey dashed line is the one-to-one line between the two Alfv\'enic Mach numbers and $\Mao$. The dotted line shows the model for $\mathcal{M}_{A,\text{turb}}$ for the $\Mao \leq 2$ (shown with vertical purple line) regime. The separation of $\Ma$ into field components shows explicitly that the turbulent component of the field is highly super-Alfv\'enic in the sub-Alfv\'enic large-scale field regime (i.e., the turbulent kinetic energy is much larger than the turbulent magnetic energy, shown with the horizontal purple line), and the large-scale field dominates the total magnetic energy, $\Mao\approx\mathcal{M}_{A,\text{total}}$, which coincides to $\dbBovol \propto \Exp{\delta v^2}_{\V}$, as discussed in \autoref{sec:strong-field_dbBo}. The transition into the super-Alfv\'enic large-scale field regime happens at a critical point in the $\Ma - \Mao$ diagram, $\Mao \approx \Matot \approx \Maturb \approx 2$, where there is energy equipartition between the turbulent and large-scale magnetic field. For $\Mao > 2$, the turbulent magnetic energy is greater than the energy in the large-scale field (vertical purple line), hence $\mathcal{M}_{A,\text{turb}} \approx \mathcal{M}_{A,\text{total}}$, and the turbulent kinetic energy is greater than the magnetic energy, corresponding to the $\Exp{\dB^2}_{\V} \propto \Exp{\delta v^2}_{\V}$ regime, as discussed in \autoref{sec:weak_field_limit}.}
        \label{fig:ma0-ma}
    \end{figure*}    
    
\section{The three Alfv\'en Mach Numbers}\label{sec:alfven_mach}

    \subsection{Definitions and results}
    
    The Alfv\'en Mach number, $\Ma$, is another part of the energy balance story, because the quantity itself is directly related to the energy equilibrium in the plasma,
    \begin{align}\label{eq:ma_energy}
        \mathcal{M}_{A}^{-2} = \Exp{\frac{B^2 / (8\pi)}{(1/2)\rho \delta v^2}}_{\V} = \Exp{\frac{\emag}{\ekin}}_{\V},
    \end{align}
    which is similar to $\sat$ in \autoref{eq:energy_balance_1}, but not exactly the same, because $\Exp{X/Y}_{\V} \neq \Exp{X}_{\V}/\Exp{Y}_{\V}$ if there are any correlations between $X$ and $Y$, as is the case for the magnetic and kinetic energy ($\sat = \MaO{-2/3}$, \autoref{sec:weak_field_limit}). 
    
    Throughout the previous section, we utilised $\Mao$ to construct our models around values of $\Bo$. We could do this easily because $\Mao$ is an input (or at least controlled, albeit with some small variation due to velocity fluctuations) in our simulations. However, in many astrophysical turbulence studies, authors prefer to use $\Ma$. For some of these studies, it is not clear if one should interpret this as the $\Ma$ with respect to just turbulent fluctuations, or the total field strength. The difference between these quantities is rarely appreciated, so we make a point by defining and relating three different canonical constructions of $\Ma$. The three definitions we use\footnote{We note there are, of course, even more definitions that one could in principle construct, for example, $\Ma = \Exp{\delta v}\sqrt{4\pi\rho_0}/\Exp{B}$, which we use to set the Alfv\'en Mach number in \autoref{sec:sims}, or one could even use component-wise constructions.} are
    \begin{align}
        \Mao &= \Exp{\frac{ \delta v\sqrt{4\pi\rho}}{B_0}}_{\V}, \label{eq:meanMa} \\  
        \Maturb &= \Exp{\frac{\delta v \sqrt{4\pi\rho}}{\delta B}}_{\V},\label{eq:turbMa} \\
        \Matot &= \Exp{\frac{\delta v\sqrt{4\pi\rho}}{B}}_{\V}, \label{eq:totalMa}
    \end{align}
    where the first of the three defines the large-scale field (or mean-field on the system scale) Alfv\'en Mach number, which compares the large-scale magnetic energy with the kinetic energy, the second is the turbulent-field Alfv\'en Mach number, which compares the turbulent magnetic energy with the kinetic energy, and the third, the total field Alfv\'en Mach number. 
    
    To understand the relation between the three quantities, we plot them in \autoref{fig:ma0-ma} ($\Matot$ in red, and $\Maturb$ in blue, both as a function of $\Mao$). The dashed, grey line shows the one-to-one line between $\Mao$ and $\Ma$. For $\Mao \lesssim 2$, $\Mao \approx \Matot$, which means the energetics of the fluid are completely dominated by the large-scale field, and not the turbulence at all. $\Maturb$ follows a power-law in $\Mao$ which prevents the fluctuating magnetic field from ever becoming stronger than $\Maturb \approx 2$. Once the turbulent field has reached $\Ma \approx 2$, it then begins weakening again, but this time $\Maturb \approx \Matot$, with $\Matot \lesssim \Mao$, i.e., transitioning into a turbulent magnetic field dominant regime as the $\vecB{B}$-field becomes tangled and more energy dense than the large-scale field. As discussed in \autoref{sec:strong-field_dbBo}, the $\Mao = 2$ transition between the sub-and-super-Alfv\'enic regimes defines exactly when the $B_0$ and $\delta B$ field are equal energy in energy, and the transition between $\dbBovol \propto \Exp{\delta v^2}_{\V}$ and $\Exp{\delta B^2}_{\V} \propto \Exp{\delta v^2}_{\V}$, as annotated in the plot.
    
    We are able to derive the relation between $\Maturb$ and $\Mao$ in the sub-Alfv\'enic regime by rearranging the coupling magnetic components, $\dbBovol = \Exp{\delta B_{\parallel}^2}^{1/2}_{\V} B_0$, on the RHS and turbulent components on the LHS of \autoref{eq:e_balance_sub_limit},
    \begin{align} \label{eq:ma_turb_01}
        \frac{2 \Exp{\delta B_{\parallel}^2}^{1/2}_{\V}}{\delta v \sqrt{4\pi\rho_0}} = \Mao.
    \end{align} 
    But now we need to use total fluctuating magnetic field, not just the parallel field, to get the complete $\Maturb$. In \aref{app:anisotropy} we directly measure the different field components and relate them to the total fields. The most strongly-anisotropic regime, in the highly-sub-Alfv\'enic turbulence, corresponds to $(1/3)\Exp{\dBpar^2}^{1/2}_{\V} \leq \Exp{\dBperp^2}^{1/2}_{\V} \leq (2/3)\Exp{\dBpar^2}^{1/2}_{\V}$ and $\Exp{\delta v_{\parallel}^2}^{1/2}_{\V} \approx (2/3)\Exp{\delta v_{\perp}^2}^{1/2}_{\V}$. For the magnetic fluctuations, we pick the average between these two values, $\Exp{\dBperp^2}^{1/2}_{\V} = (1/2) \Exp{\dBpar^2}^{1/2}_{\V}$, and then propagate both the magnetic anisotropy through the regular vector magnitude equations, which gives $\Exp{\delta B^2}^{1/2}_{\V} = \sqrt{3/2}\Exp{\dBpar^2}^{1/2}_{\V}$. Substituting this back into \autoref{eq:ma_turb_01} gives,
    \begin{align}\label{eq:ma_turb_02}
        \Maturb = \sqrt{6}\MaO{-1}, \,\,\Mao \leq 2,
    \end{align}
    which we plot with the grey dotted line in \autoref{fig:ma0-ma}. This simple model intersects with the $\Ma = \Mao$ line at the $\Matot \approx \Maturb$ transition. At lower $\Mao$ there is some deviation away from the model, which is because of the stronger than average magnetic fluctuation anisotropy present in the $\Mao \ll 1$ data. 
    
    \subsection{Hypothesis on limiting behaviour}
    
    The discussion in the preceding sections leads us to propose a hypothesis regarding the limiting behaviour of MHD turbulent systems in \autoref{fig:ma0-ma}, which we illustrate schematically in \autoref{fig:ma_schematic}.
    
    \subsubsection{$\Mao \rightarrow 0$} 
    
    As $\Mao\rightarrow 0$, the turbulent field should continue to become weaker and weaker ($\Maturb \gg 1$). The reason for this is that the magnetic fluctuations (specifically the shear Alfv\'en waves) are smoothed out by the increasing magnetic tension, \protect{$\tens$}, and are reduced in degrees of freedom since they are perfectly flux-frozen into $\Bo$. Therefore, instead of any field-line stretching, coherent magnetic field-lines are randomly walked in the perpendicular plane to $\Bo$ (field line random walk, \citealt{Jokipii1968_field_line_random_walk}). 2D planar motions cannot instigate dynamo action (\citealt{Zeldovich_planar_dynamo_theory} theorem), so it is unlikely that $\delta B$ ever grows irreversibly\footnote{Note that it may grow locally, through reversible processes such as compression, but these ought to average out over time.} again in this limit, and we find $\Exp{\delta B^2}^{1/2}_{\V_L} \propto B_0^{-1}$ (\autoref{eq:ma_turb_02}; consistent with qualitative observations in previous studies, \citealt{Haugen2004_large_scale_field_supression}), until the magnetic field only has a large-scale component and $B = B_0$. 
    
    \subsubsection{$\Mao \rightarrow \infty$}
    
    In the $\Mao \rightarrow \infty$ $(B_0\rightarrow 0)$ limit, we reach the results from the small-scale dynamo community. Very broadly speaking, in these studies, where $B_0 = 0$, and hence $\Mao \rightarrow \infty$, the equilibrium magnetic field strength asymptotes to a value that depends on the sonic Mach number and ratio of compressive to solenoidal modes in the turbulence, and plasma Reynolds numbers (if they are finite). In this limit, the maximally efficient turbulent dynamo is for the most sub-sonic, solenoidal flows and the least efficient for the highest-$\M$, most compressible flows \citep{Federrath2014,Schober2015,Federrath2016_dynamo,Chirakkara2021}. Hence, as $\Mao$ becomes larger, we should expect to observe the $\Ma(\Mao)$ curves separate and asymptote to different constant values of $\Maturb$ as a function of $\M$, which is what we find in \autoref{fig:ma0-ma}. Since the most efficient turbulent dynamo\footnote{Note this is in absence of magnetic helicity, which may significantly change the saturation of the dynamo given that there are more degrees of freedom to store magnetic energy than in non-helical turbulence and magnetic modes above the outer scale of the turbulence can be energised \citep[e.g., \S4-5 in][]{Rincon2019_dynamo_review}.} leads to saturation of $\Matot \approx 2$, (see highly sub-sonic, solenoidal experiments in \citealt{Chirakkara2021}) this defines a floor that bounds $\Ma$ from below as $\Mao \rightarrow \infty$. 
    
    \begin{figure}
        \centering
        \includegraphics[width=\linewidth]{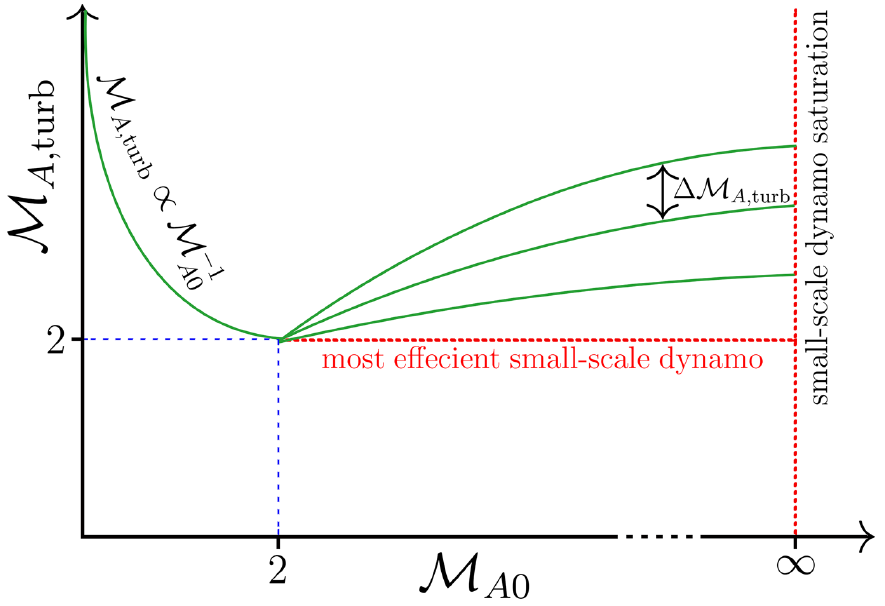}
        \caption{Schematic for the $\Maturb - \Mao$ plane, showing the small-scale dynamo saturation in the $\Mao \rightarrow \infty$ limit, and the $\Maturb \propto \MaO{-1}$ $(\Exp{\delta B^2}^{1/2}_{\V_L} \propto B_0^{-1})$ power-law in the $\Mao \rightarrow 0$ limit, derived using energy balance. The separation between $\Maturb(\Mao)$ curves, $\Delta\Maturb$, changes with different plasma parameters that control the small-scale dynamo saturation.}
        \label{fig:ma_schematic}
    \end{figure}
    
    \subsection{Sub-Alfv\'enic turbulent fields do not exist}\label{sec:Ma_discussion}
    
    We find that the fluctuating magnetic field becomes extremely weak at $\Mao < 2$, and is bounded from below by the most efficient saturation of the small-scale dynamo in the limit $\Mao\to\infty$. An immediate consequence of these two limiting behaviours is that there is no room in the $\Ma-\Mao$ plane for $\Maturb \lesssim 2$ turbulence. Hence the only sub-Alfv\'enic turbulence that is possible in this parameter space is sub-Alfv\'enic \textit{large-scale field} (or coherent field) fluid turbulence. We show that this kind of turbulence is highly super-Alfv\'enic with respect to the turbulent velocity fluctuations (shocks and vortices), $\Maturb \gg 1$, and it is only through the non-turbulent components of the plasma that the magnetic energy is able to be sufficiently stronger than the kinetic energy. As a caveat of this analysis, in this section, we volume-averaged the plasma beyond the correlation scale of the turbulence, hence this does not rule out that the turbulence can be sub-Alfv\'enic on scales much smaller than the correlation scale. In the next section we explore averaging below the correlation scale, and discuss how sub-Alfv\'enic turbulence can emerge by taking statistics below the correlation scale.
    
    \begin{figure}
        \centering
         \includegraphics[width=\linewidth]{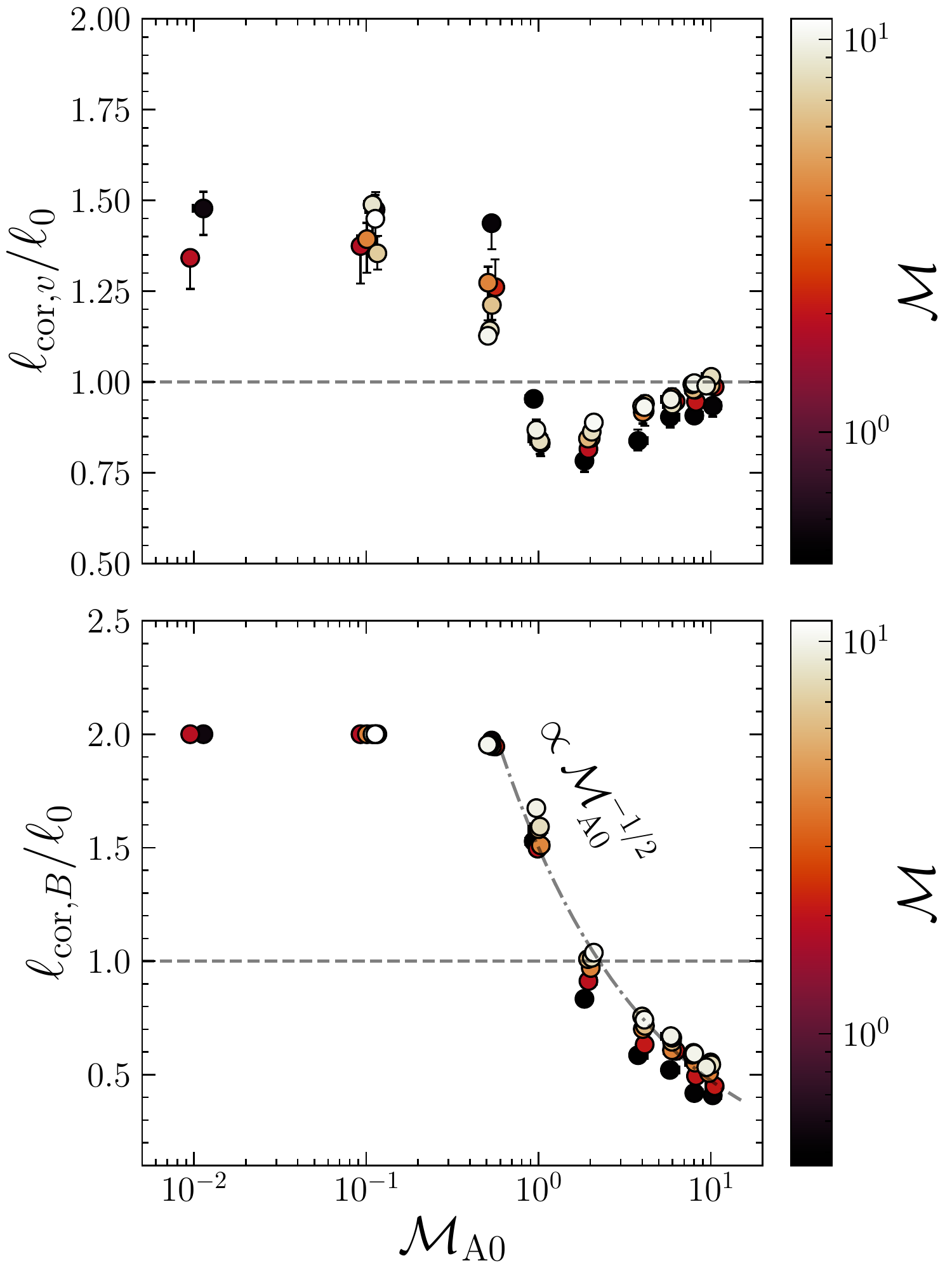}
        \caption{\textbf{Top:} The turbulent correlation scale, $\cor{v}$ (\autoref{eq:correlation_scale}), in units of turbulent driving scale, $\ell_0$, as a function of $\Mao$, for all simulations. We find $\cor{v}\approx \ell_0$, with some systematic deviation at low-$\Mao$, most likely due to the strong, global anisotropy in those simulations. \textbf{Bottom:} The same, but for the correlation scale of the magnetic field, $\cor{B}$. The scatter at each $\Mao$ is determined by $\M$, which ranges between $\M \approx 0.5 - 10$ and only weakly changes the correlation scale of the turbulence.}
        \label{fig:turb_correlation_scale}
    \end{figure}

    \begin{figure}
        \centering
        \includegraphics[width=\linewidth]{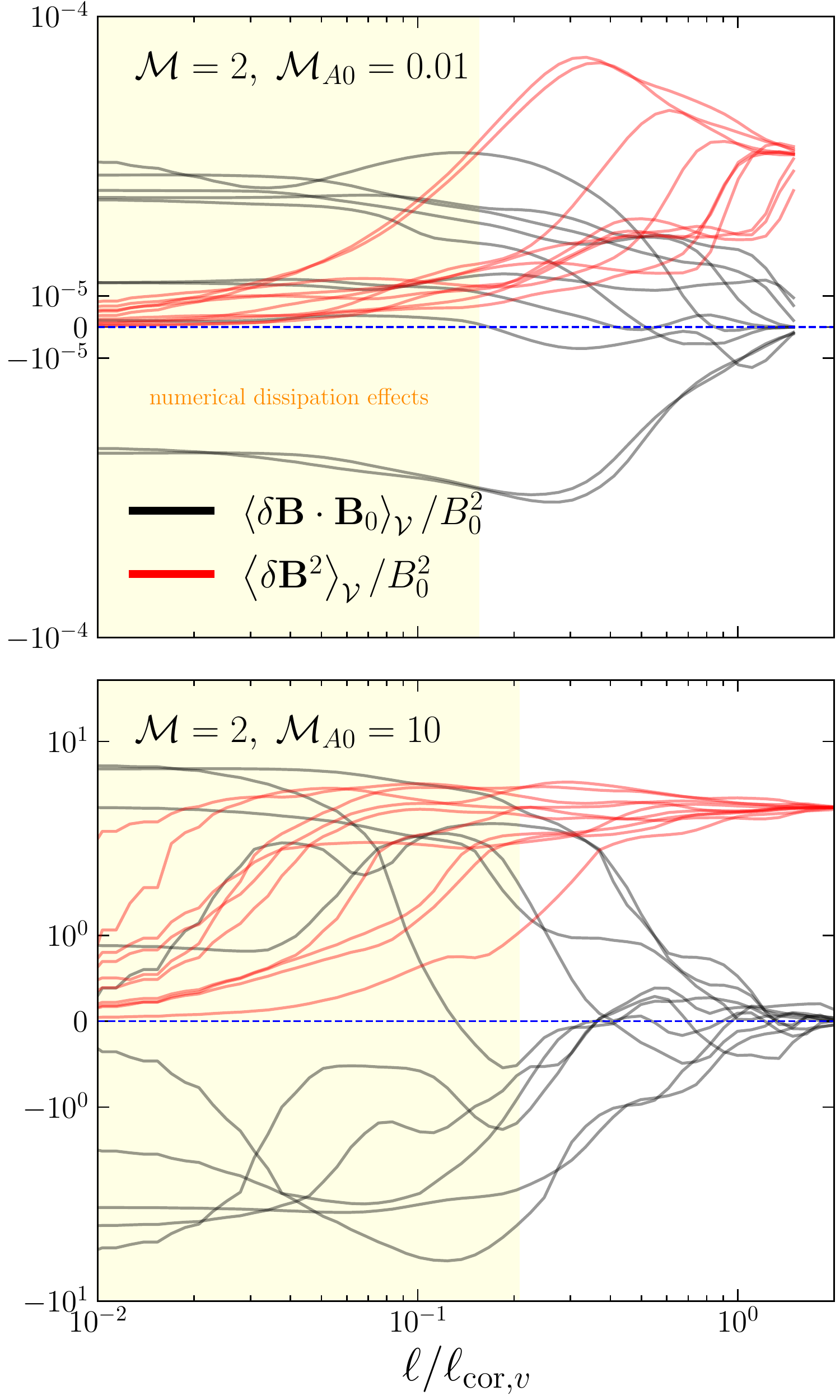}
        \caption{Coupling and mean squared values of magnetic fluctuation terms in units of large-scale field as a function of averaging scale normalised to the correlation scale of the turbulence, $\cor{v}$, for the 10 randomly sampled regions in the \texttt{M2MA001} (top) and \texttt{M2MA10} (bottom) simulations. The square magnitude of $\dB$ on scale $\ell/\cor{v}$ is shown in red, and the mean of $\dbBo$ on that scale is shown in black. The yellow shaded region indicates where numerical dissipation effects may influence the field statistics. Due to the spatial correlation of the magnetic field, on scales $\ell < \cor{v}$ (volumes $\V \lesssim \cor{v}^3)$, turbulent fields are converted into effective mean-fields. This means $\left\langle \dbBo \right\rangle_{\V} \neq 0$ on scales below $\cor{v}$, acting as an effective mean-field on that scale. On scales above $\cor{v}$ the Reynolds rule of averaging holds and $\left\langle \dbBo \right\rangle_{\V} = 0$ as expected.}
        \label{fig:scale_dependent_B}
    \end{figure}

    \section{The averaging scale}\label{sec:consideration}
    
    We have shown it is important to average \autoref{eq:energy_balance_1} in such a way that all of the terms are positive-definite, and that $\dbBovol$ balances the kinetic turbulent energy perfectly in the sub-Alfv\'enic large-scale field turbulence regime. However, these methods critically rely on an averaging scale $\V$. In this section we highlight the importance of $\V$ and show that even if one adopts the traditional ansatz that does not enforce positivity, \autoref{eq:energy_balance}, the coupling term can make non-zero contributions to the turbulent energy.

    The fundamental reason for this is that the volume averaging scale for $\Exp{\hdots}_{\V}$ is important. In simulations, we regularly report volume averaged statistics over a few turbulent correlation scales, $\cor{v}$, i.e., $\V = \bigcup_{i = 1}^N(\ell_{\text{cor},v}^3)_i$, where $N$ is a few, or directly at the full size of a simulation box, $\V = \V_L$. However, for many observations of the ISM, the region sampled is far smaller than the turbulent correlation scale. For example, magnetic fields in star-forming clouds are observed to be correlated on scales up to $\sim 100\,\rm{pc}$ \citep{Li14b}, comparable to the scale height and the outer scale of turbulence in galactic discs \citep[e.g.][]{Karlsson2013_turbulence_scale,Goncalves2014_driving_scale,Krumholz2018_metallicity_SF}. Dust polarisation observations using \textit{Herschel} (e.g., the \textit{Herschel} Gould Belt Survey -- \citealt{Andre2010}) generally sample\footnote{In observations we can define a sampling scale as the maximum spatial separation of observational data in the plane of the sky.} much smaller fields of view (of order $10\,\rm{pc}$); over size scales typically probed by such observations, there is no sign of a flattening in the velocity dispersion-size relation \citep{Federrath2021, Yun2021_internal_MC_scaling_law, Zhou2021}, clear evidence that the region being studied is much smaller than the correlation scale.

    To explore the implications of this, we directly compute the correlation scales of the turbulence $\cor{v}$ and magnetic field $\cor{B}$ in our simulations, and plot them as a function of $\Mao$, coloured by $\M$ in \autoref{fig:turb_correlation_scale}. We compute both of them in the textbook manner, directly from the energy spectra, $\mathscr{P}_v(k)$, and $\mathscr{P}_B(k)$ as
    \begin{align}\label{eq:correlation_scale}
        \frac{\ell_{\text{cor}}}{\ell_0} = \frac{L}{\ell_0}\frac{\displaystyle\int_0^{\infty} \d{k}\, k^{-1} \mathscr{P}(k)}{\displaystyle\int_0^{\infty} \d{k}\,\mathscr{P}(k)},
    \end{align}
    where $\mathscr{P}(k)$ is replaced by $\mathscr{P}_v(k)$ for the turbulence correlation scales, and $\mathscr{P}_B(k)$ for the magnetic field correlation scales. 
    
    We focus first on the top panel of \autoref{fig:turb_correlation_scale}, the turbulence correlation scale. The super-Alfv\'enic large-scale field experiments have $\cor{v}\approx \ell_0$, with a small dip at $\Mao = 2$ as the turbulence transitions between $\Bo$ and $\dB$ dominated (as discussed in \autoref{sec:energy_balance}). The sub-Alfv\'enic turbulence has correlations scales above the driving scale, most likely due to the system-scale vortices that develop in the flow \citep{Beattie2020c,Beattie2021b}. If we interpret this experiment at face value, this means that if individual clouds are sub-Alfv\'enic, we should expect correlated turbulent velocities beyond the extent of the entire sub-Alfv\'enic region, even if the driving scale is not itself larger than the individual clouds. For the ISM in general, which is probably trans-Alfv\'enic-to-super-Alfv\'enic ($\Mao \approx 2$) and trans-sonic on average \citep{Gaesnsler_2011_trans_ISM,Krumholz2020,Liu2021_cor_scales,Seta2021b}, we should expect turbulent motions to be correlated out to the driving scale of the largest turbulent motions. This, of course, is a natural repercussion of one of the central tenets of turbulence: the energy cascade from large (galactic, in this context) to small \citep[molecular cloud and smaller,][]{Armstrong1995_power_law} scales. 
    
    Similar to the turbulence correlation scales, the magnetic correlation scales of the sub-Alfv\'enic simulations are on larger scales than the driving scale, indicating, as we showed in \autoref{sec:alfven_mach}, that the fluctuating magnetic field is negligible in the sub-Alfv\'enic regime, and is strongly suppressed by the large-scale field. The super-Alfv\'enic simulations show a decaying power-law $\cor{B}\propto \ell_0\MaO{-1/2}$, which demonstrates that as the large-scale field weakens, the causally connected regions in the magnetic field move to smaller and smaller scales \citep[qualitatively consistent with previous expectations, e.g.,][]{Lazarian2006_cr_scattering}. This is likely due to the strong turbulent motions tangling the magnetic field \citep[e.g., \S 4.1.1. in ][]{Sampson2022_SCR_diffusion}, increasing the net curvature \citep{Yuen2020} and facilitating a smaller scale field\footnote{In the sense that the ratio of the magnetic energy in the low $k$ modes to the magnetic energy in the high $k$ modes is shrinking.}. This means, to place the correlation scale of the magnetic field on comparable scales of a $10\,\rm{pc}$ observation \citep[e.g.,][]{Federrath2016_brick,Panopoulou2016_polaris_b_field,Beattie2019b,Hu2019}, for $\ell_0 = 100\,\rm{pc}$ we require that $\Mao \gtrsim 100$, which on average is unrealistically high for the $\Mao$ in the disc of Milky Way analogues \citep{Wibking2021_magnetised_galaxy,Hopkins2021_testing_gev_cr_models}. Likewise, for average ISM parameters that we use from the above discussion ($\ell_0 = 100\,\rm{pc}$, $\Mao = 2$), $\cor{B}\propto \ell_0\MaO{-1/2}$ gives $\cor{B} \approx 70\,\rm{pc}$, which determines the largest scale in which the magnetic field can be casually connected via magnetic field fluctuations, when driven at $100\,\rm{pc}$.
    
    The significance of this for magnetic energy balance, in both simulations and observations, is that we are often dealing with volumes $\V \ll \V_L$, and while $\Exp{\dbBo}_{\V_L} = 0$, in general $\Exp{\dbBo}_{\V} \neq 0$ when $\V < \V_L$. More generally, for random fields $\vecB{X}$ and $\vecB{Y}$, averaged on volumes $\V<\V_L$, $\langle\Exp{\vecB{X}}_{\V} \vecB{Y}\rangle_{\V} \neq \Exp{\vecB{X}}_{\V}\Exp{\vecB{Y}}_{\V}$ and without loss of generality, $\Exp{\delta \vecB{X}}_{\V} \neq 0$, i.e., the Reynolds rule of averaging is no longer valid \citep{Germano1992,Hollins2018}. We show an explicit example of this for sub-Alfv\'enic \texttt{M2MA001} (top) and super-Alfv\'enic \texttt{M2MA10} (bottom) simulations, mimicking the trans-sonic average ISM in \autoref{fig:scale_dependent_B}. We provide a full description of the methodology for performing the experiment in \aref{app:length_scale_average_method}, but to summarise here, we use real-space spherical top-hat filters initialised at random coordinates in each the simulation, each with diameters $\ell/\cor{v}$, to compute the mean-squared (red) and mean (black) of the filtered $\dB$ and $\dbBo$ fields, in units of the large-scale field, respectively; $\cor{v}$ is computed independently, directly from the velocity power spectra, \autoref{eq:correlation_scale}. We indicate in yellow the scales for which numerical dissipation may influence the rms statistics, which can be up to $\sim30\,$grid cells for our (and other grid) simulation solvers \citep[see \S2 in][for further details]{Kitsionas2009_dissipation_comparison,Federrath2011_numerical_dissipation}.
    
    \begin{figure}
        \centering
        \includegraphics[width=\linewidth]{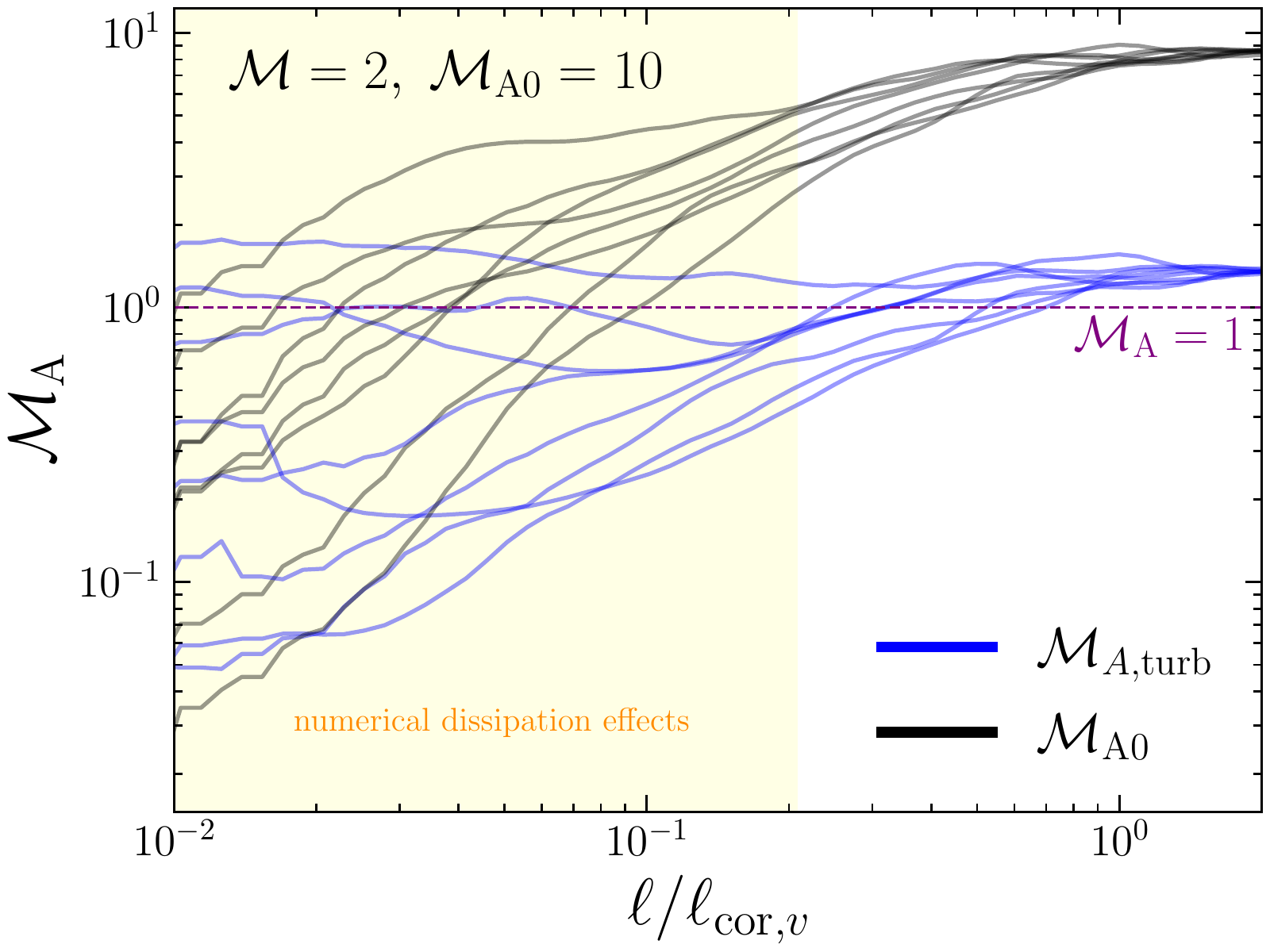}
        \caption{The same as the bottom panel of \autoref{fig:scale_dependent_B}, but for the scale-dependent Alfv\'en Mach numbers of the large-scale field, \autoref{eq:meanMa}, and the turbulent field, \autoref{eq:turbMa}.}
        \label{fig:scale_dependent_Ma}
    \end{figure}    
    
    For both simulations, we find that on scales $\ell < \cor{v}$, $\Exp{\dbBo(\ell/\cor{v})}_{\V}\neq0$, coupled with a small $\Exp{\dB^2(\ell/\cor{v})}_{\V}$ compared to the system-scale $\Exp{\dB^2}_{\V}$. This is a natural repercussion of finite spatial correlation in the plasma, where local regions in the turbulence, $\ell^3 < \cor{v}^3$, can have fields that appear ordered, even though they are part of the global fluctuating field, as evident from \autoref{fig:scale_dependent_B}. When averaging over these filtered regions the spatially correlated fluctuating field acts as an effective large-scale field on that scale. The main difference between the two simulations is the size of the fluctuating field, which is orders of magnitude smaller on all scales in the sub-Alfv\'enic simulation, as expected from our previous discussions in \autoref{sec:alfven_mach}. Our analysis also illustrates the difficulty of distinguishing between large and small-scale magnetic fields when one is making observations well below $\cor{v}$, and provides a very clear reason why $\dbBo$ may be an important quantity for magnetic field observations made over a finite field of view.  
    This finding has strong implications for the interpretation of observations. As we discussed in \autoref{sec:intro}, some ISM observations suggest that clouds are in a sub-Alfv\'enic state \citep{HuaBai2013,Federrath2016_brick,Hu2019,Heyer2020,Hwang2021,Hoang2021,Skalidis2021_obs_sub_alf}. Based on our analysis in \autoref{sec:alfven_mach}, this means that a very strong, large-scale field must be present. Naively, the small-scale dynamo, which is generally invoked to explain the magnetic field strengths in the ISM, should not be able to maintain such a system \citep[the most efficient dynamo saturates at $\Ma \approx 2$, ][]{Federrath2016_dynamo}, with all of the magnetic energy being stored in the large-scale field. An $\alpha$-$\Omega$ dynamo that can grow a large-scale, coherent magnetic field through the Parker loops $(\alpha$; \citealt{Parker1979_B_fields_book}) or differential, possibly galactic, rotation $(\Omega$; see \S2.6 in \citealt{Beck_2013_Bfield_in_gal}) may be required to grow such a field at the kpc scale, that is coupled to the ISM of the galaxy, piercing individual clouds and making them highly magnetised. 
    
    Our current analysis suggests an alternative possibility: one way of creating an effective mean-field, which may act like a large-scale field for scales below it (e.g., for a sub-Alfv\'enic plasma embedded in a super-Alfv\'enic plasma), is by taking filtered statistics of the turbulence, and hence observing the fluctuating field well below the correlation scale of the turbulence. Because this process turns fluctuating field into an effective large-scale field, it facilitates the perfect conditions for moving left in \autoref{fig:ma0-ma}, with sub-dominant magnetic field fluctuations and a strong coherent field. In \autoref{fig:scale_dependent_Ma} we show the same filtered turbulence calculation as in \autoref{fig:scale_dependent_B} but now instead with $\Mao(\ell/\cor{v})$ (black curve) and $\Maturb(\ell/\cor{v})$ (blue curve). We use the \texttt{M2MA10} simulation, which is globally super-Alfv\'enic, $\Mao(L/\cor{v}) \approx 10$, as indicated to the far right of the black curve. On scales smaller than the $\cor{v}$, a majority of the random samples exhibit $\Maturb(\ell/\cor{v}) < 1$, and likewise for $\Mao(\ell/\cor{v})$, albeit over a narrower range in $\ell/\cor{v}$. In the context of simulations, this shows that the statistics of small regions in the turbulence can be effectively sub-Alfv\'enic, even in a globally super-Alfv\'enic plasma. In the context of observations of the cold, molecular ISM, it means that even though individual clouds may be observed to be sub-Alfv\'enic, the magnetic fields in these clouds may still be the result of a small-scale dynamo process, saturating at super-Alfv\'enic values, but operating on scales much larger than the cloud being observed. Thus observing a cloud to be sub-Alfv\'enic, $\Ma \la 2$ does \textit{not} automatically mean that the field in that cloud is the product of an $\alpha$-$\Omega$ or similar large-scale dynamo; one can conclude that such a process is at work only if one recovers $\Ma \la 2$ \textit{on scales larger than the turbulent correlation length}.
    
\begin{figure}
        \centering
        \includegraphics[width=\linewidth]{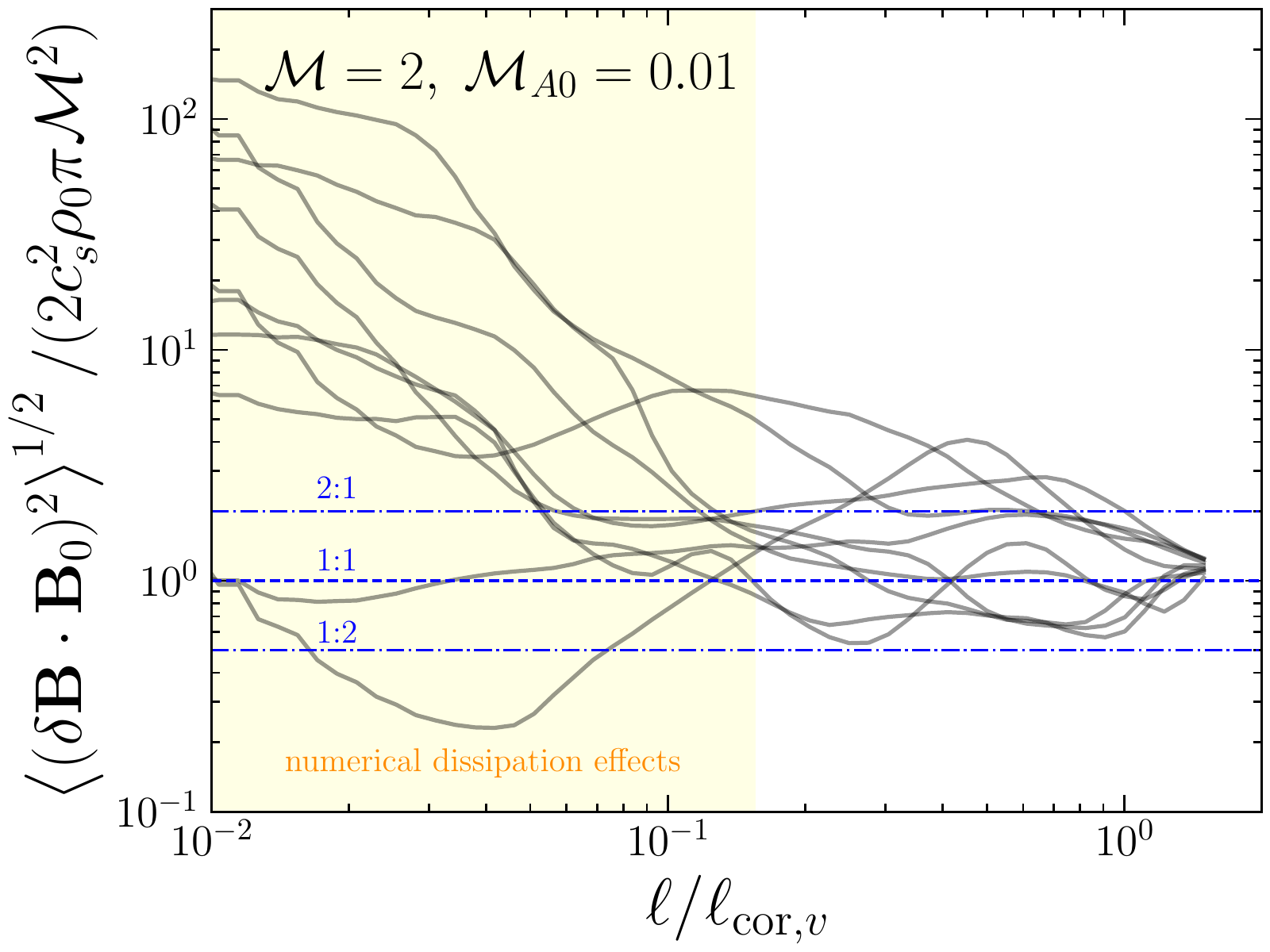}
        \caption{The same as the top panel of \autoref{fig:scale_dependent_B}, but for the scale-dependent energy balance, in the strong-field limit \autoref{eq:e_balance_sub_limit}. We show lines of 1:1 (blue, dashed), 2:1 and 1:2 (blue, dot-dashed), highlighting that between the largest scales in the simulations and the dissipation scales (indicated with the yellow shaded region) almost all of the random regions sampled fall within a factor of 2 in exact equipartition.}
        \label{fig:scale_dependent_balance}
    \end{figure}

    As a final calculation for this study, we compute the energy ratio of the coupling term to the kinetic energy, $\dbBovol / (2c_s^2\rho_0\pi\mathcal{M}^2)$ (the ratio of the left- to right-hand side of \autoref{eq:e_balance_sub_limit}) in a sub-Alfv\'enic plasma as a function of scale, just as we did in the previous paragraphs for the other rms statistics. This tests if the energy balance we presented in \autoref{sec:energy_balance} is valid over a range of scales, necessary for making it a useful relation for applications. We plot $\dbBovol / (2c_s^2\rho_0\pi\mathcal{M}^2)$ as a function of $\ell/\cor{v}$ in \autoref{fig:scale_dependent_balance}, for the same simulation as in the top panel of \autoref{fig:scale_dependent_B}. We find that on the interval between the system scale $L$ and the scale in which numerical dissipation effects exist $\ell_{\nu}$ (the largest scale which is shaded yellow) that our filtered samples of the turbulence mostly fall within the 1:2 to 2:1 interval (blue, dot-dashed lines). On average, across all samples and $\ell \in [\ell_{\nu},L]$, we find $\dbBovol / (2c_s^2\rho_0\pi\mathcal{M}^2) = 1.5 \pm 1$, capturing the exact equipartition within 0.5$\sigma$. Below $\ell_{\nu}$ most of the samples become highly-magnetised, due to the kinetic energy being dominated by numerical dissipation and the large-scale magnetic field permeating through all of the scales in the plasma.
    
\section{Summary and key findings}\label{sec:conclusion}

    Motivated by recent works on measuring and modelling magnetic fields in the ISM \citep{Beattie2020,Skalidis2020,Skalidis2021}, we provide a theoretical and numerical exposition of the root-mean-squared (rms) energy balance between the kinetic and magnetic energy, highlighting the role of the magnetic coupling term, $\dbBovol$, which describes the energy contained in magnetic fluctuations $\dB$ coupled to the large-scale magnetic field $\Bo$. We discuss the significance of this term in the context of the $1^{\rm st}$ (comparing volume-averaged energies) and $2^{\rm nd}$ (comparing magnitudes of energy fluctuations) moments of the energy balance equations, deriving its typical value directly from the $2^{\rm nd}$ moment equations, which preserve the positivity of each energy contribution. From this argument we derive a number of analytical models with no free parameters, for the coupling term and fluctuating magnetic field, $\dB$, and demonstrate that these yield outstanding agreement with the results of a large suite of MHD simulations. Our analysis demonstrates that $\dbBovol$ plays an important role in sub-to-trans-Alfv\'enic large-scale field turbulence, regardless of the sonic Mach number $\M$. This term becomes less important for $\Mao > 2$, where $\delta B^2$ becomes dominant, but the large-scale field still has an effect. In \autoref{sec:alfven_mach} we explore three different formulations of the Alfv\'enic Mach number $\Ma$, and the relations between them, showing that sub-Alfv\'enic \textit{large-scale} field turbulence, supports an extremely super-Alfv\'enic turbulent field, suggesting that the magnetic field fluctuations are smaller than velocity fluctuations in this limit. We present a heuristic for understanding the whole turbulent and large-scale field Alfv\'en Mach number parameter plane and discuss the implications for interpreting ISM observations and sub-Alfv\'enic turbulence. We list the key results of this study below:
    \begin{itemize}
        \item We provide theoretical models for the volume-averaged fluctuating and coupling magnetic fields, $\Exp{\delta \vecB{B}^2}$ and $\dbBovol$, assuming energy equipartition between $\dbBovol$ and the volume-averaged velocity fluctuations, $\Exp{\delta v^2}_{\V}$ in the sub-Alfv\'enic regime (\autoref{eq:strong_B_coupling_term}, \autoref{eq:strong_B_coupling_ratio}), and $\Exp{\dB^2}_{\V}$ and $\Exp{\delta v^2}_{\V}$ in the super-Alfv\'enic regime (\autoref{eq:weak_B_coupling_term}, \autoref{eq:weak_B_coupling_ratio}). These models are free of parameters, but rely on the numerical observation that $\Exp{\delta v^2} \approx \Exp{\delta v^4}^{1/2}$ and $\Exp{\delta B^2} \approx \Exp{\delta B^4}^{1/2}$, i.e., that the average energy scales with the magnitude of energy fluctuations, which we demonstrate in \autoref{fig:second_fourth_order}. Our models show excellent agreement with numerical compressible MHD data, over a very broad range of plasma values, in \autoref{fig:B_coupling}, \autoref{fig:coupling_ratio_ma0} and \autoref{fig:dbBo_ma}. We discuss how this provides strong support for the polarisation dispersion models (DCF-like methods) derived in \citet{Skalidis2020} and \citet{Skalidis2021}.\\
        
        \item We define large-scale field, turbulent and total Alfv\'en Mach numbers (\autoref{sec:alfven_mach}) and propose that we can completely define the whole $\Maturb-\Mao$ data plane, shown in \autoref{fig:ma_schematic}, based on the small-scale dynamo saturation as $\Mao \rightarrow \infty$, and an analytical model that we derive using energy balance for $\Mao \rightarrow 0$, \autoref{eq:ma_turb_02}, which implies $\Exp{\delta B^2}^{1/2}_{\V_L} \propto B_0^{-1}$. Critically, we show that the turbulent magnetic field never becomes sub-Alfv\'enic, and it is only through a strong, large-scale magnetic field that the turbulence can transition into this regime. We show that the turbulence becomes highly super-Alfv\'enic in the sub-Alfv\'enic large-scale field regime, and discuss the implications for sub-Aflv\'enic ISM observations in \autoref{sec:Ma_discussion}. We suggest that a contributing factor to sub-Alfv\'enic ISM observations may be from measuring a trans-to-super Alfv\'enic average ISM, which, unlike sub-Alfv\'enic turbulence, may be supported by a small-scale dynamo, well below the correlation scale of the turbulence. We show that this is true for simulations of globally super-Alfv\'enic turbulence in \autoref{fig:scale_dependent_Ma}. \\  
        
        \item In \autoref{fig:scale_dependent_B} we explicitly show that by measuring filtered magnetic field statistics below the correlation scale of the turbulence (\autoref{eq:correlation_scale}), which is roughly equal to the driving scale with some slight deviations in the sub-Alfv\'enic regime, (\autoref{fig:turb_correlation_scale}), one turns the fluctuating field into an effective mean-field. We show this is true for both the sub-Alfv\'enic and super-Alfv\'enic regime. This highlights that for quantities such as $\Exp{\dbBo}_{\V}$ or $\Exp{\dB^2}_{\V}$, if the volume averaging scale, $\V$, does not resolve the correlation scale of the turbulence, then the volume average need not be zero, as previously discussed in \citet{Germano1992} and \citet{Hollins2018}. Furthermore, in \autoref{fig:scale_dependent_balance}, we show our sub-Alfv\'enic energy balance model provides good agreement across the resolved scales available to us in the sub-Alfv\'enic regime.
    \end{itemize}

\section*{Acknowledgements}
    We thank the anonymous referee for the useful suggestions that increased the quality of our study.
    J.~R.~B.~thanks Christoph Federrath's and Mark Krumholz's research groups for many productive discussions and acknowledges financial support from the Australian National University, via the Deakin PhD and Dean's Higher Degree Research (theoretical physics) Scholarships and the Australian Government via the Australian Government Research Training Program Fee-Offset Scholarship.
    M.~R.~K. acknowledges support from the Australian Research Council's \textit{Discovery Projects} scheme, award DP190101258.
    R.~S.~acknowledges financial support from the European Research Council (ERC) under the European Unions Horizon 2020 research and innovation programme under grant agreement No. 771282.
    C.~F.~acknowledges funding provided by the Australian Research Council (Future Fellowship FT180100495), and the Australia-Germany Joint Research Cooperation Scheme (UA-DAAD). J.~R.~B. and C.~F. further acknowledge high-performance computing resources provided by the Leibniz Rechenzentrum and the Gauss Centre for Supercomputing (grants~pr32lo, pn73fi, and GCS Large-scale project~22542), and the Australian National Computational Infrastructure (grant~ek9) in the framework of the National Computational Merit Allocation Scheme and the ANU Merit Allocation Scheme.\\
    
    The simulation software, \textsc{flash}, was in part developed by the Flash Centre for Computational Science at the Department of Physics and Astronomy of the University of Rochester. Data analysis and visualisation software used in this study: \textsc{C++} \citep{Stroustrup2013}, \textsc{numpy} \citep{Oliphant2006,numpy2020}, \textsc{matplotlib} \citep{Hunter2007}, \textsc{cython} \citep{Behnel2011}, \textsc{visit} \citep{Childs2012}, \textsc{scipy} \citep{Virtanen2020},
    \textsc{scikit-image} \citep{vanderWalts2014}.

\section*{Data Availability}
    The data underlying this article will be shared on reasonable request to the corresponding author.

\bibliographystyle{mnras.bst}
\bibliography{Jan2022.bib} 

\begin{thebibliography}{}
\makeatletter
\relax
\def\mn@urlcharsother{\let\do\@makeother \do\$\do\&\do\#\do\^\do\_\do\%\do\~}
\def\mn@doi{\begingroup\mn@urlcharsother \@ifnextchar [ {\mn@doi@}
  {\mn@doi@[]}}
\def\mn@doi@[#1]#2{\def\@tempa{#1}\ifx\@tempa\@empty \href
  {http://dx.doi.org/#2} {doi:#2}\else \href {http://dx.doi.org/#2} {#1}\fi
  \endgroup}
\def\mn@eprint#1#2{\mn@eprint@#1:#2::\@nil}
\def\mn@eprint@arXiv#1{\href {http://arxiv.org/abs/#1} {{\tt arXiv:#1}}}
\def\mn@eprint@dblp#1{\href {http://dblp.uni-trier.de/rec/bibtex/#1.xml}
  {dblp:#1}}
\def\mn@eprint@#1:#2:#3:#4\@nil{\def\@tempa {#1}\def\@tempb {#2}\def\@tempc
  {#3}\ifx \@tempc \@empty \let \@tempc \@tempb \let \@tempb \@tempa \fi \ifx
  \@tempb \@empty \def\@tempb {arXiv}\fi \@ifundefined
  {mn@eprint@\@tempb}{\@tempb:\@tempc}{\expandafter \expandafter \csname
  mn@eprint@\@tempb\endcsname \expandafter{\@tempc}}}

\bibitem[\protect\citeauthoryear{Acharya et~al.,}{Acharya
  et~al.}{2022}]{Acharya2022_no_monopoles}
Acharya B.,  et~al., 2022, \mn@doi [Nature] {10.1038/s41586-021-04298-1}, 602,
  63

\bibitem[\protect\citeauthoryear{{Achikanath Chirakkara}, {Federrath},
  {Trivedi}  \& {Banerjee}}{{Achikanath Chirakkara}
  et~al.}{2021}]{Chirakkara2021}
{Achikanath Chirakkara} R.,  {Federrath} C.,  {Trivedi} P.,   {Banerjee} R.,
  2021, \mn@doi [\prl] {10.1103/PhysRevLett.126.091103}, \href
  {https://ui.adsabs.harvard.edu/abs/2021PhRvL.126i1103A} {126, 091103}

\bibitem[\protect\citeauthoryear{{Allys}, {Levrier}, {Zhang}, {Colling},
  {Regaldo-Saint Blancard}, {Boulanger}, {Hennebelle}  \& {Mallat}}{{Allys}
  et~al.}{2019}]{Allys2019_scatteringtransforms}
{Allys} E.,  {Levrier} F.,  {Zhang} S.,  {Colling} C.,  {Regaldo-Saint
  Blancard} B.,  {Boulanger} F.,  {Hennebelle} P.,   {Mallat} S.,  2019,
  \mn@doi [\aap] {10.1051/0004-6361/201834975}, \href
  {https://ui.adsabs.harvard.edu/abs/2019A&A...629A.115A} {629, A115}

\bibitem[\protect\citeauthoryear{{Andr{\'e}} et~al.,}{{Andr{\'e}}
  et~al.}{2010}]{Andre2010}
{Andr{\'e}} P.,  et~al., 2010, \mn@doi [\aap] {10.1051/0004-6361/201014666},
  \href {http://adsabs.harvard.edu/abs/2010A\%26A...518L.102A} {518, L102}

\bibitem[\protect\citeauthoryear{{Armstrong}, {Rickett}  \&
  {Spangler}}{{Armstrong} et~al.}{1995}]{Armstrong1995_power_law}
{Armstrong} J.~W.,  {Rickett} B.~J.,   {Spangler} S.~R.,  1995, \mn@doi [\apj]
  {10.1086/175515}, \href
  {https://ui.adsabs.harvard.edu/abs/1995ApJ...443..209A} {443, 209}

\bibitem[\protect\citeauthoryear{{Beattie} \& {Federrath}}{{Beattie} \&
  {Federrath}}{2020}]{Beattie2020}
{Beattie} J.~R.,  {Federrath} C.,  2020, \mn@doi [\mnras]
  {10.1093/mnras/stz3377}, \href
  {https://ui.adsabs.harvard.edu/abs/2020MNRAS.492..668B} {492, 668}

\bibitem[\protect\citeauthoryear{Beattie, Federrath, Klessen  \&
  Schneider}{Beattie et~al.}{2019}]{Beattie2019b}
Beattie J.~R.,  Federrath C.,  Klessen R.~S.,   Schneider N.,  2019, \mn@doi
  [\mnras] {10.1093/mnras/stz1853}, 488, 2493

\bibitem[\protect\citeauthoryear{{Beattie}, {Federrath}  \& {Seta}}{{Beattie}
  et~al.}{2020}]{Beattie2020c}
{Beattie} J.~R.,  {Federrath} C.,   {Seta} A.,  2020, \mn@doi [\mnras]
  {10.1093/mnras/staa2257}, \href
  {https://ui.adsabs.harvard.edu/abs/2020MNRAS.498.1593B} {498, 1593}

\bibitem[\protect\citeauthoryear{{Beattie}, {Mocz}, {Federrath}  \&
  {Klessen}}{{Beattie} et~al.}{2021a}]{Beattie2021b}
{Beattie} J.~R.,  {Mocz} P.,  {Federrath} C.,   {Klessen} R.~S.,  2021a, arXiv
  e-prints, \href {https://ui.adsabs.harvard.edu/abs/2021arXiv210910470B} {p.
  arXiv:2109.10470}

\bibitem[\protect\citeauthoryear{{Beattie}, {Mocz}, {Federrath}  \&
  {Klessen}}{{Beattie} et~al.}{2021b}]{Beattie2021}
{Beattie} J.~R.,  {Mocz} P.,  {Federrath} C.,   {Klessen} R.~S.,  2021b,
  \mn@doi [\mnras] {10.1093/mnras/stab1037}, \href
  {https://ui.adsabs.harvard.edu/abs/2021MNRAS.504.4354B} {504, 4354}

\bibitem[\protect\citeauthoryear{{Beattie}, {Krumholz}, {Federrath}, {Sampson}
  \& {Crocker}}{{Beattie} et~al.}{2022}]{Beattie2022_va_fluctuations}
{Beattie} J.~R.,  {Krumholz} M.~R.,  {Federrath} C.,  {Sampson} M.,   {Crocker}
  R.~M.,  2022, arXiv e-prints, \href
  {https://ui.adsabs.harvard.edu/abs/2022arXiv220313952B} {p. arXiv:2203.13952}

\bibitem[\protect\citeauthoryear{{Beck} \& {Wielebinski}}{{Beck} \&
  {Wielebinski}}{2013}]{Beck_2013_Bfield_in_gal}
{Beck} R.,  {Wielebinski} R.,  2013, {Magnetic Fields in Galaxies}.
p.~641, \mn@doi{10.1007/978-94-007-5612-0\_13}

\bibitem[\protect\citeauthoryear{Behnel, Bradshaw, Citro, Dalcin, Seljebotn  \&
  Smith}{Behnel et~al.}{2011}]{Behnel2011}
Behnel S.,  Bradshaw R.,  Citro C.,  Dalcin L.,  Seljebotn D.~S.,   Smith K.,
  2011, Computing in Science \& Engineering, 13, 31

\bibitem[\protect\citeauthoryear{{Bhattacharjee}, {Ng}  \&
  {Spangler}}{{Bhattacharjee} et~al.}{1998}]{Bhattacharjee1998}
{Bhattacharjee} A.,  {Ng} C.~S.,   {Spangler} S.~R.,  1998, \mn@doi [\apj]
  {10.1086/305184}, \href
  {https://ui.adsabs.harvard.edu/abs/1998ApJ...494..409B} {494, 409}

\bibitem[\protect\citeauthoryear{{Biermann}}{{Biermann}}{1950}]{Biermann1950_battery}
{Biermann} L.,  1950, Zeitschrift Naturforschung Teil A, \href
  {https://ui.adsabs.harvard.edu/abs/1950ZNatA...5...65B} {5, 65}

\bibitem[\protect\citeauthoryear{Boldyrev}{Boldyrev}{2006}]{Boldyrev2006}
Boldyrev S.,  2006, \mn@doi [Phys. Rev. Lett.] {10.1103/PhysRevLett.96.115002},
  96, 115002

\bibitem[\protect\citeauthoryear{Bouchut, Klingenberg  \& Waagan}{Bouchut
  et~al.}{2010}]{Bouchut2010}
Bouchut F.,  Klingenberg C.,   Waagan K.,  2010, \mn@doi [Numerische
  Mathematik] {10.1007/s00211-010-0289-4}, 115, 647

\bibitem[\protect\citeauthoryear{Bruno \& Carbone}{Bruno \&
  Carbone}{2013}]{Roberto2013_solar_turb_review}
Bruno R.,  Carbone V.,  2013, \mn@doi [Living Reviews in Solar Physics]
  {10.12942/lrsp-2013-2}, 10, 2

\bibitem[\protect\citeauthoryear{Brunt \& Federrath}{Brunt \&
  Federrath}{2014}]{Brunt2014}
Brunt C.~M.,  Federrath C.,  2014, \mn@doi [\mnras] {10.1093/mnras/stu888},
  442, 1451

\bibitem[\protect\citeauthoryear{{Brunt}, {Heyer}  \& {Mac Low}}{{Brunt}
  et~al.}{2009}]{Brunt2009}
{Brunt} C.~M.,  {Heyer} M.~H.,   {Mac Low} M.~M.,  2009, \mn@doi [\aap]
  {10.1051/0004-6361/200911797}, \href
  {https://ui.adsabs.harvard.edu/abs/2009A&A...504..883B} {504, 883}

\bibitem[\protect\citeauthoryear{Brunt, Federrath  \& Price}{Brunt
  et~al.}{2010a}]{Brunt2010a}
Brunt C.~M.,  Federrath C.,   Price D.~J.,  2010a, \mn@doi [\mnras]
  {10.1111/j.1365-2966.2009.16215.x}, 403, 1507

\bibitem[\protect\citeauthoryear{{Brunt}, {Federrath}  \& {Price}}{{Brunt}
  et~al.}{2010b}]{Brunt2010b}
{Brunt} C.~M.,  {Federrath} C.,   {Price} D.~J.,  2010b, \mn@doi [\mnras]
  {10.1111/j.1745-3933.2010.00858.x}, \href
  {https://ui.adsabs.harvard.edu/\#abs/2010MNRAS.405L..56B} {405, L56}

\bibitem[\protect\citeauthoryear{{Burkhart}}{{Burkhart}}{2018}]{Burkhart2018}
{Burkhart} B.,  2018, \mn@doi [\apj] {10.3847/1538-4357/aad002}, \href
  {https://ui.adsabs.harvard.edu/abs/2018ApJ...863..118B} {863, 118}

\bibitem[\protect\citeauthoryear{{Burkhart}}{{Burkhart}}{2021}]{Burkhart2021_stats_turb_review}
{Burkhart} B.,  2021, \mn@doi [\pasp] {10.1088/1538-3873/ac25cf}, \href
  {https://ui.adsabs.harvard.edu/abs/2021PASP..133j2001B} {133, 102001}

\bibitem[\protect\citeauthoryear{{Chandrasekhar} \& {Fermi}}{{Chandrasekhar} \&
  {Fermi}}{1953}]{Chandrasekhar_fermi_1953}
{Chandrasekhar} S.,  {Fermi} E.,  1953, \mn@doi [\apj] {10.1086/145731}, \href
  {https://ui.adsabs.harvard.edu/abs/1953ApJ...118..113C} {118, 113}

\bibitem[\protect\citeauthoryear{Childs et~al.,}{Childs
  et~al.}{2012}]{Childs2012}
Childs H.,  et~al., 2012, in , {High Performance Visualization--Enabling
  Extreme-Scale Scientific Insight}.
Taylor \& Francis, pp 357--372

\bibitem[\protect\citeauthoryear{{Colling}, {Hennebelle}, {Geen}, {Iffrig}  \&
  {Bournaud}}{{Colling} et~al.}{2018}]{Colling2018}
{Colling} C.,  {Hennebelle} P.,  {Geen} S.,  {Iffrig} O.,   {Bournaud} F.,
  2018, \mn@doi [\aap] {10.1051/0004-6361/201833161}, \href
  {https://ui.adsabs.harvard.edu/abs/2018A&A...620A..21C} {620, A21}

\bibitem[\protect\citeauthoryear{{Davis}}{{Davis}}{1951}]{Davis1951}
{Davis} L.,  1951, \mn@doi [Physical Review] {10.1103/PhysRev.81.890.2}, \href
  {https://ui.adsabs.harvard.edu/abs/1951PhRv...81..890D} {81, 890}

\bibitem[\protect\citeauthoryear{{Dubey} et~al.,}{{Dubey}
  et~al.}{2008}]{Dubey2008}
{Dubey} A.,  et~al., 2008, in {Pogorelov} N.~V.,  {Audit} E.,   {Zank} G.~P.,
  eds,  Astronomical Society of the Pacific Conference Series Vol. 385,
  Numerical Modeling of Space Plasma Flows. p.~145

\bibitem[\protect\citeauthoryear{{Elmegreen}}{{Elmegreen}}{2009}]{Elmegreen2009IAUS}
{Elmegreen} B.~G.,  2009, in {Andersen} J.,  {Nordstr{\"o}ara} {m} B.,   {Bland
  -Hawthorn} J.,  eds,  IAU Symposium Vol. 254, The Galaxy Disk in Cosmological
  Context. pp 289--300 (\mn@eprint {arXiv} {0810.5406}),
  \mn@doi{10.1017/S1743921308027713}

\bibitem[\protect\citeauthoryear{Falceta-Gon\c{c}alves, Kowal, Falgarone  \&
  Chian}{Falceta-Gon\c{c}alves et~al.}{2014}]{Goncalves2014_driving_scale}
Falceta-Gon\c{c}alves D.,  Kowal G.,  Falgarone E.,   Chian A. C.-L.,  2014,
  \mn@doi [Nonlinear Processes in Geophysics] {10.5194/npg-21-587-2014}, 21,
  587

\bibitem[\protect\citeauthoryear{Federrath}{Federrath}{2013}]{Federrath2013_universality}
Federrath C.,  2013, \mn@doi [\mnras] {10.1093/mnras/stt1644}, 436, 1245

\bibitem[\protect\citeauthoryear{{Federrath}}{{Federrath}}{2015}]{Federrath2015_inefficient_SFR}
{Federrath} C.,  2015, \mn@doi [\mnras] {10.1093/mnras/stv941}, \href
  {https://ui.adsabs.harvard.edu/abs/2015MNRAS.450.4035F} {450, 4035}

\bibitem[\protect\citeauthoryear{{Federrath}}{{Federrath}}{2016}]{Federrath2016_dynamo}
{Federrath} C.,  2016, \mn@doi [Journal of Plasma Physics]
  {10.1017/S0022377816001069}, \href
  {https://ui.adsabs.harvard.edu/abs/2016JPlPh..82f5301F} {82, 535820601}

\bibitem[\protect\citeauthoryear{Federrath \& Klessen}{Federrath \&
  Klessen}{2012}]{Federrath2012}
Federrath C.,  Klessen R.~S.,  2012, \mn@doi [\apj]
  {10.1088/0004-637X/761/2/156}, 761

\bibitem[\protect\citeauthoryear{{Federrath}, {Klessen}  \&
  {Schmidt}}{{Federrath} et~al.}{2008}]{Federrath2008}
{Federrath} C.,  {Klessen} R.~S.,   {Schmidt} W.,  2008, \mn@doi [\apjl]
  {10.1086/595280}, \href {http://adsabs.harvard.edu/abs/2008ApJ...688L..79F}
  {688, L79}

\bibitem[\protect\citeauthoryear{Federrath, Klessen  \& Schmidt}{Federrath
  et~al.}{2009}]{Federrath2009}
Federrath C.,  Klessen R.~S.,   Schmidt W.,  2009, \mn@doi [\apj]
  {10.1086/595280}, 692, 364

\bibitem[\protect\citeauthoryear{Federrath, Roman-Duval, Klessen, Schmidt  \&
  {Mac Low}}{Federrath et~al.}{2010}]{Federrath2010}
Federrath C.,  Roman-Duval J.,  Klessen R.,  Schmidt W.,   {Mac Low} M.~M.,
  2010, \mn@doi [\aap] {10.1051/0004-6361/200912437}, 512

\bibitem[\protect\citeauthoryear{{Federrath}, {Chabrier}, {Schober},
  {Banerjee}, {Klessen}  \& {Schleicher}}{{Federrath}
  et~al.}{2011a}]{Federrath2011_mach_dynamo}
{Federrath} C.,  {Chabrier} G.,  {Schober} J.,  {Banerjee} R.,  {Klessen}
  R.~S.,   {Schleicher} D.~R.~G.,  2011a, \mn@doi [\prl]
  {10.1103/PhysRevLett.107.114504}, \href
  {https://ui.adsabs.harvard.edu/abs/2011PhRvL.107k4504F} {107, 114504}

\bibitem[\protect\citeauthoryear{{Federrath}, {Sur}, {Schleicher}, {Banerjee}
  \& {Klessen}}{{Federrath}
  et~al.}{2011b}]{Federrath2011_numerical_dissipation}
{Federrath} C.,  {Sur} S.,  {Schleicher} D. R.~G.,  {Banerjee} R.,   {Klessen}
  R.~S.,  2011b, \mn@doi [\apj] {10.1088/0004-637X/731/1/62}, \href
  {https://ui.adsabs.harvard.edu/abs/2011ApJ...731...62F} {731, 62}

\bibitem[\protect\citeauthoryear{{Federrath}, {Schober}, {Bovino}  \&
  {Schleicher}}{{Federrath} et~al.}{2014}]{Federrath2014}
{Federrath} C.,  {Schober} J.,  {Bovino} S.,   {Schleicher} D. R.~G.,  2014,
  \mn@doi [\apjl] {10.1088/2041-8205/797/2/L19}, \href
  {https://ui.adsabs.harvard.edu/abs/2014ApJ...797L..19F} {797, L19}

\bibitem[\protect\citeauthoryear{{Federrath} et~al.,}{{Federrath}
  et~al.}{2016}]{Federrath2016_brick}
{Federrath} C.,  et~al., 2016, \mn@doi [\apj] {10.3847/0004-637X/832/2/143},
  \href {https://ui.adsabs.harvard.edu/\#abs/2016ApJ...832..143F} {832, 143}

\bibitem[\protect\citeauthoryear{{Federrath} et~al.,}{{Federrath}
  et~al.}{2017a}]{Federrath2017IAUS}
{Federrath} C.,  et~al., 2017a, in {Crocker} R.~M.,  {Longmore} S.~N.,
  {Bicknell} G.~V.,  eds,  IAU Symposium Vol. 322, The Multi-Messenger
  Astrophysics of the Galactic Centre. pp 123--128 (\mn@eprint {arXiv}
  {1609.08726}), \mn@doi{10.1017/S1743921316012357}

\bibitem[\protect\citeauthoryear{{Federrath}, {Krumholz}  \&
  {Hopkins}}{{Federrath} et~al.}{2017b}]{federrath2017_imf}
{Federrath} C.,  {Krumholz} M.,   {Hopkins} P.~F.,  2017b, in Journal of
  Physics Conference Series. p. 012007, \mn@doi{10.1088/1742-6596/837/1/012007}

\bibitem[\protect\citeauthoryear{Federrath, Klessen, Iapichino  \&
  Beattie}{Federrath et~al.}{2021}]{Federrath2021}
Federrath C.,  Klessen R.~S.,  Iapichino L.,   Beattie J.~R.,  2021, \mn@doi
  [Nature Astronomy] {10.1038/s41550-020-01282-z}

\bibitem[\protect\citeauthoryear{{Federrath}, {Roman-Duval}, {Klessen},
  {Schmidt}  \& {Mac Low}}{{Federrath}
  et~al.}{2022}]{Federrath2022_turbulence_driving_module}
{Federrath} C.,  {Roman-Duval} J.,  {Klessen} R.~S.,  {Schmidt} W.,   {Mac Low}
  M.~M.,  2022, {TG: Turbulence Generator}, Astrophysics Source Code Library,
  record ascl:2204.001 (\mn@eprint {ascl} {2204.001})

\bibitem[\protect\citeauthoryear{{Fryxell} et~al.,}{{Fryxell}
  et~al.}{2000}]{Fryxell2000}
{Fryxell} B.,  et~al., 2000, \mn@doi [\apjs] {10.1086/317361}, \href
  {https://ui.adsabs.harvard.edu/abs/2000ApJS..131..273F} {131, 273}

\bibitem[\protect\citeauthoryear{{Gaensler} et~al.,}{{Gaensler}
  et~al.}{2011}]{Gaesnsler_2011_trans_ISM}
{Gaensler} B.~M.,  et~al., 2011, \mn@doi [\nat] {10.1038/nature10446}, \href
  {https://ui.adsabs.harvard.edu/abs/2011Natur.478..214G} {478, 214}

\bibitem[\protect\citeauthoryear{{Galishnikova}, {Kunz}  \&
  {Schekochihin}}{{Galishnikova}
  et~al.}{2022}]{Galishnikova2022_tearing_instability_sat}
{Galishnikova} A.~K.,  {Kunz} M.~W.,   {Schekochihin} A.~A.,  2022, arXiv
  e-prints, \href {https://ui.adsabs.harvard.edu/abs/2022arXiv220107757G} {p.
  arXiv:2201.07757}

\bibitem[\protect\citeauthoryear{Germano}{Germano}{1992}]{Germano1992}
Germano M.,  1992, \mn@doi [Journal of Fluid Mechanics]
  {10.1017/S0022112092001733}, 238, 325–336

\bibitem[\protect\citeauthoryear{{Goldreich} \& {Sridhar}}{{Goldreich} \&
  {Sridhar}}{1995}]{Goldreich1995}
{Goldreich} P.,  {Sridhar} S.,  1995, \mn@doi [\apj] {10.1086/175121}, \href
  {https://ui.adsabs.harvard.edu/abs/1995ApJ...438..763G} {438, 763}

\bibitem[\protect\citeauthoryear{{Grisdale}, {Agertz}, {Romeo}, {Renaud}  \&
  {Read}}{{Grisdale} et~al.}{2017}]{Grisdale2017}
{Grisdale} K.,  {Agertz} O.,  {Romeo} A.~B.,  {Renaud} F.,   {Read} J.~I.,
  2017, \mn@doi [\mnras] {10.1093/mnras/stw3133}, \href
  {https://ui.adsabs.harvard.edu/abs/2017MNRAS.466.1093G} {466, 1093}

\bibitem[\protect\citeauthoryear{Harris et~al.,}{Harris
  et~al.}{2020}]{numpy2020}
Harris C.~R.,  et~al., 2020, \mn@doi [Nature] {10.1038/s41586-020-2649-2}, 585,
  357

\bibitem[\protect\citeauthoryear{{Hartlep}, {Matthaeus}, {Padhye}  \&
  {Smith}}{{Hartlep} et~al.}{2000}]{Hartlep2000_b_field_pdf}
{Hartlep} T.,  {Matthaeus} W.~H.,  {Padhye} N.~S.,   {Smith} C.~W.,  2000,
  \mn@doi [\jgr] {10.1029/1999JA000223}, \href
  {https://ui.adsabs.harvard.edu/abs/2000JGR...105.5135H} {105, 5135}

\bibitem[\protect\citeauthoryear{{Haugen}, {Brandenburg}  \& {Dobler}}{{Haugen}
  et~al.}{2004}]{Haugen2004_large_scale_field_supression}
{Haugen} N.~E.~L.,  {Brandenburg} A.,   {Dobler} W.,  2004, \mn@doi [\apss]
  {10.1023/B:ASTR.0000045000.08395.a3}, \href
  {https://ui.adsabs.harvard.edu/abs/2004Ap&SS.292...53H} {292, 53}

\bibitem[\protect\citeauthoryear{{Hennebelle} \& {Chabrier}}{{Hennebelle} \&
  {Chabrier}}{2009}]{Hennebelle2009}
{Hennebelle} P.,  {Chabrier} G.,  2009, \mn@doi [\apj]
  {10.1088/0004-637X/702/2/1428}, \href
  {https://ui.adsabs.harvard.edu/abs/2009ApJ...702.1428H} {702, 1428}

\bibitem[\protect\citeauthoryear{{Hennebelle} \& {Inutsuka}}{{Hennebelle} \&
  {Inutsuka}}{2019}]{Hennebelle2019}
{Hennebelle} P.,  {Inutsuka} S.-i.,  2019, \mn@doi [Frontiers in Astronomy and
  Space Sciences] {10.3389/fspas.2019.00005}, \href
  {https://ui.adsabs.harvard.edu/abs/2019FrASS...6....5H} {6, 5}

\bibitem[\protect\citeauthoryear{{Hennebelle}, {Commer{\c{c}}on}, {Joos},
  {Klessen}, {Krumholz}, {Tan}  \& {Teyssier}}{{Hennebelle}
  et~al.}{2011}]{Hennebelle2011}
{Hennebelle} P.,  {Commer{\c{c}}on} B.,  {Joos} M.,  {Klessen} R.~S.,
  {Krumholz} M.,  {Tan} J.~C.,   {Teyssier} R.,  2011, \mn@doi [\aap]
  {10.1051/0004-6361/201016052}, \href
  {https://ui.adsabs.harvard.edu/\#abs/2011A&A...528A..72H} {528, A72}

\bibitem[\protect\citeauthoryear{{Heyer}, {Soler}  \& {Burkhart}}{{Heyer}
  et~al.}{2020}]{Heyer2020}
{Heyer} M.,  {Soler} J.~D.,   {Burkhart} B.,  2020, \mn@doi [\mnras]
  {10.1093/mnras/staa1760}, \href
  {https://ui.adsabs.harvard.edu/abs/2020MNRAS.496.4546H} {496, 4546}

\bibitem[\protect\citeauthoryear{{Hoang} et~al.,}{{Hoang}
  et~al.}{2021}]{Hoang2021}
{Hoang} T.~D.,  et~al., 2021, arXiv e-prints, \href
  {https://ui.adsabs.harvard.edu/abs/2021arXiv210810045H} {p. arXiv:2108.10045}

\bibitem[\protect\citeauthoryear{{Hollins}, {Sarson}, {Shukurov}, {Fletcher}
  \& {Gent}}{{Hollins} et~al.}{2018}]{Hollins2018}
{Hollins} J.~F.,  {Sarson} G.~R.,  {Shukurov} A.,  {Fletcher} A.,   {Gent}
  F.~A.,  2018, arXiv e-prints, \href
  {https://ui.adsabs.harvard.edu/abs/2018arXiv180901098H} {p. arXiv:1809.01098}

\bibitem[\protect\citeauthoryear{{Hopkins}}{{Hopkins}}{2012}]{Hopkins2012_imf}
{Hopkins} P.~F.,  2012, \mn@doi [\mnras] {10.1111/j.1365-2966.2012.20731.x},
  \href {https://ui.adsabs.harvard.edu/abs/2012MNRAS.423.2037H} {423, 2037}

\bibitem[\protect\citeauthoryear{{Hopkins}}{{Hopkins}}{2013}]{Hopkins2013a}
{Hopkins} P.~F.,  2013, \mn@doi [\mnras] {10.1093/mnras/sts704}, \href
  {https://ui.adsabs.harvard.edu/abs/2013MNRAS.430.1653H} {430, 1653}

\bibitem[\protect\citeauthoryear{{Hopkins}, {Squire}, {Chan}, {Quataert}, {Ji},
  {Kere{\v{s}}}  \& {Faucher-Gigu{\`e}re}}{{Hopkins}
  et~al.}{2021}]{Hopkins2021_testing_gev_cr_models}
{Hopkins} P.~F.,  {Squire} J.,  {Chan} T.~K.,  {Quataert} E.,  {Ji} S.,
  {Kere{\v{s}}} D.,   {Faucher-Gigu{\`e}re} C.-A.,  2021, \mn@doi [\mnras]
  {10.1093/mnras/staa3691}, \href
  {https://ui.adsabs.harvard.edu/abs/2021MNRAS.501.4184H} {501, 4184}

\bibitem[\protect\citeauthoryear{{Hu} et~al.,}{{Hu} et~al.}{2019}]{Hu2019}
{Hu} Y.,  et~al., 2019, \mn@doi [Nature Astronomy] {10.1038/s41550-019-0769-0},
  \href {https://ui.adsabs.harvard.edu/abs/2019NatAs...3..776H} {3, 776}

\bibitem[\protect\citeauthoryear{Hunter}{Hunter}{2007}]{Hunter2007}
Hunter J.~D.,  2007, \mn@doi [Computing in Science \& Engineering]
  {10.1109/MCSE.2007.55}, 9, 90

\bibitem[\protect\citeauthoryear{{Hwang} et~al.,}{{Hwang}
  et~al.}{2021}]{Hwang2021}
{Hwang} J.,  et~al., 2021, \mn@doi [\apj] {10.3847/1538-4357/abf3c4}, \href
  {https://ui.adsabs.harvard.edu/abs/2021ApJ...913...85H} {913, 85}

\bibitem[\protect\citeauthoryear{{Iroshnikov}}{{Iroshnikov}}{1964}]{Iroshnikov_1965_IK_turb}
{Iroshnikov} P.~S.,  1964, \sovast, \href
  {https://ui.adsabs.harvard.edu/abs/1964SvA.....7..566I} {7, 566}

\bibitem[\protect\citeauthoryear{{Jaupart} \& {Chabrier}}{{Jaupart} \&
  {Chabrier}}{2021}]{Jaupart2021}
{Jaupart} E.,  {Chabrier} G.,  2021, arXiv e-prints, \href
  {https://ui.adsabs.harvard.edu/abs/2021arXiv211009090J} {p. arXiv:2110.09090}

\bibitem[\protect\citeauthoryear{{Jin}, {Salim}, {Federrath}, {Tasker}, {Habe}
  \& {Kainulainen}}{{Jin} et~al.}{2017}]{Jin2017}
{Jin} K.,  {Salim} D.~M.,  {Federrath} C.,  {Tasker} E.~J.,  {Habe} A.,
  {Kainulainen} J.~T.,  2017, \mn@doi [\mnras] {10.1093/mnras/stx737}, \href
  {https://ui.adsabs.harvard.edu/abs/2017MNRAS.469..383J} {469, 383}

\bibitem[\protect\citeauthoryear{Jokipii \& Parker}{Jokipii \&
  Parker}{1968}]{Jokipii1968_field_line_random_walk}
Jokipii J.~R.,  Parker E.~N.,  1968, \mn@doi [Phys. Rev. Lett.]
  {10.1103/PhysRevLett.21.44}, 21, 44

\bibitem[\protect\citeauthoryear{{Karlsson}, {Bromm}  \&
  {Bland-Hawthorn}}{{Karlsson} et~al.}{2013}]{Karlsson2013_turbulence_scale}
{Karlsson} T.,  {Bromm} V.,   {Bland-Hawthorn} J.,  2013, \mn@doi [Reviews of
  Modern Physics] {10.1103/RevModPhys.85.809}, \href
  {https://ui.adsabs.harvard.edu/abs/2013RvMP...85..809K} {85, 809}

\bibitem[\protect\citeauthoryear{{Kitsionas} et~al.,}{{Kitsionas}
  et~al.}{2009}]{Kitsionas2009_dissipation_comparison}
{Kitsionas} S.,  et~al., 2009, \mn@doi [\aap] {10.1051/0004-6361/200811170},
  \href {https://ui.adsabs.harvard.edu/abs/2009A&A...508..541K} {508, 541}

\bibitem[\protect\citeauthoryear{{K{\"o}rtgen} \& {Soler}}{{K{\"o}rtgen} \&
  {Soler}}{2020}]{Kortgen2020}
{K{\"o}rtgen} B.,  {Soler} J.~D.,  2020, \mn@doi [\mnras]
  {10.1093/mnras/staa3078}, \href
  {https://ui.adsabs.harvard.edu/abs/2020MNRAS.499.4785K} {499, 4785}

\bibitem[\protect\citeauthoryear{{K{\"o}rtgen}, {Federrath}  \&
  {Banerjee}}{{K{\"o}rtgen} et~al.}{2017}]{Kortgen2017}
{K{\"o}rtgen} B.,  {Federrath} C.,   {Banerjee} R.,  2017, \mn@doi [\mnras]
  {10.1093/mnras/stx2208}, \href
  {https://ui.adsabs.harvard.edu/abs/2017MNRAS.472.2496K} {472, 2496}

\bibitem[\protect\citeauthoryear{Kraichnan}{Kraichnan}{1965}]{Kraichnan1965_IKturb}
Kraichnan R.~H.,  1965, \mn@doi [The Physics of Fluids] {10.1063/1.1761412}, 8,
  1385

\bibitem[\protect\citeauthoryear{{Kriel}, {Beattie}, {Seta}  \&
  {Federrath}}{{Kriel} et~al.}{2022}]{Kriel2022_turb_dynamo}
{Kriel} N.,  {Beattie} J.~R.,  {Seta} A.,   {Federrath} C.,  2022, \mn@doi
  [\mnras] {10.1093/mnras/stac969}, \href
  {https://ui.adsabs.harvard.edu/abs/2022MNRAS.tmp..948K} {}

\bibitem[\protect\citeauthoryear{Kritsuk, Ustyugov  \& Norman}{Kritsuk
  et~al.}{2017}]{Kritsuk2017}
Kritsuk A.~G.,  Ustyugov S.~D.,   Norman M.~L.,  2017, \mn@doi [New Journal of
  Physics] {10.1088/1367-2630/aa7156}, 19, 065003

\bibitem[\protect\citeauthoryear{{Krumholz} \& {Burkhart}}{{Krumholz} \&
  {Burkhart}}{2016}]{Krumholz2016}
{Krumholz} M.~R.,  {Burkhart} B.,  2016, \mn@doi [\mnras]
  {10.1093/mnras/stw434}, \href
  {https://ui.adsabs.harvard.edu/abs/2016MNRAS.458.1671K} {458, 1671}

\bibitem[\protect\citeauthoryear{{Krumholz} \& {Federrath}}{{Krumholz} \&
  {Federrath}}{2019}]{Krumholz2019}
{Krumholz} M.~R.,  {Federrath} C.,  2019, \mn@doi [Frontiers in Astronomy and
  Space Sciences] {10.3389/fspas.2019.00007}, \href
  {https://ui.adsabs.harvard.edu/abs/2019FrASS...6....7K} {6, 7}

\bibitem[\protect\citeauthoryear{Krumholz \& McKee}{Krumholz \&
  McKee}{2005}]{Krumholz2005}
Krumholz M.~R.,  McKee C.~F.,  2005, \mn@doi [\apj] {10.1086/431734}, 630, 250

\bibitem[\protect\citeauthoryear{{Krumholz} \& {Ting}}{{Krumholz} \&
  {Ting}}{2018}]{Krumholz2018_metallicity_SF}
{Krumholz} M.~R.,  {Ting} Y.-S.,  2018, \mn@doi [\mnras]
  {10.1093/mnras/stx3286}, \href
  {https://ui.adsabs.harvard.edu/abs/2018MNRAS.475.2236K} {475, 2236}

\bibitem[\protect\citeauthoryear{{Krumholz}, {Crocker}, {Xu}, {Lazarian},
  {Rosevear}  \& {Bedwell-Wilson}}{{Krumholz} et~al.}{2020}]{Krumholz2020}
{Krumholz} M.~R.,  {Crocker} R.~M.,  {Xu} S.,  {Lazarian} A.,  {Rosevear}
  M.~T.,   {Bedwell-Wilson} J.,  2020, \mn@doi [\mnras]
  {10.1093/mnras/staa493}, \href
  {https://ui.adsabs.harvard.edu/abs/2020MNRAS.493.2817K} {493, 2817}

\bibitem[\protect\citeauthoryear{Landau \& Lifshitz}{Landau \&
  Lifshitz}{1959}]{Landau1959}
Landau L.,  Lifshitz E.,  1959, Fluid Mechanics: Landau and Lifshitz: Course of
  Theoretical Physics.
Butterworth-Heinemann

\bibitem[\protect\citeauthoryear{{Lazarian} \& {Beresnyak}}{{Lazarian} \&
  {Beresnyak}}{2006}]{Lazarian2006_cr_scattering}
{Lazarian} A.,  {Beresnyak} A.,  2006, \mn@doi [\mnras]
  {10.1111/j.1365-2966.2006.11093.x}, \href
  {https://ui.adsabs.harvard.edu/abs/2006MNRAS.373.1195L} {373, 1195}

\bibitem[\protect\citeauthoryear{{Lazarian}, {Yuen}, {Ho}, {Chen}, {Lazarian},
  {Lu}, {Yang}  \& {Hu}}{{Lazarian} et~al.}{2018}]{Lazarian2018_VGT}
{Lazarian} A.,  {Yuen} K.~H.,  {Ho} K.~W.,  {Chen} J.,  {Lazarian} V.,  {Lu}
  Z.,  {Yang} B.,   {Hu} Y.,  2018, \mn@doi [\apj] {10.3847/1538-4357/aad7ff},
  \href {https://ui.adsabs.harvard.edu/abs/2018ApJ...865...46L} {865, 46}

\bibitem[\protect\citeauthoryear{{Lazarian}, {Yuen}  \& {Pogosyan}}{{Lazarian}
  et~al.}{2020}]{Lazarian2020_diff_measure_VCM}
{Lazarian} A.,  {Yuen} K.~H.,   {Pogosyan} D.,  2020, arXiv e-prints, \href
  {https://ui.adsabs.harvard.edu/abs/2020arXiv200207996L} {p. arXiv:2002.07996}

\bibitem[\protect\citeauthoryear{{Lazarian}, {Yuen}  \& {Pogosyan}}{{Lazarian}
  et~al.}{2022}]{Lazarian2022_diff_measure}
{Lazarian} A.,  {Yuen} K.~H.,   {Pogosyan} D.,  2022, arXiv e-prints, \href
  {https://ui.adsabs.harvard.edu/abs/2022arXiv220409731L} {p. arXiv:2204.09731}

\bibitem[\protect\citeauthoryear{Li, Fang, Henning  \& Kainulainen}{Li
  et~al.}{2013}]{HuaBai2013}
Li H.-b.,  Fang M.,  Henning T.,   Kainulainen J.,  2013, \mn@doi [Monthly
  Notices of the Royal Astronomical Society] {10.1093/mnras/stt1849}, 436, 3707

\bibitem[\protect\citeauthoryear{{Li}, {Goodman}, {Sridharan}, {Houde}, {Li},
  {Novak}  \& {Tang}}{{Li} et~al.}{2014}]{Li14b}
{Li} H.~B.,  {Goodman} A.,  {Sridharan} T.~K.,  {Houde} M.,  {Li} Z.~Y.,
  {Novak} G.,   {Tang} K.~S.,  2014, in {Beuther} H.,  {Klessen} R.~S.,
  {Dullemond} C.~P.,   {Henning} T.,  eds, Protostars and Planets VI. p.~101
  (\mn@eprint {arXiv} {1404.2024}),
  \mn@doi{10.2458/azu\_uapress\_9780816531240-ch005}

\bibitem[\protect\citeauthoryear{{Li}, {Lopez-Rodriguez}, {Ajeddig},
  {Andr{\'e}}, {McKee}, {Rho}  \& {Klein}}{{Li}
  et~al.}{2021a}]{Li2021_critiqueSK2021}
{Li} P.~S.,  {Lopez-Rodriguez} E.,  {Ajeddig} H.,  {Andr{\'e}} P.,  {McKee}
  C.~F.,  {Rho} J.,   {Klein} R.~I.,  2021a, \mn@doi [\mnras]
  {10.1093/mnras/stab3448}, \href
  {https://ui.adsabs.harvard.edu/abs/2021MNRAS.tmp.3119L} {}

\bibitem[\protect\citeauthoryear{{Li}, {Lopez-Rodriguez}, {Ajeddig},
  {Andr{\'e}}, {McKee}, {Rho}  \& {Klein}}{{Li}
  et~al.}{2021b}]{Li2021_dbB0_foundations}
{Li} P.~S.,  {Lopez-Rodriguez} E.,  {Ajeddig} H.,  {Andr{\'e}} P.,  {McKee}
  C.~F.,  {Rho} J.,   {Klein} R.~I.,  2021b, arXiv e-prints, \href
  {https://ui.adsabs.harvard.edu/abs/2021arXiv211112864L} {p. arXiv:2111.12864}

\bibitem[\protect\citeauthoryear{{Liu}, {Qiu}  \& {Zhang}}{{Liu}
  et~al.}{2021}]{Liu2021_cor_scales}
{Liu} J.,  {Qiu} K.,   {Zhang} Q.,  2021, arXiv e-prints, \href
  {https://ui.adsabs.harvard.edu/abs/2021arXiv211105836L} {p. arXiv:2111.05836}

\bibitem[\protect\citeauthoryear{{Lu}, {Pelkonen}, {Padoan}, {Pan},
  {Haugb{\o}lle}  \& {Nordlund}}{{Lu} et~al.}{2020}]{Lu2020}
{Lu} Z.-J.,  {Pelkonen} V.-M.,  {Padoan} P.,  {Pan} L.,  {Haugb{\o}lle} T.,
  {Nordlund} {\r{A}}.,  2020, arXiv e-prints, \href
  {https://ui.adsabs.harvard.edu/abs/2020arXiv200709518L} {p. arXiv:2007.09518}

\bibitem[\protect\citeauthoryear{{Lyra} \& {Umurhan}}{{Lyra} \&
  {Umurhan}}{2019}]{Wladimir2019_planet_turb_review}
{Lyra} W.,  {Umurhan} O.~M.,  2019, \mn@doi [\pasp] {10.1088/1538-3873/aaf5ff},
  \href {https://ui.adsabs.harvard.edu/abs/2019PASP..131g2001L} {131, 072001}

\bibitem[\protect\citeauthoryear{{McKee}, {Stacy}  \& {Li}}{{McKee}
  et~al.}{2020}]{McKee2020}
{McKee} C.~F.,  {Stacy} A.,   {Li} P.~S.,  2020, \mn@doi [\mnras]
  {10.1093/mnras/staa1903}, \href
  {https://ui.adsabs.harvard.edu/abs/2020MNRAS.496.5528M} {496, 5528}

\bibitem[\protect\citeauthoryear{{Menon}, {Federrath}, {Klaassen}, {Kuiper}  \&
  {Reiter}}{{Menon} et~al.}{2021}]{Menon2020}
{Menon} S.~H.,  {Federrath} C.,  {Klaassen} P.,  {Kuiper} R.,   {Reiter} M.,
  2021, \mn@doi [\mnras] {10.1093/mnras/staa3271}, \href
  {https://ui.adsabs.harvard.edu/abs/2021MNRAS.500.1721M} {500, 1721}

\bibitem[\protect\citeauthoryear{{Mocz} \& {Burkhart}}{{Mocz} \&
  {Burkhart}}{2018}]{Mocz2018}
{Mocz} P.,  {Burkhart} B.,  2018, \mn@doi [\mnras] {10.1093/mnras/sty1976},
  \href {https://ui.adsabs.harvard.edu/\#abs/2018MNRAS.480.3916M} {480, 3916}

\bibitem[\protect\citeauthoryear{{Nam}, {Federrath}  \& {Krumholz}}{{Nam}
  et~al.}{2021}]{Nam2021}
{Nam} D.~G.,  {Federrath} C.,   {Krumholz} M.~R.,  2021, \mn@doi [\mnras]
  {10.1093/mnras/stab505}, \href
  {https://ui.adsabs.harvard.edu/abs/2021MNRAS.503.1138N} {503, 1138}

\bibitem[\protect\citeauthoryear{Oliphant}{Oliphant}{2006}]{Oliphant2006}
Oliphant T.,  2006, {NumPy}: A guide to {NumPy}, USA: Trelgol Publishing, \url
  {http://www.numpy.org/}

\bibitem[\protect\citeauthoryear{{Omukai}, {Tsuribe}, {Schneider}  \&
  {Ferrara}}{{Omukai} et~al.}{2005}]{Omukai2005_isothermal_ism}
{Omukai} K.,  {Tsuribe} T.,  {Schneider} R.,   {Ferrara} A.,  2005, \mn@doi
  [\apj] {10.1086/429955}, \href
  {https://ui.adsabs.harvard.edu/abs/2005ApJ...626..627O} {626, 627}

\bibitem[\protect\citeauthoryear{{Orkisz} et~al.,}{{Orkisz}
  et~al.}{2017}]{Orkisz2017}
{Orkisz} J.~H.,  et~al., 2017, \mn@doi [\aap] {10.1051/0004-6361/201629220},
  \href {https://ui.adsabs.harvard.edu/abs/2017A&A...599A..99O} {599, A99}

\bibitem[\protect\citeauthoryear{{Padoan} \& {Nordlund}}{{Padoan} \&
  {Nordlund}}{2011}]{Padoan2011}
{Padoan} P.,  {Nordlund} {\AA}.,  2011, \mn@doi [\apj]
  {10.1088/0004-637X/730/1/40}, \href
  {http://adsabs.harvard.edu/abs/2011ApJ...730...40P} {730, 40}

\bibitem[\protect\citeauthoryear{{Padoan}, {Nordlund}  \& {Jones}}{{Padoan}
  et~al.}{1997}]{Padoan1997_imf}
{Padoan} P.,  {Nordlund} P.,   {Jones} B.~J.~T.,  1997, Commmunications of the
  Konkoly Observatory Hungary, \href
  {https://ui.adsabs.harvard.edu/\#abs/1997CoKon.100..341P} {100, 341}

\bibitem[\protect\citeauthoryear{{Padoan}, {Pan}, {Haugb{\o}lle}  \&
  {Nordlund}}{{Padoan} et~al.}{2016}]{Padoan2016_supernova_driving}
{Padoan} P.,  {Pan} L.,  {Haugb{\o}lle} T.,   {Nordlund} {\r{A}}.,  2016,
  \mn@doi [\apj] {10.3847/0004-637X/822/1/11}, \href
  {https://ui.adsabs.harvard.edu/abs/2016ApJ...822...11P} {822, 11}

\bibitem[\protect\citeauthoryear{{Panopoulou}, {Psaradaki}  \&
  {Tassis}}{{Panopoulou} et~al.}{2016}]{Panopoulou2016_polaris_b_field}
{Panopoulou} G.~V.,  {Psaradaki} I.,   {Tassis} K.,  2016, \mn@doi [\mnras]
  {10.1093/mnras/stw1678}, \href
  {https://ui.adsabs.harvard.edu/abs/2016MNRAS.462.1517P} {462, 1517}

\bibitem[\protect\citeauthoryear{{Parker}}{{Parker}}{1970}]{Parker1970_monopoles_origin}
{Parker} E.~N.,  1970, \mn@doi [\apj] {10.1086/150442}, \href
  {https://ui.adsabs.harvard.edu/abs/1970ApJ...160..383P} {160, 383}

\bibitem[\protect\citeauthoryear{{Parker}}{{Parker}}{1979}]{Parker1979_B_fields_book}
{Parker} E.~N.,  1979, {Cosmical magnetic fields. Their origin and their
  activity}

\bibitem[\protect\citeauthoryear{{Peek} \& {Burkhart}}{{Peek} \&
  {Burkhart}}{2019}]{Peek2019_CNN}
{Peek} J.~E.~G.,  {Burkhart} B.,  2019, \mn@doi [\apjl]
  {10.3847/2041-8213/ab3a9e}, \href
  {https://ui.adsabs.harvard.edu/abs/2019ApJ...882L..12P} {882, L12}

\bibitem[\protect\citeauthoryear{Reynolds}{Reynolds}{1895}]{Reynolds1895_averaging}
Reynolds O.,  1895, \mn@doi [Philosophical Transactions of the Royal Society of
  London. (A.)] {10.1098/rsta.1895.0004}, 186, 123

\bibitem[\protect\citeauthoryear{{Rincon}}{{Rincon}}{2019}]{Rincon2019_dynamo_review}
{Rincon} F.,  2019, \mn@doi [Journal of Plasma Physics]
  {10.1017/S0022377819000539}, \href
  {https://ui.adsabs.harvard.edu/abs/2019JPlPh..85d2001R} {85, 205850401}

\bibitem[\protect\citeauthoryear{{Sampson}, {Beattie}, {Krumholz}, {Crocker},
  {Federrath}  \& {Seta}}{{Sampson} et~al.}{2022}]{Sampson2022_SCR_diffusion}
{Sampson} M.~L.,  {Beattie} J.~R.,  {Krumholz} M.~R.,  {Crocker} R.~M.,
  {Federrath} C.,   {Seta} A.,  2022, arXiv e-prints, \href
  {https://ui.adsabs.harvard.edu/abs/2022arXiv220508174S} {p. arXiv:2205.08174}

\bibitem[\protect\citeauthoryear{{Saydjari}, {Portillo}, {Slepian}, {Kahraman},
  {Burkhart}  \& {Finkbeiner}}{{Saydjari}
  et~al.}{2021}]{Saydjari2021_scattering_transform}
{Saydjari} A.~K.,  {Portillo} S. K.~N.,  {Slepian} Z.,  {Kahraman} S.,
  {Burkhart} B.,   {Finkbeiner} D.~P.,  2021, \mn@doi [\apj]
  {10.3847/1538-4357/abe46d}, \href
  {https://ui.adsabs.harvard.edu/abs/2021ApJ...910..122S} {910, 122}

\bibitem[\protect\citeauthoryear{{Schekochihin}}{{Schekochihin}}{2020}]{Schekochihin2020_bias_review}
{Schekochihin} A.~A.,  2020, arXiv e-prints, \href
  {https://ui.adsabs.harvard.edu/abs/2020arXiv201000699S} {p. arXiv:2010.00699}

\bibitem[\protect\citeauthoryear{Schekochihin, Cowley, Hammett, Maron  \&
  McWilliams}{Schekochihin
  et~al.}{2002}]{Schekochihin2002_saturation_evolution}
Schekochihin A.~A.,  Cowley S.~C.,  Hammett G.~W.,  Maron J.~L.,   McWilliams
  J.~C.,  2002, \mn@doi [New Journal of Physics] {10.1088/1367-2630/4/1/384},
  4, 84

\bibitem[\protect\citeauthoryear{Schekochihin, Cowley, Taylor, Maron  \&
  McWilliams}{Schekochihin et~al.}{2004}]{schekochihin2004simulations}
Schekochihin A.~A.,  Cowley S.~C.,  Taylor S.~F.,  Maron J.~L.,   McWilliams
  J.~C.,  2004, The Astrophysical Journal, 612, 276

\bibitem[\protect\citeauthoryear{{Schneider} et~al.,}{{Schneider}
  et~al.}{2013}]{Schneider2013}
{Schneider} N.,  et~al., 2013, \mn@doi [\apj] {10.1088/2041-8205/766/2/L17},
  \href {https://ui.adsabs.harvard.edu/\#abs/2013ApJ...766L..17S} {766, L17}

\bibitem[\protect\citeauthoryear{{Schober}, {Schleicher}, {Federrath},
  {Klessen}  \& {Banerjee}}{{Schober} et~al.}{2012}]{Schober2012}
{Schober} J.,  {Schleicher} D.,  {Federrath} C.,  {Klessen} R.,   {Banerjee}
  R.,  2012, \mn@doi [\pre] {10.1103/PhysRevE.85.026303}, \href
  {https://ui.adsabs.harvard.edu/abs/2012PhRvE..85b6303S} {85, 026303}

\bibitem[\protect\citeauthoryear{{Schober}, {Schleicher}, {Federrath}, {Bovino}
   \& {Klessen}}{{Schober} et~al.}{2015}]{Schober2015}
{Schober} J.,  {Schleicher} D.~R.~G.,  {Federrath} C.,  {Bovino} S.,
  {Klessen} R.~S.,  2015, \mn@doi [\pre] {10.1103/PhysRevE.92.023010}, \href
  {https://ui.adsabs.harvard.edu/abs/2015PhRvE..92b3010S} {92, 023010}

\bibitem[\protect\citeauthoryear{Schruba, Kruijssen  \& Leroy}{Schruba
  et~al.}{2019}]{Schruba2019}
Schruba A.,  Kruijssen J. M.~D.,   Leroy A.~K.,  2019, \mn@doi [The
  Astrophysical Journal] {10.3847/1538-4357/ab3a43}, 883, 2

\bibitem[\protect\citeauthoryear{{Seta} \& {Federrath}}{{Seta} \&
  {Federrath}}{2020}]{Seta2020b}
{Seta} A.,  {Federrath} C.,  2020, \mn@doi [\mnras] {10.1093/mnras/staa2978},
  \href {https://ui.adsabs.harvard.edu/abs/2020MNRAS.499.2076S} {499, 2076}

\bibitem[\protect\citeauthoryear{{Seta} \& {Federrath}}{{Seta} \&
  {Federrath}}{2021a}]{Seta2021}
{Seta} A.,  {Federrath} C.,  2021a, \mn@doi [Physical Review Fluids]
  {10.1103/PhysRevFluids.6.103701}, \href
  {https://ui.adsabs.harvard.edu/abs/2021PhRvF...6j3701S} {6, 103701}

\bibitem[\protect\citeauthoryear{{Seta} \& {Federrath}}{{Seta} \&
  {Federrath}}{2021b}]{Seta2021b}
{Seta} A.,  {Federrath} C.,  2021b, \mn@doi [\mnras] {10.1093/mnras/stab128},
  \href {https://ui.adsabs.harvard.edu/abs/2021MNRAS.502.2220S} {502, 2220}

\bibitem[\protect\citeauthoryear{{Seta}, {Bushby}, {Shukurov}  \&
  {Wood}}{{Seta} et~al.}{2020}]{Seta2020}
{Seta} A.,  {Bushby} P.~J.,  {Shukurov} A.,   {Wood} T.~S.,  2020, \mn@doi
  [Physical Review Fluids] {10.1103/PhysRevFluids.5.043702}, \href
  {https://ui.adsabs.harvard.edu/abs/2020PhRvF...5d3702S} {5, 043702}

\bibitem[\protect\citeauthoryear{{Sharda} et~al.,}{{Sharda}
  et~al.}{2021}]{Sharda2021_driving_mode}
{Sharda} P.,  et~al., 2021, \mn@doi [\mnras] {10.1093/mnras/stab3048}, \href
  {https://ui.adsabs.harvard.edu/abs/2021MNRAS.tmp.2752S} {}

\bibitem[\protect\citeauthoryear{{Skalidis} \& {Tassis}}{{Skalidis} \&
  {Tassis}}{2020}]{Skalidis2020}
{Skalidis} R.,  {Tassis} K.,  2020, arXiv e-prints, \href
  {https://ui.adsabs.harvard.edu/abs/2020arXiv201015141S} {p. arXiv:2010.15141}

\bibitem[\protect\citeauthoryear{{Skalidis} et~al.,}{{Skalidis}
  et~al.}{2021a}]{Skalidis2021_obs_sub_alf}
{Skalidis} R.,  et~al., 2021a, arXiv e-prints, \href
  {https://ui.adsabs.harvard.edu/abs/2021arXiv211011878S} {p. arXiv:2110.11878}

\bibitem[\protect\citeauthoryear{{Skalidis}, {Sternberg}, {Beattie}, {Pavlidou}
   \& {Tassis}}{{Skalidis} et~al.}{2021b}]{Skalidis2021}
{Skalidis} R.,  {Sternberg} J.,  {Beattie} J.~R.,  {Pavlidou} V.,   {Tassis}
  K.,  2021b, \mn@doi [\aap] {10.1051/0004-6361/202142045}, \href
  {https://ui.adsabs.harvard.edu/abs/2021A&A...656A.118S} {656, A118}

\bibitem[\protect\citeauthoryear{{Soler}, {Hennebelle}, {Martin},
  {Miville-Desch{\^e}nes}, {Netterfield}  \& {Fissel}}{{Soler}
  et~al.}{2013}]{Soler2013}
{Soler} J.~D.,  {Hennebelle} P.,  {Martin} P.~G.,  {Miville-Desch{\^e}nes}
  M.~A.,  {Netterfield} C.~B.,   {Fissel} L.~M.,  2013, \mn@doi [\apj]
  {10.1088/0004-637X/774/2/128}, \href
  {https://ui.adsabs.harvard.edu/abs/2013ApJ...774..128S} {774, 128}

\bibitem[\protect\citeauthoryear{{Sridhar} \& {Goldreich}}{{Sridhar} \&
  {Goldreich}}{1994}]{Sridhar1994_weak_turbulence}
{Sridhar} S.,  {Goldreich} P.,  1994, \mn@doi [\apj] {10.1086/174600}, \href
  {https://ui.adsabs.harvard.edu/abs/1994ApJ...432..612S} {432, 612}

\bibitem[\protect\citeauthoryear{Stroustrup}{Stroustrup}{2013}]{Stroustrup2013}
Stroustrup B.,  2013, The C++ Programming Language, 4th edn.
Addison-Wesley Professional

\bibitem[\protect\citeauthoryear{{Subramanian}}{{Subramanian}}{2016}]{Subramanian2016_origins_review}
{Subramanian} K.,  2016, \mn@doi [Reports on Progress in Physics]
  {10.1088/0034-4885/79/7/076901}, \href
  {https://ui.adsabs.harvard.edu/abs/2016RPPh...79g6901S} {79, 076901}

\bibitem[\protect\citeauthoryear{{Subramanian}}{{Subramanian}}{2019}]{Subramanian2019_origins}
{Subramanian} K.,  2019, \mn@doi [Galaxies] {10.3390/galaxies7020047}, \href
  {https://ui.adsabs.harvard.edu/abs/2019Galax...7...47S} {7, 47}

\bibitem[\protect\citeauthoryear{{Sur}, {Schleicher}, {Banerjee}, {Federrath}
  \& {Klessen}}{{Sur} et~al.}{2010}]{Sur2010}
{Sur} S.,  {Schleicher} D.~R.~G.,  {Banerjee} R.,  {Federrath} C.,   {Klessen}
  R.~S.,  2010, \mn@doi [\apjl] {10.1088/2041-8205/721/2/L134}, \href
  {https://ui.adsabs.harvard.edu/abs/2010ApJ...721L.134S} {721, L134}

\bibitem[\protect\citeauthoryear{{Virtanen} et~al.,}{{Virtanen}
  et~al.}{2020}]{Virtanen2020}
{Virtanen} P.,  et~al., 2020, \mn@doi [Nature Methods]
  {https://doi.org/10.1038/s41592-019-0686-2}, \href {https://rdcu.be/b08Wh}
  {17, 261}

\bibitem[\protect\citeauthoryear{{Waagan}, {Federrath}  \&
  {Klingenberg}}{{Waagan} et~al.}{2011}]{Waagan2011}
{Waagan} K.,  {Federrath} C.,   {Klingenberg} C.,  2011, \mn@doi [Journal of
  Computational Physics] {10.1016/j.jcp.2011.01.026}, \href
  {https://ui.adsabs.harvard.edu/abs/2011JCoPh.230.3331W} {230, 3331}

\bibitem[\protect\citeauthoryear{{Wibking} \& {Krumholz}}{{Wibking} \&
  {Krumholz}}{2021}]{Wibking2021_magnetised_galaxy}
{Wibking} B.~D.,  {Krumholz} M.~R.,  2021, arXiv e-prints, \href
  {https://ui.adsabs.harvard.edu/abs/2021arXiv210504136W} {p. arXiv:2105.04136}

\bibitem[\protect\citeauthoryear{{Wolfire}, {Hollenbach}, {McKee}, {Tielens}
  \& {Bakes}}{{Wolfire} et~al.}{1995}]{Wolfire1995_isothermal_ISM}
{Wolfire} M.~G.,  {Hollenbach} D.,  {McKee} C.~F.,  {Tielens} A.~G.~G.~M.,
  {Bakes} E.~L.~O.,  1995, \mn@doi [\apj] {10.1086/175510}, \href
  {https://ui.adsabs.harvard.edu/abs/1995ApJ...443..152W} {443, 152}

\bibitem[\protect\citeauthoryear{{Xu} \& {Lazarian}}{{Xu} \&
  {Lazarian}}{2016}]{Xu2016_dynamo}
{Xu} S.,  {Lazarian} A.,  2016, \mn@doi [\apj] {10.3847/1538-4357/833/2/215},
  \href {https://ui.adsabs.harvard.edu/abs/2016ApJ...833..215X} {833, 215}

\bibitem[\protect\citeauthoryear{{Yuen} \& {Lazarian}}{{Yuen} \&
  {Lazarian}}{2020}]{Yuen2020}
{Yuen} K.~H.,  {Lazarian} A.,  2020, \mn@doi [\apj] {10.3847/1538-4357/ab9360},
  \href {https://ui.adsabs.harvard.edu/abs/2020ApJ...898...66Y} {898, 66}

\bibitem[\protect\citeauthoryear{{Yun} et~al.,}{{Yun}
  et~al.}{2021}]{Yun2021_internal_MC_scaling_law}
{Yun} H.-S.,  et~al., 2021, \mn@doi [\apj] {10.3847/1538-4357/ac193e}, \href
  {https://ui.adsabs.harvard.edu/abs/2021ApJ...921...31Y} {921, 31}

\bibitem[\protect\citeauthoryear{Zel’dovich}{Zel’dovich}{1957}]{Zeldovich_planar_dynamo_theory}
Zel’dovich Y.~B.,  1957, Sov. Phys. JETP, 460

\bibitem[\protect\citeauthoryear{{Zhou}, {Li}  \& {Chen}}{{Zhou}
  et~al.}{2021}]{Zhou2021}
{Zhou} J.-X.,  {Li} G.-X.,   {Chen} B.-Q.,  2021, arXiv e-prints, \href
  {https://ui.adsabs.harvard.edu/abs/2021arXiv211011595Z} {p. arXiv:2110.11595}

\bibitem[\protect\citeauthoryear{{Zweibel} \& {McKee}}{{Zweibel} \&
  {McKee}}{1995}]{Zweibel1995_energy_equipartition}
{Zweibel} E.~G.,  {McKee} C.~F.,  1995, \mn@doi [\apj] {10.1086/175216}, \href
  {https://ui.adsabs.harvard.edu/abs/1995ApJ...439..779Z} {439, 779}

\bibitem[\protect\citeauthoryear{van~der Walt et~al.,}{van~der Walt
  et~al.}{2014}]{vanderWalts2014}
van~der Walt S.,  et~al., 2014, \mn@doi [PeerJ] {10.7717/peerj.453}, 2, e453

\makeatother
\end{thebibliography}

\onecolumn
\def\arraystretch{1.35}
\LTcapwidth=\textwidth
\begin{ThreePartTable}
\footnotesize\setlength{\tabcolsep}{5pt}
\begin{TableNotes}
\item \textbf{\textit{Notes:}} All simulations listed are run with grid resolutions of $16^3$, $36^3$, $72^3$, $144^3$ and $288^3$. All statistics are spatially averaged over the entire domain, $\V = \V_{L}$, and are computed for 51 time realisations, across 5 correlation times of the Ornstein-Uhlenbeck forcing function. From the distributions in time, we report the values for the $16^{\rm th}$, $50^{\rm th}$, and $84^{\rm th}$ percentiles. This process minimises the possibility of using statistics that are undergoing temporally intermittent turbulent events \citep{Beattie2021b}. Column (1): the simulation ID, used throughout this study. Column (2): the turbulent Mach number, $\M \equiv \Exp{(\delta v / c_s)^2}^{1/2}_{\V_{L}}$. Column (3): the correlation scale of the turbulence, $\cor{v}$, in units of the driving scale, $\ell_0$, defined directly from the power spectra in \autoref{eq:correlation_scale}. Column (4): the Alfv\'en Mach number of the mean magnetic field, $\Mao \equiv \Exp{ (\delta v \sqrt{4\pi\rho_0} ) / B_0}$, with fluctuations coming from $\delta v$, since $\partial_{x_i}\vecB{B}_0 = \partial_{t} \vecB{B}_0 = 0$. Column (5): the mean magnetic field strength in units of $c_s \rho_0^{1/2}$. Column (6): the volume-averaged square of the turbulent magnetic field, proportional to the turbulent magnetic energy, in units of thermal energy. Column (7): the volume-averaged root-mean-squared of the magnetic coupling term $\dbBo$, in units of thermal energy. Column (8): the Alfv\'en Mach number of the turbulent magnetic field, \autoref{eq:turbMa}. Column (9): the Alfv\'en Mach number of the total magnetic field, \autoref{eq:totalMa}. 
\end{TableNotes}
\begin{longtable}{l r@{}l r@{}l r@{}l c r@{}l r@{}l r@{}l r@{}l}
\caption{Main simulation parameters and derived quantities used throughout this study.}\\
\hline
\hline
& \multicolumn{4}{c}{Turbulence} & \multicolumn{3}{c}{Large-scale $\vecB{B}$-Field} & \multicolumn{6}{c}{Fluctuating $\vecB{B}$-Field} & \multicolumn{2}{c}{Total $\vecB{B}$-Field}\\
\hline\\
\multicolumn{1}{c}{Simulation ID} & \multicolumn{2}{c}{$\M$} & \multicolumn{2}{c}{$\frac{\ell_{\text{cor}, v}}{\ell_0}$} & \multicolumn{2}{c}{$\Mao$} & \multicolumn{1}{c}{$\frac{B_0}{c_s \rho_0^{1/2}}$} & \multicolumn{2}{c}{$\Exp{\frac{\delta B^2}{c_s^2 \rho_0}}_{\V}$} & \multicolumn{2}{c}{$\Exp{\left(\frac{\dB\cdot\Bo}{c_s^2 \rho_0}\right)^2 }^{1/2}_{\V}$} & \multicolumn{2}{c}{$\Maturb$} & \multicolumn{2}{c}{$\Matot$}\\
\multicolumn{1}{c}{(1)} & \multicolumn{2}{c}{(2)} & \multicolumn{2}{c}{(3)} & \multicolumn{2}{c}{(4)} & \multicolumn{1}{c}{(5)} & \multicolumn{2}{c}{(6)} & \multicolumn{2}{c}{(7)} & \multicolumn{2}{c}{(8)} & \multicolumn{2}{c}{(9)}\\[0.5em]
\hline
\texttt{M05MA001}&$0.567$ &$_{-0.07}^{+0.02}$&$1.48$ &$_{-0.07}^{+0.05}$&$0.0113$ &$_{-0.001}^{+0.0003}$&$177.0$&$0.000191$ &$_{-0.005}^{+0.004}$&$2.32$ &$_{-0.5}^{+0.4}$&$392.0$ &$_{-40.0}^{+20.0}$&$0.0113$ &$_{-0.001}^{+0.0003}$\\
\texttt{M05MA01}&$0.566$ &$_{-0.07}^{+0.02}$&$1.47$ &$_{-0.07}^{+0.05}$&$0.113$ &$_{-0.01}^{+0.004}$&$17.7$&$0.0181$ &$_{-0.04}^{+0.03}$&$2.23$ &$_{-0.4}^{+0.3}$&$35.6$ &$_{-2.0}^{+2.0}$&$0.113$ &$_{-0.01}^{+0.004}$\\
\texttt{M05MA05}&$0.534$ &$_{-0.04}^{+0.004}$&$1.44$ &$_{-0.07}^{+0.02}$&$0.534$ &$_{-0.04}^{+0.004}$&$3.54$&$0.233$ &$_{-0.08}^{+0.05}$&$1.16$ &$_{-0.1}^{+0.05}$&$7.28$ &$_{-0.2}^{+0.3}$&$0.544$ &$_{-0.05}^{+0.005}$\\
\texttt{M05MA1}&$0.469$ &$_{-0.05}^{+0.01}$&$0.953$ &$_{-0.005}^{+0.02}$&$0.938$ &$_{-0.1}^{+0.03}$&$1.77$&$1.2$ &$_{-0.3}^{+0.1}$&$1.0$ &$_{-0.07}^{+0.08}$&$2.75$ &$_{-0.2}^{+0.1}$&$0.974$ &$_{-0.08}^{+0.07}$\\
\texttt{M05MA2}&$0.462$ &$_{-0.03}^{+0.03}$&$0.783$ &$_{-0.03}^{+0.06}$&$1.85$ &$_{-0.1}^{+0.1}$&$0.886$&$3.17$ &$_{-0.3}^{+0.3}$&$0.871$ &$_{-0.07}^{+0.03}$&$1.67$ &$_{-0.08}^{+0.2}$&$1.59$ &$_{-0.09}^{+0.2}$\\
\texttt{M05MA4}&$0.473$ &$_{-0.03}^{+0.06}$&$0.838$ &$_{-0.03}^{+0.03}$&$3.79$ &$_{-0.2}^{+0.5}$&$0.443$&$3.12$ &$_{-0.2}^{+0.2}$&$0.415$ &$_{-0.009}^{+0.02}$&$1.98$ &$_{-0.2}^{+0.2}$&$2.1$ &$_{-0.2}^{+0.4}$\\
\texttt{M05MA6}&$0.483$ &$_{-0.03}^{+0.06}$&$0.904$ &$_{-0.03}^{+0.03}$&$5.8$ &$_{-0.4}^{+0.8}$&$0.295$&$2.74$ &$_{-0.1}^{+0.09}$&$0.256$ &$_{-0.006}^{+0.01}$&$2.32$ &$_{-0.1}^{+0.4}$&$2.51$ &$_{-0.2}^{+0.5}$\\
\texttt{M05MA8}&$0.502$ &$_{-0.02}^{+0.06}$&$0.907$ &$_{-0.02}^{+0.05}$&$8.03$ &$_{-0.3}^{+1.0}$&$0.222$&$2.47$ &$_{-0.2}^{+0.1}$&$0.192$ &$_{-0.01}^{+0.009}$&$2.76$ &$_{-0.1}^{+0.3}$&$2.99$ &$_{-0.2}^{+0.3}$\\
\texttt{M05MA10}&$0.514$ &$_{-0.03}^{+0.06}$&$0.934$ &$_{-0.03}^{+0.02}$&$10.3$ &$_{-0.5}^{+1.0}$&$0.177$&$2.24$ &$_{-0.07}^{+0.1}$&$0.145$ &$_{-0.006}^{+0.01}$&$3.05$ &$_{-0.2}^{+0.4}$&$3.18$ &$_{-0.1}^{+0.4}$\\
\texttt{M05MA100}&$0.637$ &$_{-0.05}^{+0.09}$&$0.889$ &$_{-0.03}^{+0.04}$&$127.0$ &$_{-10.0}^{+20.0}$&$0.0177$&$0.721$ &$_{-0.3}^{+0.1}$&$0.00852$ &$_{-0.001}^{+0.0008}$&$7.9$ &$_{-1.0}^{+3.0}$&$7.92$ &$_{-1.0}^{+3.0}$\\
\texttt{M05MA1000}&$0.672$ &$_{-0.05}^{+0.07}$&$0.835$ &$_{-0.04}^{+0.03}$&$1340.0$ &$_{-100.0}^{+100.0}$&$0.00177$&$0.36$ &$_{-0.3}^{+0.2}$&$0.000613$ &$_{-0.0002}^{+0.0001}$&$11.8$ &$_{-3.0}^{+8.0}$&$11.8$ &$_{-3.0}^{+8.0}$\\
\texttt{M2MA001}&$1.9$ &$_{-0.06}^{+0.04}$&$1.34$ &$_{-0.09}^{+0.02}$&$0.0095$ &$_{-0.0003}^{+0.0002}$&$709.0$&$0.000879$ &$_{-0.003}^{+0.02}$&$18.7$ &$_{-2.0}^{+9.0}$&$498.0$ &$_{-30.0}^{+20.0}$&$0.00943$ &$_{-0.0003}^{+0.0002}$\\
\texttt{M2MA01}&$1.85$ &$_{-0.02}^{+0.06}$&$1.37$ &$_{-0.1}^{+0.03}$&$0.0926$ &$_{-0.001}^{+0.003}$&$70.9$&$0.0989$ &$_{-0.03}^{+0.1}$&$18.4$ &$_{-1.0}^{+4.0}$&$41.1$ &$_{-3.0}^{+2.0}$&$0.0917$ &$_{-0.001}^{+0.003}$\\
\texttt{M2MA05}&$2.24$ &$_{-0.09}^{+0.07}$&$1.26$ &$_{-0.06}^{+0.08}$&$0.561$ &$_{-0.02}^{+0.02}$&$14.2$&$6.81$ &$_{-0.3}^{+0.3}$&$20.6$ &$_{-2.0}^{+3.0}$&$5.28$ &$_{-0.3}^{+0.3}$&$0.575$ &$_{-0.03}^{+0.01}$\\
\texttt{M2MA1}&$1.98$ &$_{-0.09}^{+0.1}$&$0.845$ &$_{-0.02}^{+0.02}$&$0.989$ &$_{-0.05}^{+0.05}$&$7.09$&$25.3$ &$_{-1.0}^{+0.3}$&$16.3$ &$_{-1.0}^{+0.7}$&$2.47$ &$_{-0.2}^{+0.2}$&$1.01$ &$_{-0.06}^{+0.05}$\\
\texttt{M2MA2}&$1.95$ &$_{-0.2}^{+0.2}$&$0.815$ &$_{-0.04}^{+0.04}$&$1.95$ &$_{-0.2}^{+0.2}$&$3.54$&$40.6$ &$_{-0.9}^{+0.6}$&$12.3$ &$_{-1.0}^{+0.7}$&$1.73$ &$_{-0.07}^{+0.4}$&$1.54$ &$_{-0.1}^{+0.1}$\\
\texttt{M2MA4}&$2.07$ &$_{-0.1}^{+0.2}$&$0.92$ &$_{-0.04}^{+0.02}$&$4.13$ &$_{-0.3}^{+0.5}$&$1.77$&$37.0$ &$_{-0.3}^{+0.3}$&$5.88$ &$_{-0.3}^{+0.4}$&$2.39$ &$_{-0.4}^{+0.1}$&$2.39$ &$_{-0.4}^{+0.3}$\\
\texttt{M2MA6}&$2.09$ &$_{-0.2}^{+0.2}$&$0.946$ &$_{-0.05}^{+0.02}$&$6.26$ &$_{-0.6}^{+0.6}$&$1.18$&$33.8$ &$_{-0.4}^{+0.5}$&$3.83$ &$_{-0.1}^{+0.1}$&$2.45$ &$_{-0.3}^{+0.5}$&$2.68$ &$_{-0.5}^{+0.6}$\\
\texttt{M2MA8}&$2.05$ &$_{-0.2}^{+0.09}$&$0.945$ &$_{-0.02}^{+0.04}$&$8.21$ &$_{-0.7}^{+0.4}$&$0.886$&$21.5$ &$_{-0.3}^{+0.3}$&$2.39$ &$_{-0.1}^{+0.1}$&$3.44$ &$_{-0.3}^{+0.2}$&$3.91$ &$_{-0.3}^{+0.2}$\\
\texttt{M2MA10}&$2.11$ &$_{-0.2}^{+0.1}$&$0.987$ &$_{-0.03}^{+0.03}$&$10.5$ &$_{-1.0}^{+0.7}$&$0.709$&$18.4$ &$_{-0.1}^{+0.3}$&$1.74$ &$_{-0.04}^{+0.1}$&$4.07$ &$_{-0.6}^{+0.3}$&$4.65$ &$_{-0.9}^{+0.4}$\\
\texttt{M2MA100}&$2.36$ &$_{-0.2}^{+0.1}$&$1.01$ &$_{-0.02}^{+0.03}$&$118.0$ &$_{-8.0}^{+6.0}$&$0.0709$&$3.3$ &$_{-0.5}^{+0.08}$&$0.0729$ &$_{-0.006}^{+0.003}$&$16.1$ &$_{-2.0}^{+3.0}$&$17.0$ &$_{-3.0}^{+4.0}$\\
\texttt{M2MA1000}&$2.37$ &$_{-0.1}^{+0.08}$&$0.999$ &$_{-0.05}^{+0.01}$&$1180.0$ &$_{-70.0}^{+40.0}$&$0.00709$&$0.276$ &$_{-0.3}^{+0.3}$&$0.00214$ &$_{-0.0007}^{+0.0005}$&$61.9$ &$_{-20.0}^{+40.0}$&$62.8$ &$_{-20.0}^{+40.0}$\\
\texttt{M4MA01}&$4.03$ &$_{-0.5}^{+0.06}$&$1.39$ &$_{-0.09}^{+0.05}$&$0.101$ &$_{-0.01}^{+0.002}$&$142.0$&$0.838$ &$_{-0.4}^{+0.2}$&$106.0$ &$_{-20.0}^{+9.0}$&$36.5$ &$_{-3.0}^{+3.0}$&$0.1$ &$_{-0.01}^{+0.002}$\\
\texttt{M4MA05}&$4.08$ &$_{-0.1}^{+0.04}$&$1.27$ &$_{-0.1}^{+0.04}$&$0.51$ &$_{-0.01}^{+0.005}$&$28.4$&$24.4$ &$_{-0.6}^{+0.4}$&$74.1$ &$_{-7.0}^{+7.0}$&$4.94$ &$_{-0.4}^{+0.3}$&$0.513$ &$_{-0.01}^{+0.007}$\\
\texttt{M4MA1}&$4.12$ &$_{-0.4}^{+0.5}$&$0.832$ &$_{-0.04}^{+0.04}$&$1.03$ &$_{-0.09}^{+0.1}$&$14.2$&$84.3$ &$_{-2.0}^{+2.0}$&$61.9$ &$_{-10.0}^{+7.0}$&$2.49$ &$_{-0.1}^{+0.3}$&$1.02$ &$_{-0.1}^{+0.05}$\\
\texttt{M4MA2}&$4.02$ &$_{-0.3}^{+0.2}$&$0.846$ &$_{-0.02}^{+0.04}$&$2.01$ &$_{-0.1}^{+0.08}$&$7.09$&$125.0$ &$_{-1.0}^{+1.0}$&$43.4$ &$_{-6.0}^{+2.0}$&$1.93$ &$_{-0.1}^{+0.2}$&$1.49$ &$_{-0.1}^{+0.2}$\\
\texttt{M4MA4}&$4.03$ &$_{-0.4}^{+0.2}$&$0.916$ &$_{-0.03}^{+0.02}$&$4.03$ &$_{-0.4}^{+0.2}$&$3.54$&$109.0$ &$_{-0.5}^{+0.2}$&$20.3$ &$_{-0.5}^{+2.0}$&$2.38$ &$_{-0.3}^{+0.1}$&$2.41$ &$_{-0.3}^{+0.1}$\\
\texttt{M4MA6}&$3.97$ &$_{-0.4}^{+0.2}$&$0.959$ &$_{-0.02}^{+0.02}$&$5.96$ &$_{-0.6}^{+0.3}$&$2.36$&$76.3$ &$_{-0.6}^{+0.3}$&$11.7$ &$_{-0.4}^{+0.7}$&$2.94$ &$_{-0.3}^{+0.4}$&$3.29$ &$_{-0.2}^{+0.4}$\\
\texttt{M4MA8}&$4.05$ &$_{-0.5}^{+0.08}$&$0.979$ &$_{-0.03}^{+0.02}$&$8.1$ &$_{-1.0}^{+0.2}$&$1.77$&$68.5$ &$_{-0.4}^{+0.6}$&$8.38$ &$_{-0.3}^{+0.3}$&$3.41$ &$_{-0.6}^{+0.3}$&$4.15$ &$_{-1.0}^{+0.4}$\\
\texttt{M4MA10}&$3.91$ &$_{-0.2}^{+0.2}$&$0.994$ &$_{-0.01}^{+0.03}$&$9.78$ &$_{-0.6}^{+0.5}$&$1.42$&$59.7$ &$_{-0.2}^{+0.2}$&$6.27$ &$_{-0.3}^{+0.2}$&$3.69$ &$_{-0.1}^{+0.3}$&$4.23$ &$_{-0.3}^{+0.9}$\\
\texttt{M4MA100}&$4.21$ &$_{-0.08}^{+0.2}$&$1.02$ &$_{-0.02}^{+0.03}$&$105.0$ &$_{-2.0}^{+5.0}$&$0.142$&$6.29$ &$_{-0.4}^{+0.4}$&$0.203$ &$_{-0.01}^{+0.01}$&$21.0$ &$_{-2.0}^{+3.0}$&$24.2$ &$_{-2.0}^{+5.0}$\\
\texttt{M4MA1000}&$4.31$ &$_{-0.2}^{+0.1}$&$1.05$ &$_{-0.01}^{+0.02}$&$1080.0$ &$_{-50.0}^{+30.0}$&$0.0142$&$0.391$ &$_{-0.3}^{+0.4}$&$0.00519$ &$_{-0.001}^{+0.001}$&$115.0$ &$_{-40.0}^{+20.0}$&$128.0$ &$_{-50.0}^{+30.0}$\\
\texttt{M6MA01}&$6.96$ &$_{-0.8}^{+0.5}$&$1.35$ &$_{-0.04}^{+0.05}$&$0.116$ &$_{-0.01}^{+0.008}$&$213.0$&$3.86$ &$_{-0.7}^{+0.3}$&$325.0$ &$_{-60.0}^{+20.0}$&$29.0$ &$_{-3.0}^{+4.0}$&$0.115$ &$_{-0.01}^{+0.008}$\\
\texttt{M6MA05}&$6.44$ &$_{-0.1}^{+0.2}$&$1.21$ &$_{-0.04}^{+0.02}$&$0.536$ &$_{-0.01}^{+0.02}$&$42.5$&$58.0$ &$_{-0.7}^{+1.0}$&$169.0$ &$_{-10.0}^{+10.0}$&$5.02$ &$_{-0.3}^{+0.3}$&$0.535$ &$_{-0.01}^{+0.01}$\\
\texttt{M6MA1}&$6.01$ &$_{-0.8}^{+0.8}$&$0.844$ &$_{-0.02}^{+0.04}$&$1.0$ &$_{-0.1}^{+0.1}$&$21.3$&$157.0$ &$_{-4.0}^{+1.0}$&$121.0$ &$_{-20.0}^{+10.0}$&$2.73$ &$_{-0.2}^{+0.2}$&$0.937$ &$_{-0.2}^{+0.1}$\\
\texttt{M6MA2}&$5.81$ &$_{-0.3}^{+0.3}$&$0.844$ &$_{-0.03}^{+0.03}$&$1.94$ &$_{-0.09}^{+0.1}$&$10.6$&$217.0$ &$_{-0.5}^{+2.0}$&$83.7$ &$_{-5.0}^{+9.0}$&$2.1$ &$_{-0.1}^{+0.2}$&$1.58$ &$_{-0.09}^{+0.09}$\\
\texttt{M6MA4}&$6.23$ &$_{-0.4}^{+0.1}$&$0.94$ &$_{-0.01}^{+0.02}$&$4.15$ &$_{-0.3}^{+0.08}$&$5.32$&$202.0$ &$_{-2.0}^{+0.8}$&$42.8$ &$_{-5.0}^{+2.0}$&$2.55$ &$_{-0.2}^{+0.2}$&$2.53$ &$_{-0.2}^{+0.3}$\\
\texttt{M6MA6}&$5.96$ &$_{-0.6}^{+0.4}$&$0.946$ &$_{-0.01}^{+0.02}$&$5.96$ &$_{-0.6}^{+0.4}$&$3.54$&$168.0$ &$_{-1.0}^{+1.0}$&$25.8$ &$_{-2.0}^{+3.0}$&$2.82$ &$_{-0.2}^{+0.3}$&$3.04$ &$_{-0.2}^{+0.3}$\\
\texttt{M6MA8}&$5.96$ &$_{-0.5}^{+0.2}$&$0.979$ &$_{-0.01}^{+0.01}$&$7.95$ &$_{-0.6}^{+0.2}$&$2.66$&$139.0$ &$_{-0.9}^{+0.3}$&$17.4$ &$_{-0.6}^{+1.0}$&$3.39$ &$_{-0.4}^{+0.2}$&$3.73$ &$_{-0.4}^{+0.2}$\\
\texttt{M6MA10}&$5.99$ &$_{-0.5}^{+0.2}$&$0.99$ &$_{-0.02}^{+0.02}$&$9.99$ &$_{-0.8}^{+0.4}$&$2.13$&$124.0$ &$_{-1.0}^{+0.5}$&$12.6$ &$_{-0.4}^{+1.0}$&$3.82$ &$_{-0.6}^{+0.8}$&$4.11$ &$_{-0.7}^{+1.0}$\\
\texttt{M8MA01}&$8.7$ &$_{-0.9}^{+0.09}$&$1.49$ &$_{-0.1}^{+0.02}$&$0.109$ &$_{-0.01}^{+0.001}$&$284.0$&$4.53$ &$_{-0.7}^{+0.4}$&$488.0$ &$_{-90.0}^{+40.0}$&$32.0$ &$_{-4.0}^{+4.0}$&$0.108$ &$_{-0.01}^{+0.001}$\\
\texttt{M8MA05}&$8.35$ &$_{-0.3}^{+0.2}$&$1.14$ &$_{-0.02}^{+0.02}$&$0.522$ &$_{-0.02}^{+0.01}$&$56.7$&$90.2$ &$_{-0.6}^{+1.0}$&$278.0$ &$_{-30.0}^{+20.0}$&$5.01$ &$_{-0.4}^{+0.3}$&$0.505$ &$_{-0.03}^{+0.01}$\\
\texttt{M8MA1}&$8.16$ &$_{-1.0}^{+0.6}$&$0.837$ &$_{-0.03}^{+0.04}$&$1.02$ &$_{-0.1}^{+0.08}$&$28.4$&$257.0$ &$_{-4.0}^{+2.0}$&$205.0$ &$_{-20.0}^{+10.0}$&$2.74$ &$_{-0.3}^{+0.3}$&$0.92$ &$_{-0.1}^{+0.07}$\\
\texttt{M8MA2}&$8.15$ &$_{-0.5}^{+0.4}$&$0.863$ &$_{-0.03}^{+0.03}$&$2.04$ &$_{-0.1}^{+0.1}$&$14.2$&$395.0$ &$_{-1.0}^{+1.0}$&$148.0$ &$_{-8.0}^{+7.0}$&$2.25$ &$_{-0.2}^{+0.2}$&$1.59$ &$_{-0.2}^{+0.1}$\\
\texttt{M8MA4}&$7.99$ &$_{-0.4}^{+0.3}$&$0.933$ &$_{-0.03}^{+0.01}$&$4.0$ &$_{-0.2}^{+0.1}$&$7.09$&$344.0$ &$_{-2.0}^{+0.8}$&$72.7$ &$_{-8.0}^{+4.0}$&$2.41$ &$_{-0.1}^{+0.2}$&$2.3$ &$_{-0.2}^{+0.2}$\\
\texttt{M8MA6}&$7.97$ &$_{-0.9}^{+0.4}$&$0.938$ &$_{-0.02}^{+0.02}$&$5.98$ &$_{-0.7}^{+0.3}$&$4.73$&$278.0$ &$_{-1.0}^{+2.0}$&$42.4$ &$_{-1.0}^{+7.0}$&$2.84$ &$_{-0.5}^{+0.2}$&$2.93$ &$_{-0.6}^{+0.3}$\\
\texttt{M8MA8}&$7.82$ &$_{-0.4}^{+0.4}$&$0.994$ &$_{-0.02}^{+0.01}$&$7.82$ &$_{-0.4}^{+0.4}$&$3.54$&$247.0$ &$_{-1.0}^{+0.8}$&$31.5$ &$_{-2.0}^{+2.0}$&$3.2$ &$_{-0.3}^{+0.2}$&$3.46$ &$_{-0.3}^{+0.5}$\\
\texttt{M8MA10}&$8.06$ &$_{-1.0}^{+0.2}$&$1.01$ &$_{-0.03}^{+0.02}$&$10.1$ &$_{-1.0}^{+0.3}$&$2.84$&$197.0$ &$_{-0.5}^{+1.0}$&$22.3$ &$_{-0.5}^{+1.0}$&$3.72$ &$_{-0.4}^{+0.6}$&$3.89$ &$_{-0.5}^{+0.6}$\\
\texttt{M10MA01}&$11.3$ &$_{-1.0}^{+0.3}$&$1.45$ &$_{-0.09}^{+0.07}$&$0.113$ &$_{-0.01}^{+0.003}$&$354.0$&$6.76$ &$_{-1.0}^{+1.0}$&$720.0$ &$_{-200.0}^{+100.0}$&$33.7$ &$_{-7.0}^{+8.0}$&$0.112$ &$_{-0.01}^{+0.003}$\\
\texttt{M10MA05}&$10.2$ &$_{-0.3}^{+0.3}$&$1.13$ &$_{-0.02}^{+0.02}$&$0.509$ &$_{-0.02}^{+0.01}$&$70.9$&$143.0$ &$_{-2.0}^{+2.0}$&$420.0$ &$_{-40.0}^{+20.0}$&$4.88$ &$_{-0.5}^{+0.6}$&$0.496$ &$_{-0.02}^{+0.02}$\\
\texttt{M10MA1}&$9.72$ &$_{-0.9}^{+0.9}$&$0.868$ &$_{-0.03}^{+0.03}$&$0.972$ &$_{-0.09}^{+0.09}$&$35.4$&$338.0$ &$_{-4.0}^{+3.0}$&$309.0$ &$_{-40.0}^{+20.0}$&$2.79$ &$_{-0.1}^{+0.4}$&$0.888$ &$_{-0.1}^{+0.06}$\\
\texttt{M10MA2}&$10.5$ &$_{-0.8}^{+0.3}$&$0.888$ &$_{-0.02}^{+0.02}$&$2.1$ &$_{-0.2}^{+0.07}$&$17.7$&$549.0$ &$_{-1.0}^{+0.9}$&$215.0$ &$_{-10.0}^{+9.0}$&$2.39$ &$_{-0.3}^{+0.1}$&$1.56$ &$_{-0.1}^{+0.2}$\\
\texttt{M10MA4}&$10.3$ &$_{-0.4}^{+0.4}$&$0.931$ &$_{-0.03}^{+0.03}$&$4.12$ &$_{-0.2}^{+0.2}$&$8.86$&$529.0$ &$_{-2.0}^{+2.0}$&$112.0$ &$_{-6.0}^{+8.0}$&$2.42$ &$_{-0.2}^{+0.4}$&$2.32$ &$_{-0.2}^{+0.4}$\\
\texttt{M10MA6}&$9.78$ &$_{-1.0}^{+0.3}$&$0.952$ &$_{-0.03}^{+0.02}$&$5.87$ &$_{-0.7}^{+0.2}$&$5.91$&$398.0$ &$_{-0.7}^{+2.0}$&$65.2$ &$_{-3.0}^{+7.0}$&$2.7$ &$_{-0.4}^{+0.3}$&$2.77$ &$_{-0.4}^{+0.3}$\\
\texttt{M10MA8}&$10.0$ &$_{-0.9}^{+0.4}$&$0.997$ &$_{-0.02}^{+0.02}$&$8.03$ &$_{-0.7}^{+0.3}$&$4.43$&$346.0$ &$_{-1.0}^{+1.0}$&$46.7$ &$_{-1.0}^{+4.0}$&$3.27$ &$_{-0.4}^{+0.4}$&$3.41$ &$_{-0.6}^{+0.4}$\\
\texttt{M10MA10}&$9.42$ &$_{-0.5}^{+0.2}$&$0.99$ &$_{-0.01}^{+0.01}$&$9.42$ &$_{-0.5}^{+0.2}$&$3.54$&$288.0$ &$_{-3.0}^{+2.0}$&$34.3$ &$_{-4.0}^{+3.0}$&$3.37$ &$_{-0.2}^{+0.3}$&$3.5$ &$_{-0.3}^{+0.5}$\\
\hline
\hline
\insertTableNotes
\label{tb:simtab}
\end{longtable}
\end{ThreePartTable}
\twocolumn

\appendix


\section{Numerical convergence}
\label{app:convergence}
    All of the results presented in this study are numerically converged in grid resolution. We ensure that this is the case by computing all of our statistics on simulations with discretisations $18^3$, $36^3$, $72^3$, $144^3$ and $288^3$. To highlight the convergence trends, we fit a general logistic function to the data,
    \begin{align}
        f(N_{\rm cells}) = \frac{f^+_{\infty}}{1 + \exp\left\{ \alpha (N_{\rm cells} - N_0)\right\}} + f^-_{\infty},
    \end{align}
    where $\alpha$ controls the rate in which the function converges to either $f^-_{\infty}$, for monotonically decreasing functions ($\alpha > 0$), and $f^+_{\infty} + f^-_{\infty}$ for monotonically increasing functions ($\alpha < 0$), and $N_0$ shifts the function along the $N_{\rm cells}$ axis. 
    
\begin{figure}
        \centering
        \includegraphics[width=\linewidth]{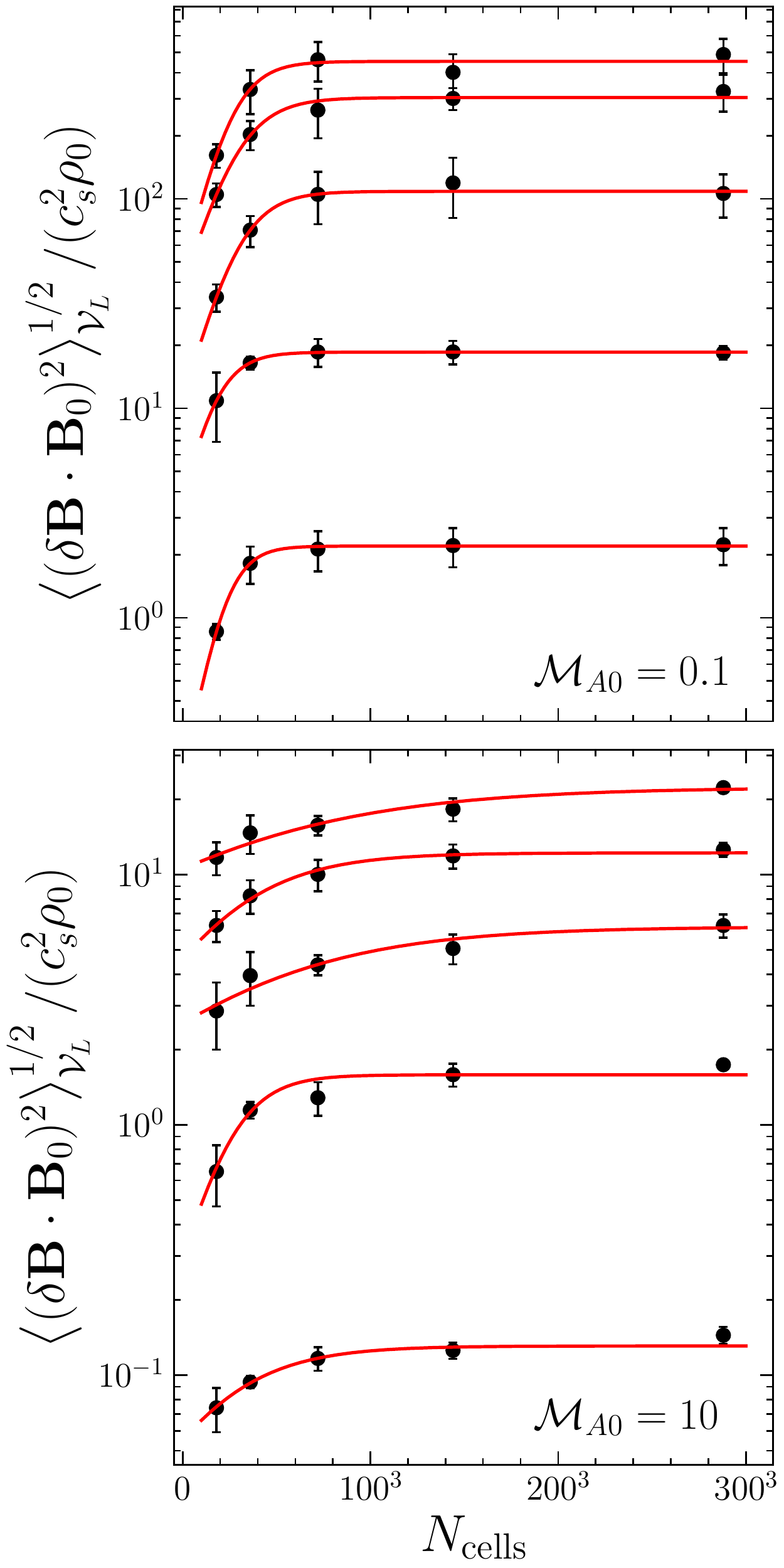}
        \caption{The coupling term, as discussed in \autoref{sec:energy_balance}, as a function of numerical grid resolution for the \texttt{MA01} and \texttt{MA10} simulations, from \autoref{tb:simtab}. Each curve shows a different set of $\M$, varying from $\M=0.5$, corresponding to data lowest on the coupling term axis, up to $\M = 10$, at the top.}
        \label{fig:dbbo_convergence}
    \end{figure}

    We plot $\dbBovol$ as a function of grid cells, $N_{\rm cells}$, for the $\Mao = 0.1$ (top) and $\Mao = 10$ (bottom) simulation ensembles, listed in \autoref{tb:simtab}, in \autoref{fig:dbbo_convergence}, showcasing representative examples for the super- and sub-Alfv\'enic simulations. Naturally, the simulations are organised into constant $\M$, with low-$\M$ corresponding to the smallest $\dbBovol$ and vice versa for high-$\M$. For all simulations in \autoref{fig:dbbo_convergence}, $144^3$ grid resolution is sufficient for numerical convergence. This is because, unlike power-spectra or other two-point statistics, which require $> 4096^3$ resolutions to converge properly \citep{Federrath2013_universality,Federrath2021} the variance of the fields convergence quickly. This is because the low-$k$ modes contain most of the power in the stochastic fields, and hence $\gtrsim144^3$ is sufficient to resolve the largest contributions to the $2^{\rm nd}$ moments of the fields. In the main text of this study, we use simulations discretised with $288^3$ cells, to ensure that the above condition is met.
    
\section{Anisotropy of the magnetic and velocity fluctuations}\label{app:anisotropy}
\begin{figure}
        \centering
        \includegraphics[width=\linewidth]{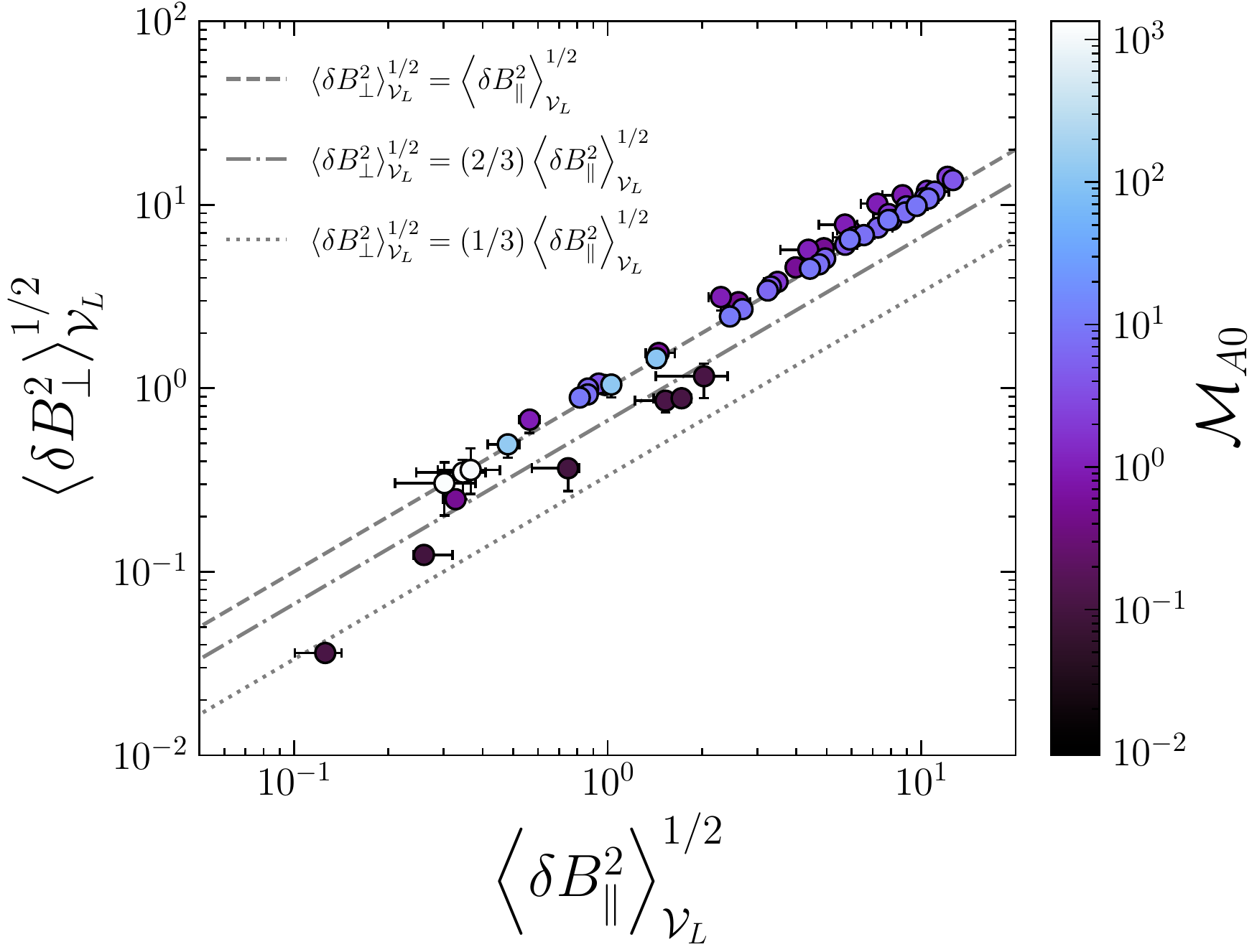}
        \caption{The rms perpendicular magnetic field fluctuations, $\Exp{\dBperp^2}^{1/2}_{\V}$ as a function of the rms parallel magnetic field fluctuations, $\Exp{\dBpar^2}^{1/2}_{\V}$, coloured by $\Mao$. We show contours at $\Exp{\delta B_{\perp}^2}^{1/2}_{\V} = \Exp{\delta B_{\parallel}^2}^{1/2}_{\V}$, $\Exp{\dBperp^2}^{1/2}_{\V} =(2/3)\Exp{\dBpar^2}^{1/2}_{\V}$ and $\Exp{\dBperp^2}^{1/2}_{\V} =(1/3)\Exp{\dBpar^2}^{1/2}_{\V}$, illustrating how the rms magnetic field fluctuations become anisotropic in the $\Mao < 1$ regime.}
        \label{fig:b_par-b_perp}
    \end{figure}

\begin{figure}
        \centering
        \includegraphics[width=\linewidth]{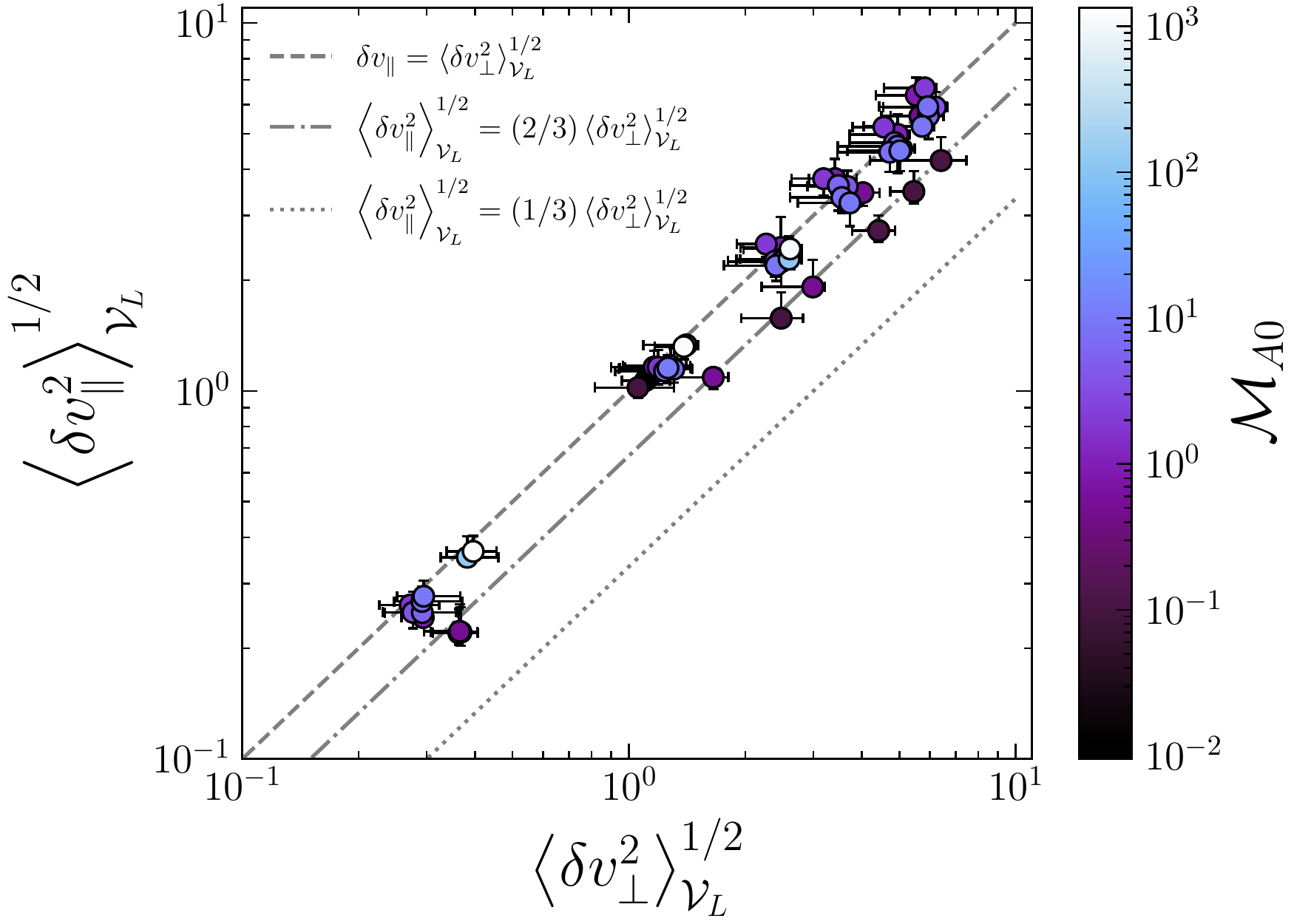}
        \caption{The same as \autoref{fig:b_par-b_perp}, but for the rms velocity fluctuations perpendicular and parallel to $\Bo$. Note, compared to \autoref{fig:b_par-b_perp} the anisotropy is inverted between the parallel and perpendicular directions.}
        \label{fig:M_perp-M_par}
    \end{figure}

    In \autoref{sec:alfven_mach} we modify the energy balance relations based on the anisotropy of the rms magnetic and velocity fluctuations. If the turbulence is isotropic (with respect to the large-scale field) we have $\sqrt{3}\Exp{\delta B_{\parallel}^2}^{1/2}_{\V} = \Exp{\delta B^2}^{1/2}_{\V}$. However, sub-Alfv\'enic mean-field turbulence is highly-anisotropic on all scales, in the velocity, magnetic field and density statistics \citep{Beattie2020}. Here we specifically plot the fluctuations, showing $\Exp{\dBperp^2}^{1/2}_{\V}$ as a function of $\Exp{\dBpar^2}^{1/2}_{\V}$ in \autoref{fig:b_par-b_perp} and $v_{\parallel}$ as a function of $\Exp{\delta v_{\perp}^2}^{1/2}_{\V}$ in \autoref{fig:M_perp-M_par}, all averaged over $5$ correlations times of the turbulent forcing function, discussed in \autoref{sec:sims}.
    
    Naturally, the most anisotropic rms statistics come from the strongly sub-Alfv\'enic simulations, which can be as extreme as $\Exp{\dBperp^2}^{1/2}_{\V} = (1/3)\Exp{\dBpar^2}^{1/2}_{\V}$ and $(1/3)\Exp{\delta v_{\perp}^2}^{1/2}_{\V} \leq \Exp{\delta v_{\parallel}^2}^{1/2}_{\V} \leq (2/3)\Exp{\delta v_{\perp}^2}^{1/2}_{\V}$. In the mean-field coordinate system that we work in through this study, $\Exp{\delta B^2}^{1/2}_{\V} = \sqrt{ \Exp{\delta B_{\perp,1}^2}_{\V} + \Exp{\delta B_{\perp,2}^2}_{\V} + \Exp{\delta B_{\parallel}^2}_{\V} }$, and likewise for the velocity. For $(1/3)\Exp{\dBpar^2}^{1/2}_{\V} \leq \Exp{\dBperp^2}^{1/2}_{\V} \leq (2/3)\Exp{\dBpar^2}^{1/2}_{\V}$, the total magnitude is then related to $\Exp{\dBpar^2}^{1/2}_{\V}$ via the inequality.
    \begin{align}
        \frac{\sqrt{11}}{3}\Exp{\dBpar^2}^{1/2}_{\V} \leq \Exp{\delta B^2}^{1/2}_{\V} \leq \frac{\sqrt{17}}{3}\Exp{\dBpar^2}^{1/2}_{\V},
    \end{align}
    and likewise for $\Exp{\delta v^2}^{1/2}_{\V}$ and $\Exp{\delta v_{\perp}^2}^{1/2}_{\V}$,
    \begin{align}
        \frac{\sqrt{19}}{3}\Exp{\delta v_{\perp}^2}^{1/2}_{\V} \leq \Exp{\delta v^2}^{1/2}_{\V} \leq \frac{\sqrt{22}}{3}\Exp{\delta v_{\perp}^2}^{1/2}_{\V},
    \end{align}   
    Because the coefficients are so close to unity ($\sqrt{11}/3\approx 1.11$, $\sqrt{17}/3\approx 1.37)$ this demonstrates that in the sub-Alfv\'enic mean-field regime it is the parallel magnetic field fluctuations and perpendicular velocity fluctuations that dominate the respective total fluctuations.

    \begin{figure}
        \centering
        \includegraphics[width=\linewidth]{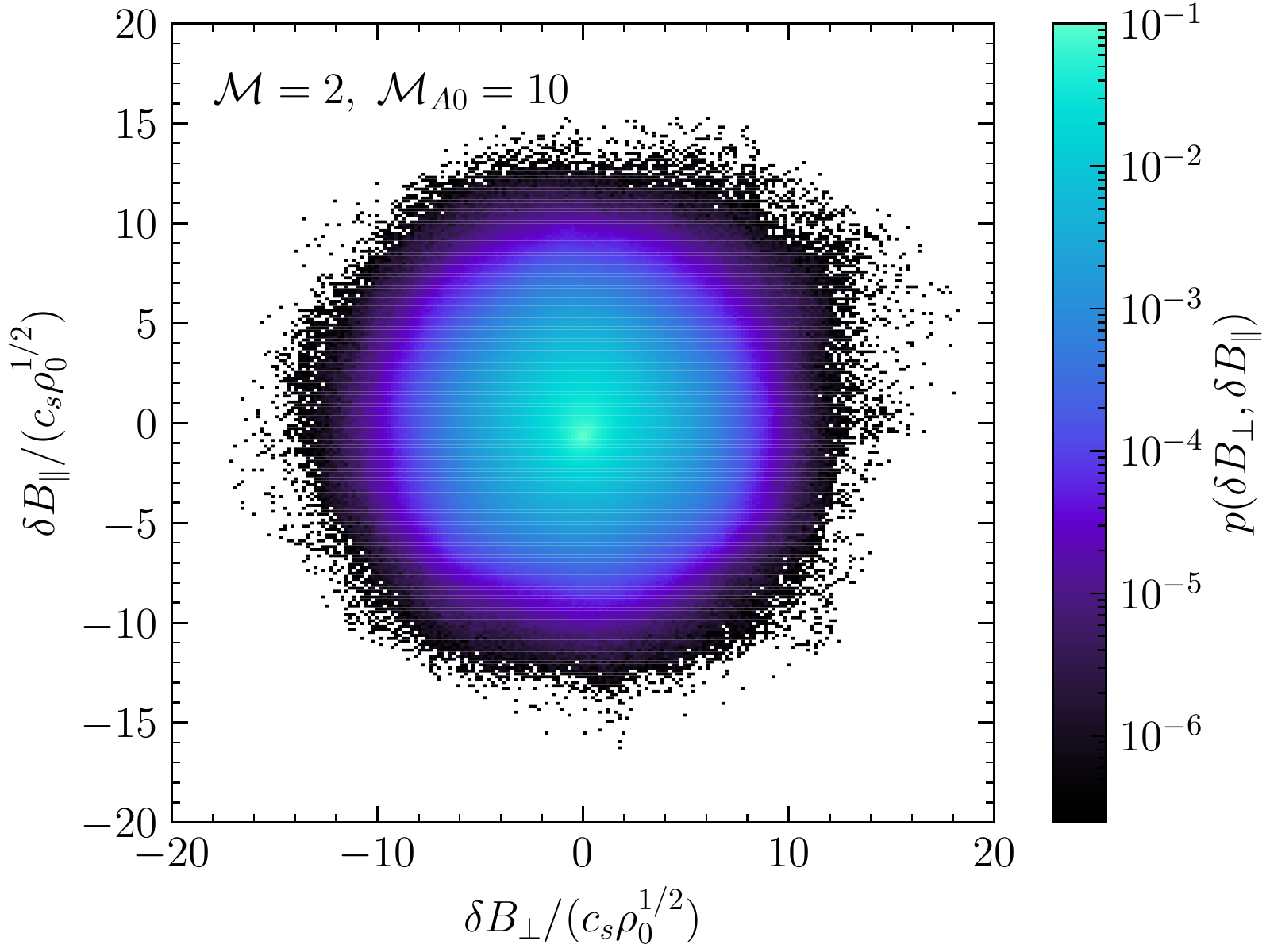}
        \caption{The same as \autoref{fig:joint_mag_pdf}, but for the \texttt{M2MA10} simulation, highlighting the global, isotropic nature of the magnetic field fluctuations in the super-Alfv\'enic regime.}
        \label{fig:m2ma10_b_pdf}
    \end{figure}
    
    In \autoref{fig:m2ma10_b_pdf} we plot the joint magnetic field fluctuation distribution for the super-Alfv\'enic simulation, \texttt{M2MA10}, to contrast the sub-Alfv\'enic case in \autoref{fig:joint_mag_pdf}. The strong anisotropy in the sub-Alfv\'enic joint PDF disappears in the super-Alfv\'enic data, and the fluctuations become spherically symmetric and hence isotropic.

\section{Averaging as a function of length scale}\label{app:length_scale_average_method}
    In \autoref{sec:consideration} we compute $\Exp{\dbBo}$ and $\Exp{\dB^2}^{1/2}$ as a function of length scale in the turbulence for the \texttt{M2MA001} and \texttt{M2MA10} simulations, as well as $\Mao$ and $\Maturb$ for the \texttt{M2MA10} simulation. To do this we pick a random coordinate $(x_1,x_2,x_3)$, in the three-dimensional simulation, and expand a set of concentric $i$ spheres, 
    \begin{align}
        \mathcal{S}_i = \left\{(x,y,z) \in \mathcal{V}_{\mathcal{L}} \;|\; (x-x_1)^2 + (y-x_2)^2 + (z-x_3)^2 =  (\ell_i/2L)^2 \right\}
    \end{align}
    over a range of diameters, $\ell_i/L \in [0,1]$. $\mathcal{S}_i$ is then our filter, and for each $\ell_i/L$ we compute the convolved field variables,
    \begin{align}
        f^*(\ell_i/L) = \int_{\V_{\mathcal{L}}} \d{f} \, \mathcal{S}_i f(x,y,z),
    \end{align}
    where $f(x,y,z)$ is either $\dbBo$, $\dB^2$, all of the components for $\Mao$ and $\Maturb$ as per our definitions in \autoref{sec:alfven_mach}, or for $\dbBovol / (2c_s^2\rho_0\pi\mathcal{M}^2)$, and $f^{*}$ is corresponding length dependent field variable. Next, we compute the volume-averages, $\Exp{f^*(\ell_i/L)}_{\V}$, where $\V$ is the volume $(4/3)\pi(\ell_i/2L)^3$, for each $\ell_i/L$. Finally we independently compute the velocity power spectra and correlation scale, $\cor{v}$, of the simulation boxes using \autoref{eq:correlation_scale}, allowing us to transform all of the length scale units into correlation scales. We show $\Exp{\dbBo(\ell_i/\cor{v})}_{\V}$ and $\Exp{\dB^2(\ell_i/\cor{v})}_{\V}$ for a representative sub-Alfv\'enic and super-Alfv\'enic simulation in \autoref{fig:scale_dependent_B}.

\bsp	
\label{lastpage}
\end{document}